# PROBING THE OUTFLOWING MULTIPHASE GAS

# ~1 kpc BELOW THE GALACTIC CENTER

Blair D. Savage[1], Tae-Sun Kim[1,2], Andrew J. Fox[3], Derck Massa[4], Rongmon Bordoloi[5], Edward. B. Jenkins[6], Nicolas Lehner[7], Joss Bland-Hawthorn[8], Felix J. Lockman[9], Svea Hernandez[10], & Bart P. Wakker[1]

ABSTRACT

Comparison of ISM absorption in the UV spectrum of LS 4825, a B1 Ib-II star d=21±5 kpc from the Sun toward $l = 1.67°$ and b = -6.63°, with ISM absorption toward an aligned foreground star at d < 7.0±1.7 kpc, allows us to isolate and study gas associated with the Milky Way nuclear wind. Spectra from the Space Telescope Imaging Spectrograph (STIS) show low ion absorption out to d < 7 kpc ( e.g., O I, C II, Mg II, Si II, Fe II, S II) only between 0 and 40 km s$^{-1}$, while absorption at d > 7 kpc, ~1 kpc below the galactic plane, is complex and spans -290 to + 94 km s$^{-1}$. The intermediate and high ions Si III, C IV, Si IV and N V show extremely strong absorption with multiple components from -283 to 107 km s$^{-1}$ implying that the ISM ~1 kpc below the galactic center has a substantial reservoir of plasma, and more gas containing C IV and N V than in the Carina OB1 association at z = 0 kpc. Abundances and physical conditions are presented for many absorption components. The high ion absorption traces cooling transition temperature plasma probably driven by the outflowing hot gas, while the extraordinary large thermal pressure, p/k ~ $10^5$ cm$^{-3}$ K$^{-1}$ in an absorption component at -114 km s$^{-1}$ probably arises from the ram pressure of the outflowing hot gas. The observations are consistent with a flow whose ionization structure in the high ions can be understood through a combination of non-equilibrium radiative cooling and turbulent mixing.

*Subject headings:* Galaxy: center - Galaxy- halo - ISM: jets and outflows - ISM: kinematics and dynamics - ISM: abundances

## 1. INTRODUCTION

The Galactic Center (GC) is the host to the Galactic black hole, Sgr A* (Schodel et al. 2002; Ghez et al. 2003) along with three extraordinary clusters of stars, Central cluster (Krabbe et al. 1991), the Arches cluster (Nagata et al. 1995), and the Quintuplet cluster (Okuda et al. 1990; Glass et al. 1990). The three clusters are massive (M ~$10^4$ -$10^5$ M$_\odot$) and contain hundreds of O and WN stars (Figer et al. 1999, 2002; Paumard et al. 2006). In addition, there are numerous isolated WR and O supergiants in the field of the GC (Mauerhan et al. 2015).

[1] Department of astronomy, University of Wisconsin, Madison, 475 North Charter Street, Madison, WI 53706, USA
[2] Observatorio Astronomio di Trieste, Via G.B. Tiepolo 11, 34143, Trieste, Italy
[3] Space Telescope Science Institute, 3700 San Martin Drive, Baltimore, MD, 21218, USA
[4] Space Science Institute, 4750 Walnut St., Suite 205, Boulder, CO 80301, USA
[5] MIT-Kavli Center for Astrophysics and Space Research, 77 Massachusetts Ave., Cambridge, MA 02139, USA
[6] Princeton University Observatory, Princeton, NJ 08544, USA
[7] Department of Physics, University of Notre Dame, Notre Dame, IN 46556, USA
[8] Institute of Astronomy, School of Physics, University of Sydney, NSW 2006, Australia
[9] Green Bank Observatory, P.O. Box 2, Green Bank, WV 24944, USA
[10] Department of Astrophysics, Radboud UniversityNijmegen, PO Box 9010 6500 GL Nijmegen, the Netherlands

The GC shows a dramatic excess of non-thermal radio, microwave, mid-IR, X-ray, and $\gamma$-ray emission extending several kpc in z-distance above and below the GC mid-plane (Sofue & Handa 1984; Finkbeiner 2004; Bland-Hawthorn & Cohen 2003; Snowden et al. 1997; Su et al. 2010).  The spectacular $\gamma$-ray emission bubbles (Fermi bubbles), discovered by the Fermi Space Telescope,  extend $\pm 50°$ in galactic latitude ($\pm 10$ kpc), $\pm 20°$  in galactic longitude (Su et al. 2010; Dobler et al. 2010; Crocker & Aharonian 2011; Ackermann et al. 2014).  Figure 1 displays the 1.5 keV X-ray image of the GC adapted from Snowden et al. (1997) by Wong (2002) along with the and $\gamma$-ray image adapted from Ackermann et al. (2014)

With a star formation rate density ~100 times larger than average in the Milky Way (Figer et al. 2004), it is reasonable to expect the GC to be the launch site for a galactic wind powered by the energy deposited by hot stellar winds and subsequent supernova explosions with additional contributions from the violent activity associated with the nuclear black hole Sgr A*. The models for the origin of the Fermi bubbles recently proposed range from those suggesting that the origin occurred at a time Sgr A* was $10^3$-$10^4$ times more active than today with its accretion flow launching a strong wind (Mu et al. 2014) to a suggestion that the Fermi bubble arises in the termination shock of a nuclear starburst driven wind powered by the GC hot stars (Lacki 2014).  Other AGN related explanations are given by Gou & Matthews (2012), Yang et al. (2012) and Zubovas et al. (2011).  Other starburst explanations are given by Crocker et al. (2015) and Carretti et al. (2013).  A study of the energetics and time scales of the interaction of the FB with the Milky Way's hot halo by Miller & Bregman (2016) suggests that the FB likely formed by a SgrA* accretion event rather than sustained nuclear star formation activity.

A study of the behavior of the highly extended distribution of H I away from the galactic plane by Lockman & McClure-Griffiths (2016) reveals a sharp transition from a thick and thin H I layer for Galactic center distances R >2.4 kpc to a thin Gaussian layer with FWHM 125 pc for R < 2.4 kpc.  For $10 < |b| < 20$ an anti-correlation between H I and $\gamma$–ray emission suggests that the boundary of the extended H I layer may coincide with the outer walls of the Fermi Bubble. Sofue (2017a) has studied the relationship of the H I hole, the expanding 3 kpc ring and the North Polar Spur and explores their possible origins in a GC explosion ~10 My ago.  In the Sofue (2017a) model illustrated in his Figure 7, the Fermi Bubble is considered to be produced by an explosive event a few My ago that is related to the innermost expanding ring in the GC described in Sofue (2017b).

Because of the large foreground extinction, there have been very few studies of the GC at ultraviolet (UV) wavelengths. This is unfortunate because the UV provides great sensitivity for the detection of gas swept up by nuclear outflows and also yields information on cooling transition temperature gas containing N V, C IV and Si IV that might directly probe gas entrained in the outflow.   While measurements close to the galactic plane are not possible because of the very large foreground extinction, it is possible to find lines of sight with galactic latitude $|b| > 5°$ away from the galactic plane to stars and to more distant AGNs that probe the plasma in the galactic halo at kiloparsec distances above and below the galactic nucleus.

We have initiated a project with the Hubble Space Telescope (HST) to probe the GC in the directions l = 330° to 360° and b = -60° to 60° with UV absorption line spectroscopy of AGNs and stars (program IDs 12936 and 13448).  In an early result from our program, Fox et al. (2015) reported UV absorption along the line of sight to PDS 456, an AGN with the lowest latitude (b = 11.2°) and longitude (l =10.4°) and smallest GC impact parameter ($\rho$ = 2.3 kpc) of any AGN in our sample.  The line of sight samples plasma in the biconical region of enhanced ROSAT 1.5 keV GC emission (Snowden et al. 1997) where the $\gamma$-ray emission associated with the northern base of the Fermi Bubble (FB) is strong (Su et al. 2010). The observations revealed two high velocity metal absorption components at $v_{LSR}$ = -235 and +250 km s$^{-1}$ suggesting the detection of the front and back-sides of an expanding biconical outflow from the GC.

In a subsequent paper Bordoloi et al. (2017) studied the UV absorption toward five lines of sight passing through the northern Fermi Bubble and found that all five exhibited blue shifted high velocity components while only 9 of 42 lines of sight outside the Fermi Bubble exhibited blue shifted high velocity components. The blue shifted Galactic standard of rest (GSR) velocities for the high velocity absorbers in directions inside the Fermi Bubble change from $v_{GSR}$ = -265 to -91 km s$^{-1}$ from Galactic latitude b = 11° to b = 50°. A simple kinematic model was developed to constrain the inferred GC outflow velocities to ~ 1000 to 1300 km s$^{-1}$ and age of the outflow to ~ 6 to 9 Myr. A future paper will expand the observations to include many additional lines of sight inside and outside the southern region of the Fermi Bubble.

When preparing the Fox et al. (2015) manuscript and considering additional HST observations we became aware of the extremely interesting HST archival spectroscopic observations for the sight line to the star LS 4825 (CD -30 15464), a luminous B1 Ib-II star lying 21±5 kpc away in the direction l = 1.67° and b = -6.63° (Ryans et al. 1997). LS 4825 was first identified by Stephenson & Sanduleak (1971) in their catalog of Luminous Stars (LS) in the Southern Milky Way. The line of sight to LS 4825 passes under the GC at a distance away from the galactic nucleus of z = -0.97 kpc. The star is at z = -2.4 kpc. In this paper we report an analysis of the ISM UV absorption toward LS 4825 along with the absorption toward the closely aligned foreground star, HD 167402, a luminous B0 II/B0.5 Ib star in the direction l = 2.26° and b = -6.39° at a distance estimated to be 7.0±1.7 kpc. Table 1 lists the properties of LS 4825 and HD 167402. The combination of the observations of LS 4825 and HD 167402 allows us to separate in the spectrum of LS 4825 the foreground absorption (d < 7.0 kpc) from the more distant absorption (d > 7.0 kpc). Figure 1 shows the positions of LS 4825, HD 167402 and PDS 456 on the ROSAT 1.5 keV view of the GC that reveals the bi-conical GC X-ray emission structures and their connection to the γ-ray emission illustrated in the lower figure.

## 2. OBSERVATIONS
### 2.1. *STIS Archival Observations*

The archival observations of LS 4825 analyzed in this paper were obtained with the Space Telescope Imaging Spectrograph (STIS) operating in the echelle E140M and E230M grating modes utilizing the 0.2″x0.06″ entrance slit and the various central set up wavelengths producing the wavelength coverage listed in Table 2. The observations were obtained for the HST GO program 8096 ("The ISM Near to and Beyond the Galactic Center", PI, F. Keenan). See Kimble et al (1998) for a discussion of the properties of STIS and its inflight performance.

We also made use of HST STIS E140M archival observations of the seven early B comparison stars listed in Table 3 in order to evaluate stellar/interstellar line blending problems for B stars and to evaluate the reliability of the classification and luminosity estimate for LS 4825 that is used to determine its distance.

The extracted and combined observations of LS 4825 and six of the comparsion stars were included in the Catalog of STIS Ultraviolet Echelle Spectra of Stars of Ayres (2010). The spectra are from the StarCAT project and were downloaded from http://casa.colorado.edu/~ayres/StarCat. The observations were extracted and combined as carefully as possible with all the details given in Ayres (2010). The spectrum of HD 52089 was obtained from the Multi-mission Archive at Space Telescope (MAST). The spectrum was extracted and combined by D. Massa.

**Table 1.  Properties of LS 4825 and the closely aligned foreground star HD 167402**

| Name | V | B-V | E(B-V) | MK | l | b | $d^a$ | z |
|------|---|-----|--------|-----|-----|-----|-----|-----|
| | | | | | ($^{\circ}$) | ($^{\circ}$) | (kpc) | (kpc) |
| LS 4825 | 11.99 | 0.05 | 0.24 | B1 Ib-II | 1.67 | -6.63 | 21±5 | -2.4 |
| HD 167402 | 8.95 | -0.01 | 0.21 | B0 II/B0.5 Ib[b] | 2.26 | -6.39 | 7.0±1.7 | -0.78 |

[a]The distance estimates are obtained by using log d = (V- 3.1E(B-V) $-M_V$+5)/ 5.   The absolute visual magnitude, $M_V$, versus spectral types and classes conversions given in Table 1 of Bowen et al. (2008) are adopted along with the intrinsic (B-V) colors from Wegner (1994).

**LS 4825**: $(B-V)_o$ = -0.19  for B1 Ib-II stars implying E(B-V) = 0.24.  For Bl Ib and B1 II stars $M_V$ = -5.95 and -5.10, respectively.  Assuming $M_V$ = -5.52  we obtain d = 22.5± 5 kpc  where most of the error is due to the uncertainty in $M_V$. Our distance estimate is so close to the original Ryans et al. (1997) estimate of 21±5 kpc.  We retain their number in this paper.

**HD 167402**: $(B-V)_o$ = -0.22 and -0.21, respectively, for B0 II and B0.5 Ib stars. We adopt $(B-V)_o$ = -0.21 for the estimate of E(B-V) = 0.20.    For a B0 II and B0.5Ib classification, $M_V$ = -5.82 and -6.05, respectively.  For the distance estimate we adopt $M_V$ = -5.93.  Note that our distance estimate of 7.0 ±1.7 kpc for HD 167402 differs from the value d = 9.0±3.2.  given in Bowen et al. (2008).   Their larger distance arises from an estimate of E(B-V) = 0.08 rather than the value 0.21 we have adopted.   The Bowen et al. (2008) values of V and E(B-V) are derived from the Tycho Starmapper (ESA 1997 Catalog) two color photometry measurements through a transformation of the Tycho colors into the Johnson B and V colors.  The value of B = 8.97 derived from the Tycho photometry is consistent with the high precision measurement of Shield et al (1983) of B = 8.94 which we adopt.  However, the Tycho value of V = 9.09 does not agree with the Shield et al. (1983) value of 8.95. The resulting 0.14 difference in V and E(B-V) difference explains the significantly larger distance estimate of Bowen et al. (2008).  We believe the problem is with the Tycho photometry because the Shield et al. (1983) B and V measurements are in good agreement with with earlier high precision photometric observations of Klare & Neckel (1977) and Dachs et al. (1982).  The smaller distance of 7.0 kpc places HD 167402  on the near side of the GC which is consistent with the analysis in this manuscript and the Na I and Ca II observations of Ryans et al. (1997).

[b] This estimate of MK for HD 167402 from Tripp et al. (1993) is based on an analysis of the UV spectrum obtained by the International Ultraviolet Explorer.

**Table 2.  STIS Archival Observations of LS 4825 from GO Program 8096**

| Grating | Slit | Central λ | Wavelength | t | Archive ID |
|---------|------|-----------|------------|---|------------|
| | ($''$) | Å | Range (Å) | (sec) | |
| E140M | 0.2x0.06 | 1425 | 1145 -1705 | 2070 | O5A001010 |
| E140M | 0.2x0.06 | 1425 | 1145-1705 | 11,780 | O5A001020 |
| E230M | 0.2x0.06 | 2707 | 2280-3115 | 2707 | O5A002010 |
| E230M | 0.2x0.06 | 1978 | 1606-2366 | 1945 | O5A002020 |
| E230M | 0.2x0.06 | 1978 | 1606-2366 | 1945 | O5A002030 |

STIS E140M observations were also obtained of the foreground comparison star HD 167402 as part of HST GO proposal 13488 ("The Closest Galactic Wind: UV Properties of the Milky Way's Nuclear Outflow", PI, A. Fox).  A 1955 sec integration was obtained using the 0.2x0.06 arc sec aperture and a central wavelength of 1425 A. These setup parameters are identical to those used for LS 4825. The resulting STIS observation (ID = OC8Y03010) was extracted using the standard CalSTIS pipeline. A combined spectrum was formed by manually combining the echelle orders, ensuring that the order edges were continuous. The wavelength coverage is from 1145 to 1705 Å.

The E140M and E230M spectral resolutions ($\lambda/\Delta\lambda$) are approximately 45,800 and 30,000, corresponding to velocity resolutions of 6.6 and 10 km s$^{-1}$, respectively. The processed spectra processed are given in heliocentric wavelengths and are estimated to have an overall wavelength uncertainty corresponding to a velocity uncertainty of ~1.4 km s$^{-1}$.

When discussing the ISM absorbers we will give wavelengths and velocities in the local standard of reference (LSR) frame. For the directions to LS 4825 and the foreground comparsion star, HD 167402, $v_{LSR}$-$v_{Helio}$ = 8.6 km s$^{-1}$.

## 2.2. Green Bank Telescope

Deep H I 21 cm spectra were taken under program GBT/14B-299 with the Green Bank Telescope(GBT) for the directions to LS 4825 and HD 167402 with an angular resolution of 9.2′ and velocity resolution of 1.2 km s$^{-1}$. Multiple scans of the sightlines were obtained with the VEGAS spectrometer in the frequency switching mode at either 3.6 or 4.0 MHz, resulting in an unconfused velocity range of approximately 800 km s$^{-1}$ about zero velocity with a channel spacing of 0.604 km s$^{-1}$. The spectra were Hanning smoothed, calibrated and corrected for stay radiation using the procedures discussed by Boothroyd et al. (2011). A fourth-order polynomial was removed from emission free portions of the final average spectra. The resulting spectrum of LS 4825 has a brightness temperature 1$\sigma$ noise of 8.5 mK in a 1.3 km s$^{-1}$ channel giving a 1$\sigma$ sensitivity to a line spanning 25 channels (32.5 km s$^{-1}$ wide) of N(H I) = 1.0x10$^{17}$ cm$^{-2}$. The corresponding 1$\sigma$ noise and 1$\sigma$ column density sensitivity for HD 167402 are 20 mK and 2.4x10$^{17}$ cm$^{-2}$, respectively for an emission line with FWHM = 32.5 km s$^{-1}$ which corresponds to b = 19.5 km s$^{-1}$. The 9.2′ angular resolution of the GBT H I observations introduces a considerable amount of antenna beam smearing for a line of sight with complex ISM structures.

## 3. THE STELLAR PROPERTIES OF LS 4825
### 3.1. Optical Observations LS 4825

LS 4825 was identified as a very distant luminous early B star as part of a program by the Queens University of Belfast hot star group to investigate the composition of hot stars near the GC. The program identified a number of normal early type stars, post-AGB stars, Be/disk stars and possible distant B type supergiants (Smartt 1996; Venn et al. 1998) in the direction of the GC. Moderate resolution optical studies of LS 4825 implied it to be a B1 Ib supergiant due to its low gravity, symmetric Balmer lines and near normal photospheric abundances although a post-AGB origin was not ruled out.

A comprehensive higher resolution optical study of LS 4825 by Ryans et al. (1997) yielded $T_{eff}$ = 23,500±1000 K from the optical Si III/Si IV transitions and log g(cm s$^{-2}$) = 3.2±0.3 from the observed Balmer line profiles compared to the Kurucz (1991) line blanketed model atmosphere codes. The atomic line data for LS 4825 confirmed the near normal stellar abundances through a differential comparison to the B1 Ib standard star HD 24398 ($\zeta$ Per).

Ryans et al. (1997) classified LS 4825 as B1 Ib but noted that the gravity was slightly higher than they derived for the standard B1 Ib comparsion star HD 24398 and suggested that LS 4825 may be slightly less luminous and likely be in the luminosity interval Ib-II. We therefore list their classification in this paper as B1 Ib-II. Using the Walborn (1972) absolute visual magnitude ($M_V$) calibration of -5.0 for a B1 II star and -6.0 for a B1 Ib star they found $M_V$= -5.5±0.5 for LS 4825. With this absolute magnitude and the Seaton (1979) reddening relation and E(B-V)= 0.28, they found that LS 4825 lies at a distance of 21±5 kpc. With galactic coordinates of l = 1.67˚ and b = -6.63˚ the line of sight traces gas under the GC. Ryans et al (1997) noted that their ISM absorption spectrum of Ca II and Na I revealed 13 absorption components ranging in LSR velocity from -206 to +93 km s$^{-1}$. In contrast the closely aligned foreground star, HD 167402 at 7.0 kpc reveals eight Ca II absorption components from -11 to 43 km s$^{-1}$ (Sembach, Danks & Savage 1993). Ryans et al. proposed it is likely the extra Na I and Ca II absorbers toward LS 4825 are tracing gas close to the GC. Comparison

### 3.2. *Ultraviolet Observations*

The UV high resolution observations of LS 4825 obtained by STIS provide independent additional information about the properties of the star and the distance estimate. The UV resonance lines of C IV, Si IV and N V are sensitive to the stellar luminosity and gravity. The same is true of the Si III λ1417 photospheric line (Massa 1989). With the new UV measurements, we can check the validity of the classification and distance estimate for LS 4825 obtained by Ryans et al. (1997).

In Figure 2 we compare the UV stellar spectrum of LS 4825 in the high ionization resonance lines of C IV, Si IV and N V with those for the seven early B stars listed in Table 3. The spectra for LS 4825 are plotted at the top of each of the high ionization panels with the spectra for the seven comparison stars ordered according to increasing line strength of the C IV wind line from top to bottom. The high ionization ISM absorption is extremely strong and broad in the spectrum of LS 4825 and relatively weak in the comparison star spectra given they are mostly relatively nearby. The C IV lines in the relatively low luminosity star HD 44743 (β CMa, B1 II-III s) show no evidence for a wind while C IV in the higher luminosity star HD 58510 (B1 Ib-II) has a strong P-Cygni profile indicating high speed outflow. The other comparison stars have C IV profiles revealing outflows with different strengths and terminal velocities. The behavior of the stellar C IV resonance line does not show a simple correspondence to the luminosity classification of the stars shown. This illustrates the difficulty of determining the spectral type and luminosity class of hot B stars. However, we note that the stars with luminosity class II to I all show evidence for high speed outflow in the C IV lines. The comparison star in Figure 2 showing the closest correspondence to the stellar spectrum for LS 4825 is BD +48 3437 which is classified B1 Iab by Morgan, Code & Whitford (1955). In Figure 3 we directly compare the STIS spectra of LS 4825 (shown as the black spectrum) and BD +48 3437 (shown as the red spectrum) from 1150 to 1650 Å. The excellent correspondence between the absorption in these two stars is evident in all the lines except for the ISM resonance lines that are much stronger in the spectrum of LS 4825. The great strength of the ISM lines in the spectrum of LS 4825 is most clearly revealed in the C IV λλ 1548, 1550 absorption doublet.

Early B supergiants with high ionization P-Cygni line profiles exhibit relatively narrow discrete absorption components (DACs) that originate in the stellar wind (Prinja et al. 2002). The spectrum of HD 99857 (B0.5 Ib) shown in Figure 2 reveals a DAC in C IV λλ1548, 1550 at approximately 1540.1 and 1542.7 Å corresponding to v = -1570 km s$^{-1}$ km s$^{-1}$. The spectrum of LS 4825 also has a C IV DAC visible in Figure 2 at approximately 1542.5 and 1545.1 Å corresponding to v = -1100 km s$^{-1}$. We do not consider that feature to arise in the interstellar medium.

The panel for Si IV in Figure 2 also displays the stellar Si III λ1417 line that Massa (1989) showed is very sensitive to the luminosity of early B stars. The different widths of this line for the stars displayed in the figure are due to the different rotational velocities of the comparison stars. We measured the logarithmic equivalent width, log(EW), LSR radial velocity, $v_{LSR}$, and the observed line width, $b_a$, for the Si III λ1417 line in each of the stars listed in Table 3 using the STIS observations displayed in Figure 2. The values are listed in Table 3. The apparent optical depth method of Savage & Sembach (1991) was used to measure the velocities and line widths as described in footnotes to the table. Figures 3c and 6 in Massa (1989) show that normal B0.5 to B2 stars with luminosity classifications III, IV and V all have Si III λ1417 line strengths, log EW(mÅ) < 2.5 while the more luminous stars classified as II, Ia , Ib or Iab (or some combination) all have log EW(mÅ) > 2.55. This luminosity sensitivity is confirmed by Prinja (1990, see Fig. 2) with a larger set of stars . The value log EW(mÅ) = 2.72±0.09 for Si III λ1419 in LS 4825 implies a luminosity ranging from II to Iab. Our simple assessment of the UV observations of the stellar lines in LS 4825 allows us to arrive at the conclusion that the luminosity of LS 4825 is greater than class III and somewhere in the range from II to Ib to Iab. We therefore agree with the classification B1 Ib-II of Ryans et al. (1997). Using slightly different values of E(B-V) and the absolute visual magnitude, $M_V$, given in the footnote to Table 1, we obtain d =22.5 ±5 kpc which agrees with the Ryans et al. (1997) distance estimate of 21±5 kpc. This places LS 4825 well beyond the GC while the closely aligned comparison star HD 167402 is at d = 7.0±1.7 kpc. For the rest of this paper we have listed the Ryans et al. (1997) distance estimate of 21±5 kpc for LS 4825.

**Table 3. The Early B Type Comparsion Stars for LS 4825[a]**

| Name | MK[b] | V[b] | B-V[b] | E(B-V)[c] | $v_{LSR}$(radial)[d] (km s$^{-1}$) | $b_a$[e] (km s$^{-1}$) | log EW(mÅ)[f] (Si III λ1417) |
|---|---|---|---|---|---|---|---|
| LS 4825 | B1 Ib-II | 11.99 | 0.05 | 0.24 | -11 | 69 | 2.72±0.09 |
| HD 44743 (β CMa) | B1 II-III | 1.97 | -0.24 | 0.00 | 3 | 37 | 2.49±0.06 |
| HD 52089(ε CMa) | B1.5 II | 1.50 | -0.21 | 0.00 | 1 | 38 | 2.59±0.05 |
| HD 151805 | B1 Ib | 9.01 | 0.00 | 0.19 | -10 | 117 | 2.59±0.09 |
| BD +48 3437 | B1 Iab | 8.73 | 0.10 | 0.29 | -17 | 54 | 2.64±0.06 |
| HD 88115 | B1.5 IIn | 8.31 | 0.05 | 0.12 | -1 | 122 | 2.80±0.04 |
| HD 99857 | B0.5 Ib | 7.45 | 0.11 | 0.33 | -21 | 97 | 2.63±0.05 |
| HD 58510 | B1 Ib-II | 7.21 | 0.05 | 0.24 | 23 | 63 | 2.65±0.04 |

[a]The comparsions stars starting with HD 44743 are ordered from top to bottom according to the increasing strength of the stellar C IV absorption which is sensitive to mass outflow. The stellar spectrum of LS 4825 most closely resembles that for the B1 Iab star BD +48 3437.

[b] The spectral types, V magnitude, and B-V are from Bowen et al. (2008) with the exception of HD 44743 and BD +48 3437 which are from Morgan Code & Whitford (1955), LS 4825 which are from Ryans et al. (1999), and HD 99857 which are from Diplas & Savage (1994).

[c] The color excess, E(B-V), listed used the values of B-V and the intrinsic colors from Wegner (1994).

[d] The LSR stellar radial velocity was determined from the Si III λ1417 line using the AOD method (Savage & Sembach 1991) with $v_{LSR} = \int v N_a(v) dv$ where $N_a(v)$ is the apparent column density per unit velocity [cm$^{-2}$ (km s$^{-1}$)$^{-1}$].

[e]The stellar line breadth was determined from the Si III λ1417 line using the AOD method where $b_a = 2(\int (v-<v>)^2 N_a(v) dv)/N_a$ with $N_a = \int N_a(v) dv$.

[f] Log of the equivalent width, EW(mÅ), of the luminosity sensitive stellar Si III λ1417 line.

## 4. THE ISM ULTRAVIOLET ABSORPTION LINE SPECTRA OF LS 4825 and HD 167402

The UV spectra of LS 4825 and HD 167402 are rich in information about the properties of the ISM extending over 21±5 and 7.0±1.7 kpc, respectively, in two closely aligned directions toward the GC. The ISM spectrum of LS 4825 is complex and stellar line blending is a complication for the analysis of the high ion absorption because of the star's spectral type (B1 Ib-II). In contrast, stellar line blending is less of a problem for HD 167402. With a spectral type of B0 II/B0.5 Ib, the strong P-Cygni stellar wind lines N V, C IV, and Si IV provide relatively smooth stellar absorption against which the high ion ISM absorption can be measured.

### 4.1. *Absorption Line Measurements*

In order to obtain measurements aimed toward determining the kinematical and the physical state of the gas in the observed ISM absorption line systems we have employed the VPFIT code Version 10 discussed in Carswell et al. (2002) and Kim et al. (2016). For each detected absorption line component we measure LSR velocity, $v_{LSR}$, the component Doppler parameter, b(km s$^{-1}$), and the column density, log N(cm$^{-2}$), and their errors assuming the observed absorption can be fitted by single or multiple Voigt profile components. We choose the minimum number of components required to provide an acceptable fit to the absorption with a reduced Chi Square, $\chi_v^2$ ~1. For each ion all the detected absorption lines are simultaneously fitted if they are not strongly contaminated by other absorbers or have significant continuum placement uncertainties. Profile fit results for each star for the high and moderate ionization absorption are given in Table 4 (for LS 4825) and Table 5 (for HD 167402). The listings are ordered by ion type according to decreasing ionization potential for the high ions from N V, C IV, Si IV and Si III. The listings give the rest vacuum wavelength of the absorption lines simultaneously fitted, the ion, $v_{LSR}$(km s$^{-1}$) ±σ, b(km s$^{-1}$) ±σ, and log N(cm$^{-2}$)± σ. Unless otherwise stated all the different ions were fitted freely, i.e. the velocities of components for a particular ion were not fixed to agree with those for other ions. The parameter fit errors derived by VPFIT allow for statistical errors, the effect of absorber component overlap errors, and the continuum placement errors. For all absorbers toward LS 4825 and HD 167402 the profile fit results are given in the appendix table ordered by the velocity of the absorption components.

The atomic parameters (vacuum wavelength in Å, oscillator strength and spontaneous transition probability in s$^{-1}$) are from the table provided by the VPFIT program. Most of these values are from Morton (2003) with a few from other sources. We note that VPFIT uses the Morton (2003) C I, C I* and C I** oscillator strengths rather than those from Jenkins & Tripp (2001). This makes no difference in the C I excitation analysis discussed in Sections 6.2.2 and 6.2.3. For the large number of Ni II lines between 1317 and 1751 Å, the oscillator strengths are from Morton (2003) except for Ni II λλ1307.132, 1317.217 which are from Jenkins & Tripp (2006), and Ni II λλ 1703.411, 1502.148 where the oscillator strengths are from Berengut et al. (2011). When fitting the absorption lines we perform a simultaneous fit to all the available absorption lines for each ion. For example, our fits to C I absorption simultaneously fit the C I λλ 1260.735, 1277.245, 1280.135, 1328.833, 1560.309, 1656.928 lines and include the effects of overlapping absorption by the many lines of C I * and C I**. For an ion with so many absorption lines such as C I, C I*, and C I** or Ni II or Fe II we do not illustrate all the fit results in the appendix.

**Table 4.  Properties of the High Ionization and Intermediate Absorption Components in the Spectrum of LS 4825 and HD 167402 Based on Profile Fits to the Absorption[a]**

| λ | Ion | $v_{LSR}$ (km s⁻¹) | b (kms⁻¹) | log N[b] | Notes[c] |
|---|-----|------|-----|-------|--------|
| LS 4825.................................................................................................. | | | | | |
| 1238, 1242 | N V | -90±1 | 35±1 | 14.21±0.01 | |
| 1238, 1242 | N V | 12±2 | 42±3 | 13.67±0.03 | |
| 1548, 1550 | C IV | -223±3 | 21±3 | 13.75±0.13 | |
| 1548, 1550 | C IV | -179±2 | 28±6 | 14.02±0.10 | |
| 1548, 1550 | C IV | -92±1 | 41±2 | >14.91±0.03 | 1 |
| 1548, 1550 | C IV | 22±1 | 33±1 | 14.49±0.01 | |
| 1548, 1550 | C IV | 107±1 | 19±1 | 13.54±0.02 | |
| 1393, 1402 | Si IV | -273±7 | 28±11 | 12.39±0.14 | |
| 1393, 1402 | Si IV | -235±1 | 8±3 | 12.34±0.13 | |
| 1393, 1402 | Si IV | -212±1 | 9±1 | 12.86±0.04 | |
| 1393, 1402 | Si IV | -183±1 | 12±3 | 12.95±0.08 | |
| 1393, 1402 | Si IV | -160±1 | 9±2 | 12.94±0.09 | |
| 1393, 1402 | Si IV | -102±1 | 29±1 | >14.22±0.02 | 1 |
| 1393, 1402 | Si IV | -47±1 | 12±1 | 13.23±0.04 | |
| 1393, 1402 | Si IV | 21±1 | 30±1 | 13.94±0.01 | |
| 1393, 1402 | Si IV | 104±1 | 14±1 | 12.78±0.04 | |
| 1206 | Si III | -283±2 | 4±2 | >12.57±0.29 | 2 |
| 1206 | Si III | -216±3 | 15±6 | >13.82±0.79 | 2 |
| 1206 | Si III | -180 to 90 | saturated | saturated | 2 |
| 1206 | Si III | 106±1 | 12±2 | >13.32±0.22 | 2 |
| HD 167402.................................................................................................. | | | | | |
| 1238, 1242 | N V | 31±1 | 42±2 | 13.73±0.02 | |
| 1238, 1242 | N V | 96±3 | 16±4 | 12.88±0.10 | |
| 1548, 1550 | C IV | 24±1 | 30±1 | 14.13±0.01 | |
| 1548, 1550 | C IV | 78±2 | 21±2 | 13.40±0.05 | |
| 1393, 1402 | Si IV | 2±1 | 11±2 | 12.79±0.18 | |
| 1393, 1402 | Si IV | 29±1 | 23±2 | 13.49±0.04 | |
| 1393, 1402 | Si IV | 72±1 | 12±2 | 12.75±0.09 | |
| 1393, 1402 | Si IV | 98±4 | 15±5 | 12.39±0.14 | |
| 1206 | Si III | -7±7 | 22±6 | 13.23±0.20 | |
| 1206 | Si III | 48±4 | 23±11 | >14.30±0.83 | 2 |
| 1206 | Si III | 100±2 | 8±2 | 12.69±0.16 | |

[a]The results listed in this table are based on profile fits to the absorption with the best continuum estimate as shown in Figure 4.

[b] The first listed column density error is the fit error produced by the VPFIT code. It includes the statistical errors and an approximate estimate of the continuum uncertainty produced by the VPFIT code. Additional continuum fit errors and the effects of line saturation are estimated through the AOD results listed in Table 5.

[c]Notes.  (1) The C IV and Si IV absorption in the very strong components near -90 to -100 km s⁻¹ is saturated.   The amount of saturation can be evaluated for the C IV and Si IV doublet lines by using the AOD results listed in Table 5. (2) These Si III absorption components are very strong and saturated and therefore listed as lower limits.

**Table 5.** **Properties of the High Ionization Absorption in the Spectrum of**
**LS 4825 and HD 167402 Based on AOD Measurements[a]**

| Ion | $[v_-, v_+]$ (km s$^{-1}$) | $\langle v(w) \rangle$ km s$^{-1}$ | $\langle v(s) \rangle$ km s$^{-1}$ | log $N_a$(w) | log $N_a$(s) | log $N_{ac}$ | Note[b] |
|---|---|---|---|---|---|---|---|
| LS 4825 .............................................................................................................................. | | | | | | | |
| N V | -150 to -30 | -92±2 | -89±1 | 14.21±0.02 | 14.21±0.02 | 14.21±0.02±0.04 | |
| C IV | -150 to -30 | -93±5 | -94±12 | 14.84±0.01 | 14.66±0.01 | 15.11±0.01±0.02±0.13 | 1, 2 |
| Si IV | -150 to -30 | -98±4 | -95±29 | 14.28±0.01 | 14.12±0.01 | 14.50±0.01±0.05±0.11 | 3, 2 |
| N V | -30 to 80 | 25±7 | 14±4 | 13.64±0.08 | 13.60±0.04 | 13.61±0.04±0.15 | |
| C IV | -30 to 80 | 20±1 | 22±7 | 14.55±0.01 | 14.46±0.01 | 14.64±0.01±0.02±0.04 | 4, 2 |
| Si IV | -30 to 80 | 24±3 | 21±8 | 14.04±0.02 | 13.98±0.02 | 14.10±0.02±0.06±0.03 | 2 |
| C IV | -250 to -150 | -190±2 | -195±2 | 14.26±0.03 | 14.24±0.03 | 14.25±0.03±0.02 | |
| Si IV | -250 to -150 | -188±3 | -187±2 | 13.47±0.04 | 13.46±0.03 | 13.46±0.03±0.09 | |
| N V | -250 to 150 | -77±6 | -70±4 | 14.37±0.03 | 14.34±0.02 | 14.35±0.02±0.08 | |
| C IV | -250 to 150 | -68±7 | -70±24 | 15.10±0.01 | 14.97±0.01 | 15.26±0.01±0.02±0.08 | 5, 2 |
| Si IV | -250 to 150 | -62±13 | -59±25 | 14.53±0.01 | 14.42±0.01 | 14.66±0.01±0.08±0.06 | 6, 2 |
| HD 167402 ...................................................................................................................... | | | | | | | |
| C IV | -40 to 110 | 32±1 | 33±1 | 14.20±0.01 | 14.19±0.01 | 14.20±0.01 | 7 |
| N V | -40 to 110 | 41±2 | 38±1 | 13.78±0.02 | 13.77±0.01 | 13.77±0.01 | 7 |
| Si IV | -40 to 110 | 35±1 | 35±1 | 13.65±0.01 | 13.64±0.01 | 13.65±0.01 | 7 |

[a] AOD results are given for the weak (w) and strong (s) absorption doublet lines of C IV, N V and Si IV integrated over the LSR velocity range from $v_-$ to $v_+$. Average LSR velocity, $\langle v \rangle$, and values of log $N_a$ are listed along with log $N_{ac}$ which is the adopted AOD column density corrected for saturation using the correction procedure derived from the weak and strong line values of log $N_a$ listed in Table 4 of Savage & Sembach (1991). The second error on the values of log $N_{ac}$ for LS 4825 are the approximate continuum placement errors determined from log $N_a$ measurements using the high and low continua illustrated in Figure 4. The C IV absorption is so strong and the continuum is relatively well determined so the continuum placement errors for C IV are small.

[b] Notes: (1) The 0.18 dex difference between the AOD column densities for the weak and strong C IV lines imply a saturation correction of 0.27 dex for the column density derived from the weaker C IV λ1550 line (see Table 4 in Savage & Sembach 1991). (2) When the saturation corrections are larger than 0.05 dex we add a third saturation correction error estimated to be 0.5 of the saturation correction. The different errors are added in quadrature when only one error is listed in the body of the paper. (3) The AOD saturation correction to the column density for the Si IV λ1403 line is 0.22 dex. (4) The AOD saturation correction to the C IV column density for the λ1550 line is 0.09 dex. (5) The AOD saturation correction to the column density Si IV column density is 0.16 dex. (6) The AOD saturation correction to the column density for the Si IV λ1403 line is 0.13 dex. (7) The nearly identical values of log $N_a$(w) and $N_a$(s) imply little or no saturation for the very well measured HD 167402 absorbers.

The stellar continuum placement for the high ions toward HD 167402 is relatively well determined because the stellar high ion lines toward this star are relatively broad and smooth. In the case of LS 4825 there is much more structure in the stellar lines and extra care was required to determine the continuum placement. For LS 4825 the continuum placement problem was greatly aided by finding that BD +48 3437 has a stellar spectrum very similar to that for LS 4825 (see Figures 2 and 3). The comparison and the final adopted continuum for LS 4825 recognized that ISM absorption also occurs toward the comparison star. Figures 4a, 4b and 4c illustrate the details of the comparison for the high ion lines of C IV λλ1548, 1550, Si IV λλ1393, 1402, and N V λλ1238, 1242 lines toward LS 4825. In addition to the stellar spectra, we show our judgement of the best continua along with higher and lower continua which are used to estimate the continuum placement uncertainties for the high ion absorption. For C IV in Figure 4a the best continuum shown as the dashed line appears well described by the C IV absorption observed toward BD +48 3437 after making allowance for the ISM C IV absorption in the spectrum of the comparison star. For Si IV $\lambda\lambda$1393, 1402 in Figure 4b we judged that the best continuum shown as the dashed line to be reasonably well matched by the stellar Si IV absorption observed toward BD +48 3437 except in the negative velocity wing of the absorption where we have lowered the continuum so that Si IV has the same negative velocity extent as the much stronger and better defined C IV absorption.

Hot stars with high ionization P-Cygni line profiles sometimes exhibit relatively narrow discrete absorption components (DACs) that originate in the stellar wind (Prinja et al. 2002). The spectrum of HD 99857 (B0.5 Ib) shown in Figure 2 reveals a DAC in C IV $\lambda\lambda$1548, 1550 at approximately 1540.1 and 1542.7 Å corresponding to v = -1570 km s$^{-1}$ km s$^{-1}$. The spectrum of LS 4825 also has a C IV DAC visible in Figure 2 at approximately 1542.5 and 1545.1 Å corresponding to v = -1100 km s$^{-1}$. We do not consider the -1100 km s$^{-1}$ feature to arise in the interstellar medium.

In the left panels of Figures 5 and 6 we illustrate the observed fluxes and the selected best continua for the high ion lines observed toward LS 4825 and HD 167402, respectively. In the right panels we illustrate the continuum normalized profiles along with the profile fits to the absorption components with the derived component properties for each ion listed in Table 4 for LS 4825 and HD 167402.

The high ionization absorption seen in the spectrum of LS 4825 is saturated over the velocity ranges from -150 to -30 and -30 to 80 km s$^{-1}$. In the case of the absorption doublets of C IV, Si IV and N V we have access to absorption lines that differ in oscillator strength by a factor of two. We therefore apply the apparent optical depth (AOD) method of Savage & Sembach (1991) to evaluate the amount of the saturation. Figure 7 displays AOD apparent column density as a function of velocity $N_a(v)$ in ions cm$^{-2}$ (km s$^{-1}$)$^{-1}$ for S II $\lambda$1250, Si IV $\lambda\lambda$1402, 1393, C IV $\lambda\lambda$1550,1548 and N V $\lambda\lambda$1242, 1238 for LS 4825 and HD 167402. The S II $\lambda$1250 $N_a(v)$ AOD profile is displayed to allow a direct comparison of a low ion $N_a(v)$ profile with those for the high ions.

The $N_a(v)$ profiles for HD 167402 for the weak and strong components of the high ionization absorption lines reveal excellent agreement. This implies that there is little or no saturation in these absorption lines and that integrations of $N_a(v)$ over different velocity ranges should give a good measure of the high ion column density over that velocity range. Table 5 lists results of integral $\int N_a(v) \, dv = N_a$ (ions cm$^{-2}$) over various velocity ranges covering the observed absorption components toward LS 4825 and HD 167402. The large high ion absorption line strength in the spectrum of LS 4825 reveals in the $N_a(v)$ plot substantial differences between the $N_a(v)$ profiles for C IV and Si IV in the velocity ranges from -130 to -40 km s$^{-1}$ and from -30 to 80 km s$^{-1}$ where the $N_a(v)$ results for the stronger components of the double absorption (Si IV $\lambda$1393 and C IV $\lambda$1550) displayed in red lie below the $N_a(v)$ values shown in black for the weaker components (Si IV $\lambda$1402 and C IV $\lambda$1548). These differences are due to substantial line saturation effects over the two velocity ranges. In the case of N V in the component between -130 and -40 km s$^{-1}$ there is no line saturation.

The difference in the values of log $N_a(w)$ for the weak and log $N_a(s)$ strong line of a high absorption doublet, $\Delta(w-s)$ can be used to determine a correction to the saturation by following the procedures recommended by Savage & Sembach (1991) and listed in their Table 4. For LS 4825 for the velocity range -30 to 80 km s$^{-1}$ Table 5 reveals that $\Delta(w-s) = 0.09$ dex for C IV and 0.06 dex for Si IV. These relatively small differences imply saturation corrections to the weak line column densities of 0.09 and 0.06 dex for C IV and Si IV. The final saturation corrected column densities of log $N_{ac}$ = 14.64 and 14.10 for C IV and Si IV should be reliable. For the stronger C IV and Si IV absorption between -150 and -30 km s$^{-1}$ the saturation corrections are larger with Δ(w-s) = 0.18 and 0.16 dex implying saturation corrections of 0.27 and 0.22 dex for C IV and Si IV according to Table 4 of Savage & Sembach (1991). The final saturation corrected column densities of log $N_{ac}$(C IV) = 15.11 and log $N_{ac}$(Si IV) = 14.50 are uncertain because of the large value of the saturation correction. The errors on the column densities listed in Table 5 include the statistical error, the continuum placement error derived by integrating $N_a(v)$ over the various velocity ranges using the best, high and low continua displayed in Figure 4, and our estimate of the saturation correction error which we take to be 0.5 times the AOD logarithmic saturation correction.

The stellar continuum placement problem is less of an issue for the low ion absorption than for the high ion absorption because the ISM absorption lines are generally narrow and the stellar lines are broad. Therefore, we do not also show the original flux plots and our adopted continua for the low ions. However, when judging the continuum for the low ion absorption toward LS 4825, occasional reference was made to the behavior of the stellar spectrum of BD +48 3437 (see Figure 3).

In the left and right panels of Figure 8 we show for selected low and high ion absorption lines the continuum normalized absorption in the direction of LS 4825 and HD 167402, respectively. The full set of low ion absorption profiles is shown in the Appendix. Tables 6 and 7 summarize the kinematics of the component structure observed in the high ionization and low ionization lines toward each star.

We note that low signal-to-noise FUSE observations of O VI λ1032, 1038 exist for LS 4825. However, complex and strong multicomponent $H_2$ absorption strongly contaminates the O VI λ1032 absorption and completely masks the O VI λ1038 absorber. Also, the stellar continuum placement in the region of O VI is very uncertain. Therefore, in this paper we have chosen to not attempt to analyze the FUSE observations. In the future, an analysis of the extremely strong and complex $H_2$ absorption may produce interesting results about $H_2$ and might possibly permit some information to be obtained about O VI λ1032 at selected velocities where the contamination is not important. FUSE observations of O VI λ1032 absorption in HD 167402 has been studied by Bowen et al. (2008) who report very strong O VI absorption in components at v = 38 and 138 km s$^{-1}$ with b = 65±2 and 62±2 km s$^{-1}$ and log N(O VI) = 14.75±0.05 and 14.65±0.05. The component at 38 km s$^{-1}$ is associated with the Si IV, C IV and N V we see toward HD 167402 listed in Table 4. The component at 138 km s$^{-1}$ is not seen in Si IV, C IV and N V and implies the existence of a hotter phase of a plasma containing O VI in front of HD 167402. The high velocity component toward HD 167402 is supported by the presence of O VI HVC toward HD 168491 (l = 5.82°, b = -6.31°, d = 6.8 kpc) with v = 153 km s$^{-1}$, b = 53±3 km s$^{-1}$ and log N(O VI) = 14.14± 0.03.

## 4.2. *H I 21 cm Emission and H I Lyman-α Absorption*

Figure 9 illustrates the H I 21 cm emission toward LS 4825 and HD 167402 observed with the GBT. Table 6 lists the H I column densities and line widths obtained from emission profile Gaussian fits to the H I GBT measurements which assume optically thin H I profiles. The intermediate and high velocity H I component fit profiles are shown as the dotted curves in

Figure 9. We do not display in Figure 9 the H I observations for v < -300 km s$^{-1}$ and v > 200 km s$^{-1}$ where there is no significant emission.

Total H I column densities are also listed based on fitting damped absorption to the H I Ly $\alpha$ profiles observed by STIS. For each star the line center of the damped profile was assumed to be the same as that observed in the weak low ionization lines of C I and O I tracing the principal absorption at low velocity.

We note that the Australia Telescope Compact Array (ATCA) H I galactic center survey with an angular resolution of 2.42′ has revealed a population of small high velocity H I clouds, loops and filaments in the directions l and b = ±5° (McClure-Griffiths et al. 2013) with |z| heights up to 700 pc which is the |z| limit of the survey. These clouds are believed to be associated with an outflow from the central star-forming region at the Galactic center. Some of the structures are just resolved by the ATCA. This implies that there could be significant angular blurring of the observations obtained by the ~4x lower angular resolution GBT for clouds associated with the GC.

A study of the behavior of the highly extended distribution of H I away from the galactic plane by Lockman & McClure-Griffiths (2016) reveals a sharp transition from a thick and thin H I layer for Galactic center distances R >2.4 kpc to a thin Gaussian layer with FWHM 125 pc for R < 2.4 kpc. For 10 < |b| < 20 an anti-correlation between H I and γ−ray emission suggests that the boundary of the extended H I layer may coincide with the outer walls of the Fermi Bubble. Sofue (2017a) has studied the relationship of the H I hole, the expanding 3 kpc ring and the North Polar Spur and explores their possible origins in a GC explosion ~10 My ago. In the Sofue (2017a) model illustrated in his Figure 7, the Fermi Bubble is considered to be produced by an explosive event a few My ago that is related to the innermost expanding ring in the GC described in Sofue (2017b).

**Table 6. H I Toward LS 4825 and HD 167402[a]**

| Object | species | $v_{LSR}$ (km s$^{-1}$) | b (km s$^{-1}$) | log N(X) | Note |
|---|---|---|---|---|---|
| LS 4825 | H I Ly $\alpha$ | = 5 | 25.0±4.8 | 21.20±0.04 | 1 |
| LS 4825 | H I 21 cm | -104.9±0.6 | 13.6±0.8 | 18.28±0.02 | 2, 3 |
| LS 4825 | H I 21 cm | -88.6±0.2 | 7.9±0.4 | 18.38±0.03 | 2 |
| LS 4825 | H I 21 cm | -70.3±0.2 | 29.1±0.2 | 19.43±0.01 | 2 |
| LS 4825 | H I 21 cm | -18.5±0.3 | 7.7±0.2 | 19.34±0.02 | 2 |
| LS 4825 | H I 21 cm | -14.3±0.1 | 3.6±0.1 | 19.25±0.02 | 2, 4 |
| LS 4825 | H I 21 cm | 4.5±0.1 | 12.4±0.1 | 20.67±0.01 | 2 |
| LS 4825 | H I 21 cm | 6.3±0.1 | 4.3±0.1 | 20.75±0.01 | 2 |
| LS 4825 | H I 21 cm | 8.8±0.1 | 1.2±0.1 | 19.61±0.01 | 2, 4 |
| LS 4825 | H I 21 cm | 13.5±0.1 | 1.6±0.1 | 19.29±0.01 | 2, 4 |
| LS 4825 | H I 21 cm | 27.6±0.1 | 7.8±0.1 | 19.35±0.02 | 2 |
| LS 4825 | H I 21 cm | 40.1±0.4 | 13.4±0.3 | 19.55±0.02 | 2 |
| LS 4825 | H I 21 cm | 89.6±1.1 | 29.9±1.6 | 18.75±0.02 | 2 |
| LS 4825 | H I 21 cm | 98.3±0.6 | 4.0±1.0 | 17.46±0.13 | 2 |
| LS 4825 | H I 21 cm | -20 to 40 | … | 21.07± 0.01 | 5 |
| | | | | | |
| HD 167402 | H I Ly $\alpha$ | = 5 | 20.9±6.0 | 21.15±0.03 | 1 |
| HD 167402 | H I 21 cm | -73.7±4.1 | 27.8±3.7 | 18.95±0.08 | 2 |
| HD 167402 | H I 21 cm | -51.5±0.4 | 11.3±1.0 | 18.79±0.09 | 2 |
| HD 167402 | H I 21 cm | -14.7±0.1 | 2.9±0.1 | 19.23±0.01 | 2 |
| HD 167402 | H I 21 cm | 4.4±0.1 | 3.0±0.1 | 20.11±0.02 | 2 |
| HD 167402 | H I 21 cm | 5.6±0.1 | 17.7±0.2 | 20.51±0.01 | 2 |
| HD 167402 | H I 21 cm | 6.0±0.1 | 6.8±0.1 | 20.52±0.01 | 2 |
| HD 167402 | H I 21 cm | 6.8±0.1 | 1.0±0.1 | 19.22±0.01 | 2, 4 |
| HD 167402 | H I 21 cm | 9.1±0.1 | 2.1±0.1 | 20.19±0.01 | 2, 4 |
| HD 167402 | H I 21 cm | 13.3±0.1 | 2.2±0.1 | 18.86±0.07 | 2, 4 |
| HD 167402 | H I 21 cm | 25.1±0.5 | 35.4±0.2 | 20.42±0.01 | 2 |
| HD 167402 | H I 21 cm | 40.1±0.1 | 5.9±0.1 | 19.27±0.01 | 2 |
| HD 167402 | H I 21 cm | -14 to 41 | … | 21.09±0.01 | 6 |

[a] In velocity ranges where there is no H I 21 cm emission evident in Figure 9, the 3σ upper limits to log N(H I) are approximately 17.5 and 17.9 dex for LS 4825 and HD 167402, respectively, for H I emission with FWHM = 32.5 km s$^{-1}$ which corresponds to b = 19.5 km s$^{-1}$.

Notes:

(1) from the H I damped Ly $\alpha$ line assuming the absorption is centered on the velocity of the weak low ionization lines of O I and C I at v = 5 km s$^{-1}$ toward each star.

(2) From Gaussian profile fits to the 21 cm emission assuming no self-absorption.

(3) If the H I emission wing from -105 to -130 km s$^{-1}$ is interpreted as belonging to a H I emission component near -105 km s$^{-1}$ we obtain the listed fit parameters by only fitting an assumed symmetric Gaussian component over the velocity range from -105 to -140 km s$^{-1}$. This H I component may be associated with the many UV absorption lines from -90 to -114 km s$^{-1}$.

(4) These very narrow components implying very cold H I were required to fit the extremely strong core of the H I emission and may be introduced by the simplistic assumption that the emission components have Gaussian line profiles. H I self-absorption could also be contributing to the problem.

(5) log [sum of values of N(H I)] over the 21 cm emission components with v between -20 and 41 km s$^{-1}$. The 0.13 dex difference between the Ly $\alpha$ value of log N(H I) and the 21 cm value may be due to H I 21 cm self-absorption and/or 21 cm beam smearing effects.

(6) log [sum of values of N(H I)] over components with v between -16 and 41 km s$^{-1}$. The 0.06 dex difference between the Ly $\alpha$ value of log N(H I) and the 21 cm value may be due to H I 21 cm self-absorption and/or 21 cm beam smearing effects.

**Table 7. A Summary of the ISM Absorption Components in the Spectrum of LS 4825**

| ~$v_{LSR}$[a] (km s$^{-1}$) | Ions detected | Comment |
|---|---|---|
| .........................**Low ions**......................................................................................... | | |
| -290 to -230 | C II, Mg II | weak absorption components |
| --212 to -205 | O I, C II, Mg II, Si II, Fe II, Al II | C II is saturated |
| -190±1 | C II | |
| -164 to -149 | O I, N I, C II, Mg II, Si II, Fe II, Al II,  Mg I | C II, Mg II, Al II  & O I are saturated |
| -114 to -96 | C I, CI*, C I**, C II, C II*, Mg II, Si II, Fe II, S II, Al II, Mn II, O I, Mg I, Na I, Ca II | C II, C II*,  Mg II & O I  are saturated |
| -78±2 | H I (21 cm), C I, C I*, C I**, C II, C II*, Mg II, Si II, Fe II,  S II, Al II, Mn II, O I , Mg I, Na I, Ca II | C II, C II*, Mg II, Si II, & O I are saturated |
| -62 to -45 | C II, C II*, Mg II, Si II, Fe II, S II, O I, Mg I, Ca II | C II, C II*, Mg II, Si II  & O I are saturated |
| 4±3 | H I, O I,  C I, C I*, C I**, C II, C II*, Mg II, Si II, Fe II, S II, Al II, Ni II, Mn II, Zn II, Ge II, Cu II,  Mg I, S I, Cl I | C II, C II*, Si II, & Al II are saturated, H I has damping wings |
| 34±5 | C II , C II*,  Mg II, Si II, Fe II, Al II, Ni II, Mn II, Zn II,  Cu II, O I, Mg I, S I | C II, C II*, Mg II, Si II  & Al II are saturated |
| 92±2 | C II, C II*, Mg II, Si II, Fe II, O I, Mg I, Ca II | O I, C II  are saturated |
| ....................**molecules**........................................................................................... | | |
| 5±1 | CO J= 0, 1 | |
| ....................**Intermediate and High ions**[b]....................................................................... | | |
| -283±2 | Si III (4) | Si III is saturated |
| -273±7 | Si IV (28) | |
| -235 to -211 | C IV(21), Si IV(8), Si IV(9), Si III (15) | Si III is saturated |
| -183  to -160 | C IV(28), Si IV (12), Si IV (9),  Si III | Si III is saturated |
| -102 to -90 | C IV(41), Si IV (29), N V(35), Si III | C IV, Si IV  & Si III are saturated |
| -47±1 | Si IV(12), Si III | Si III is saturated |
| 12±2 | N V(42) | Uncertain continuum |
| 21±1 | C IV(33), Si IV (23), Si III | C IV & Si III are  saturated |
| 106±2 | C IV (19), Si IV(14), Si III (12) | Si III is saturated |

[a] The approximate LSR velocities of the absorption components are listed.  For the high ions the velocity range of the different ionic absorption components are listed.

[b] For each high and intermediate ion component we list the b value of the absorption in parenthesis following the ion type. Si IV has a more complex and narrower component structure than C IV.

**Table 8. A Summary of the ISM Absorption Components in the Spectrum of HD 167402**

| ~$v_{LSR}$[a] (km s$^{-1}$) | Ions detected | comment |
|---|---|---|
| ....................**Low ions**............................................................................................. | | |
| 5±2 | H I, O I, C I, C I*, C I**, C II, C II*, Mg II, Si II, Fe II, S II, Mn II, Ni II, Cl I,  Cu II Ge II, Ga II | C II, Si II, S II  are saturated, H I has damping wings. |
| 38 to 47 | O I, C II, C II* Mg II, S II, N I, Si II, Fe II, Al II, Ni II, | O I, C II, Mg II, Si II, & Al II are saturated |
| ....................**Molecules**........................................................................................... | | |
| 6±1 | CO J = 0, 1 | |
| ....................**Intermediate and High Ions**[b]...................................................................... | | |
| -7±7 | Si III (22) | |
| 2±1 | Si IV (11) | |
| 24 to 31 | C IV(30), Si IV(23),  N V(42) | |
| 48±4 | Si III (23) | Si III is saturated |
| 72 to 78 | C IV(21), Si IV(12) | |
| 96 to 100 | N V(16), Si IV(15), Si III(8) | |

[a] The approximate LSR velocities of the absorption components are listed.  For the low ions, the values are mostly based on the well observed weak absorption  components.  For the high ions the velocity range of the different ionic absorption components are listed.

[b] For each high and intermediate ion component we list the b value of the absorption in parenthesis following the ion type. Si IV has a more complex and narrower component structure than C IV.

## 5. THE ISM ABSORPTION TOWARD LS 4825 vs HD 167402

Tables 7 and 8 provide a summary of the velocity structure of the absorption toward LS 4825 and HD 167402. The absorption toward LS 4825 at d = 21±5 kpc extends over the velocity range from -290 to 107 km s$^{-1}$ and is kinematically very complex with 9 components observed in the low ionization lines and 8 in the high ionization lines. In contrast the absorption toward HD 167402 at d = 7.0±1.7 kpc extends from 4 to 100 km s$^{-1}$ and is kinematically relatively simple. Two absorption components are seen in the low ionization lines and 3 in the high ionization lines.

**Table 9. Na I and Ca II Absorption Toward LS 4825 at d = 21 kpc
and Foreground Comparison stars at d < 7.0 kpc**

| Name | MK | l | b | $\Delta^a$ | d | z | Na I$^b$ | | Ca II$^b$ | |
|------|-----|------|------|------|------|------|------|------|------|------|
| | | (°) | (°) | (°) | (kpc) | (kpc) | #comp$^c$ | $v_{LSR}$ range$^d$ | #comp$^c$ | $v_{LSR}$ range$^d$ |
| LS 4825 | B1 Ib-II | 1.67 | -6.63 | … | 21 | -2.4 | 9 | -112 to **8** to 47 | 9 | -206 to **6** to 93 |
| HD 167402 | B0 II/B0.5 Ib | 2.26 | -6.39 | 0.637 | 7.0 | -0.78 | 7 | -11 to **7** to 41 | 8 | -12 to **-3** to 43 |
| LS 4784 | B2 IV | 1.70 | -6.12 | 0.511 | 4.5 | -0.48 | … | … | 4 | -15 to **0** to 50 |
| HD 164019 | O9.5 IVp | 1.91 | -2.62 | 4.017 | 3.8 | -0.17 | … | … | 7 | -29 to **6** to 30 |
| HD 165582 | B1 Ib | 357.5 | -6.96 | 4.138 | 5.2 | -0.63 | 5 | -24 to **-1** to 35 | 5 | -54 to **1** to 21 |
| HD 168941 | O9.5 II-III | 5.82 | -6.31 | 4.162 | 5.8 | -0.64 | 4 | -2 to **7** to 72 | 6 | -48 to **8** to 71 |

$^a$ Angular distance in degrees of the star from LS 4825.

$^b$Na I and Ca II component results for LS 4825 and LS 4782 are from Ryans et al. (1997). Results for the other stars are from Sembach et al. (1993). The total velocity spread in the absorption toward LS 4825 at d = 21 kpc is much larger than observed for any of the five foreground stars with distances ranging from 3.8 to 7.0 kpc.

$^c$The number of ISM Na I and Ca II components is listed. The number is dependent on the resolution and S/N of the observations. The measurements from Sembach et al. (1993) have high S/N and a spectral resolution of 5 km s$^{-1}$. The LS 4825 and LS 4784 measurement from Ryan et al. (1993) have good S/N but poorer spectral resolution of ~ 12 km s$^{-1}$.

$^d$ The range of LSR velocities of the Na I and Ca II components is listed along with the velocity of the strongest component in the bold font.

The difference between the absorption toward LS 4825 and HD 167402 is undoubtedly due to the fact that the 21±5 kpc line of sight to LS 4825 extends under the GC while that toward HD 167402 only reaches d = 7.0±1.7 kpc. Although there can be substantial differences in ISM absorption toward closely aligned stars, support for the view that HD 167402 is providing a representative view of the foreground ISM absorption is provided by Na I and Ca II observations of other stars aligned with LS 4825. In Table 9 we compare the Na I and Ca II absorption component velocities toward LS 4825 and five foreground stars with d < 7.0 kpc including HD 167402. The Na I and Ca II observations of HD 167402, HD 164019, HD 165582 and HD 168941 are from Sembach et al. (1997) while the values for LS 4784 are inferred from Figure 4 of Ryans et al. (1997). The two foreground stars most closely aligned with LS 4825 including HD 167402 and LS 4784 at d = 4.5 kpc have a very similar Ca II component structure with components spanning the velocity range from approximately -15 to 50 km s$^{-1}$ with the strongest component near 0 km s$^{-1}$. A similar behavior is found for Ca II in the other three comparison stars that are less well aligned with LS 4825 although the velocity range of the absorption is somewhat broader. From Table 9 we conclude that HD 167402 at d = 7.0 kpc is probably providing a representative view of the properties of the ISM along the general line of sight to LS 4825 up to a distance of 7.0 kpc. We note the errors on the distances for the stars listed in Table 9 are large. Hopefully better distance estimates will become available after several more years of observations with GAIA.

The lower velocity Na I and Ca II components toward LS 4825 and the other stars listed in Table 9 are therefore likely produced in various structures with d < 7.0 kpc. The strongest components near v = 5 km s$^{-1}$ are probably associated with the outer boundaries of the extensive cold H I cloud first studied by Riegel & Crutcher (1972) in H I 21 cm self-absorption that extends from l ∼ 245° to 25° and b ∼ ±6°. There have been many subsequent studies of the distance and physical conditions in this cloud using optical and UV absorption line observations of hot stars (Bates et al. 1995; Kemp et al. 1996; Montgomery et al. 1995) which places the cloud at d < 0.2 kpc. That distance is consistent with the 3D maps of the distribution of Na I absorption toward stars with d < 300 pc of Welsh et al. (2011) which reveal for l ∼ 2° and b ∼ -5° that strong absorption by Na I occurs at d ∼ 120 pc and extends to ∼ 180 pc.

Near the end of our investigation we became aware of the compact planetary nebula SwSt 1 with its WC9 central star HD 167362 in the direction l = 1.5906° and b = -6.7167° at an uncertain estimated distance of ∼ 2 kpc (de Marco et al. 2001). SwSt1 lies 0.1205° (= 7.23′) from LS 4825. HST images of de Marco et al. (2001) reveal a very small diameter for the nebular Hβ emission of (1.3x0.9)±0.2 arc sec with nebular [O III] λλ4959, 5007 being slightly smaller. The HST STIS E140M and FUSE spectrum of the central star analyzed by Sterling et al. (2005) and shown in their Figure 2 reveals interstellar absorption near v$_{LSR}$ = 6 and 38 km s$^{-1}$ in S II, Fe II and Si II. Nebular absorption is also found in S II, S III*, S III**, O I*, O I**, Fe II, Fe III, Fe III*, Fe III**, Fe III***, and Ge III with v$_{LSR}$ ranging from -38 to -32 km s$^{-1}$. The STIS E140M spectrum of HD 167362 analyzed by us reveals a strong P-Cygni C IV profile with a maximum wind velocity of -1760 km s$^{-1}$. Nebular and ISM C IV λ1550.77 Å absorption are seen in this profile with moderate strength absorption components near v$_{LSR}$ = -34 and 15 km s$^{-1}$. The asymmetric ISM C IV absorption centered near 15 km s$^{-1}$ extends from -9 to 68 km s$^{-1}$ and has a strength and shape similar to the C IV λ1550.77 absorption seen toward HD 167402 and is probably not associated with the ejecta from HD 167362. There is no evidence in the HD 167362 STIS spectrum for additional ISM absorption with v$_{LSR}$ < -45 km s$^{-1}$ or v$_{LSR}$ > 75 km s$^{-1}$. Therefore, it appears that intermediate or high velocity gas ejected from SwSt 1 either recently or in its past history is not affecting the absorption along the line of sight to LS 4825 or to HD 167402.

Various other structures at low z that the sight line to LS 4825 passes under include the Sagittarius spiral arm at d ∼ 1.5 kpc, Scutum arm at d ∼ 4 kpc, the expanding 3 kpc arm (Oort 1977) at d ∼ 6 kpc which has v ∼ -53 km s$^{-1}$ in the direction l = -2 and b = 0 (Bania 1980), and the galactic worm candidate GW 0.5 -5.9 (see Fig. 2 in Koo et al. 1992) with an estimated distance of 7 kpc and an H I velocity range from -1 to 26 km s$^{-1}$. An extensive discussion of the properties of the ISM UV absorption toward HD 167402 and three other galactic center stars at distances of ∼ 6 kpc and galactic longitude ranging from 2.3° to 5.8° and latitude ranging from -6.4° to -12.5° is found in Tripp et al. (1993). These found that three of the four stars (not including HD 167402) exhibited UV absorption at negative velocities over the range from -40 to -100 km s$^{-1}$ forbidden by galactic rotation and attributed the absorption to the possible detection of the expanding 3 kpc arm at large distances from the galactic plane. Figure 12 in Tripp et al (1992) shows the expected velocities of absorbing gas along the line of sight to each star as a function of distance assuming co-rotation of disk and halo gas and the galactic rotation curve of Clemens (1985). For the direction to HD 167402, the expected velocities range from 0 to only 27 km s$^{-1}$ at d = 7.0 kpc. The galactic rotation effects are small for a line of sight with l

= 2.3 for d < 7 kpc and very rapidly increase to $v_{max}$ ~ 240 km s$^{-1}$ over the next 1.5 kpc. However, the co-rotation assumption will certainly break down within ~1.5 kpc of the center of the galaxy.

In a related study, Danly et al. (1992) compared the relative velocity behavior of H I 21 cm emission and UV absorption toward many stars in the galactic halo including stars toward the galactic center. Toward HD 167402, Danly et al. found a substantial amount of H I emission from -35 to -115 km s$^{-1}$ that was not seen in the UV absorption line observations of the International Ultraviolet Explorer. The H I column density in this negative velocity gas has log N(H I) = 19.57. Danly et al. believed the negative velocity gas to be possibly associated with the expanding 3 kpc arm but the lack of detection of it in absorption would place the arm beyond HD 167402 with d = 7±1.7 kpc.

Close to the Galactic plane near the GC Burton & Liszt (1978) detected H I emission from a structure they attributed to a ±100 pc thick H I disk tilted 22° from the Galactic plane with a radius of 1.5 kpc. Projected on the sky, the disk extends from (l, b) = 8°, -3° to 352°, 3°. In addition to rotation, the disk has an expansion velocity of 170 km s$^{-1}$ (Burton & Liszt 1993). At the longitude of LS 4815 with l = 1.67° the structure extends to b ~ -3° and has velocities of ~ -100 km s$^{-1}$. If the disk has a high |z| extension it could be detected in absorption toward LS 4825 with b = -6.63°. Its presence indicates there are considerable non-circular motions associated with the gas near the GC.

A simple summary of the implications of the absorption line kinematical comparison between LS 4825 and HD 167402 for the line of sight to LS 4825 follows:
1. For v = -30 to -300 km s$^{-1}$, 100% of the low and high ion absorption toward LS 4825 occurs at d > 7.0 kpc.
2. For v = -30 to 50 km s$^{-1}$, ~10 to 20% of the low ionization absorption toward LS 4825 is produced at d > 7.0 kpc.
3. For v = -30 to 80 km s$^{-1}$, ~ 70% of the Si IV and C IV absorption toward LS 4825 occurs at d > 7.0 kpc.
4. The low ionization component toward LS 4825 at v ~ 92 km s$^{-1}$ occurs at d > 7.0 kpc.
5. The high ionization component toward LS 4825 at v ~ 104 to 107 km s$^{-1}$ occurs at d > 7.0 kpc.

Therefore, absorption beyond 7.0 kpc toward LS 4825 occurs over the full velocity range from -300 to 107 km s$^{-1}$ although there is contamination from closer-by gas for the velocity range -30 to 70 km s$^{-1}$. The actual position of the gas beyond d = 7.0 kpc with z < - 0.7 kpc could be anywhere up to the 21±5 kpc distance to LS 4825 with z = -2.4 kpc. The galactic exponential scale heights of C IV and Si IV have been estimated to be ~ 3 kpc (Savage & Wakker 2009). Therefore, C IV and Si IV absorption are expected to occur out to the 21 kpc distance of LS 4825 with z = - 2.4 kpc. While deviations from co-rotation of disk and halo gas will likely be strong over the center of the galaxy from approximately d = 7 to 10 kpc, it is likely that the generally distributed low halo gas for d > 10 kpc in this direction will approximately co-rotate with disk gas. Applying the galactic rotation curve discussed in Tripp et al (1993) for the LS 4825 line of sight implies a velocity for the gas ranging from + 40 to -2 km s$^{-1}$ for d ranging from 10 to 21 kpc. Very distant highly ionized halo gas toward LS 4825 with d from ~11 to 21 kpc will therefore likely blend directly with highly ionized gas in the lower halo over the first 7 kpc along the line of sight. This distant halo gas could explain the excess C IV and Si IV from 0 to 50 km s$^{-1}$ toward LS 4825 compared to HD 167402. However, the more likely location of the gas with -300 <v < -30 km s$^{-1}$ is somewhere between d = 7.0 kpc and 11 kpc corresponding to z = -0.78 to

-1.3 kpc which would place the gas in the region strongly influenced by events occurring in the central region of the galaxy.

## 6. ISM ABSORPTION IN THE NUCLEAR REGION OF THE MILKY WAY

In our previous discussions of the spectrum of LS 4825, we infer that the high and low ion gas with v ranging from -290 to -30 km s$^{-1}$ along with the low ion component at 92 km s$^{-1}$ and the high ion component at 107 km s$^{-1}$ occur beyond 7.0 kpc. Over the velocity range from -30 to 80 km s$^{-1}$ ~ 70% of the high ion absorption occurs beyond 7.0 kpc. We believe that much of the absorption beyond d = 7.0 kpc occurs at ~ -0.7 to -1.5 kpc below the nuclear region of the galaxy.

### 6.1. *The High Ionization Absorption*

Most of the extremely strong high ionization absorption observed in the spectrum of LS 4825 occurs beyond the foreground star HD 167402. It therefore is interesting to compare the integrated properties of that absorption with other lines of sight in the Milky Way disk and halo with the largest observed high ion column densities. In Table 10 we list the total N V, C IV and Si IV column densities and the velocity range of the absorption for LS 4825 and other galactic lines of sight with the largest observed high ion column densities. The other lines of sight include the galactic center AGN PDS 456, the foreground comparison star HD 167402, three distant galactic center stars (HD 156359, HD 165322, and HD 177989), three AGNs (3C 273, Mrk 509, and NGC5548), and five stars in the Carina OB1 association which sample gas in the disk associated with the prominent H II region containing ~ 15 O-type stars. The observations for these lines of sight were obtained from the literature referenced at the bottom of Table 10 to represent those cases where the C IV and Si IV column densities were the largest observed in general ISM of the Milky Way. Objects with extremely large circumstellar high ion absorption are not included in Table 10. The total C IV column density for LS 4825 exceeds those for all the objects listed in Table 10 by more than 0.44 dex. The line of sight to LS 4825 also stands out as having an extremely large velocity range in the high ion absorption from -223 to 107 km s$^{-1}$. The only comparable object is PDS 456, which also passes near the galactic center. Table 10 provides strong evidence about the very unusual nature of the line of sight to LS 4825. The observations are clearly strongly affected by having a line of sight that passes only ~ 1 kpc below the galactic center. It is interesting that the C IV absorption along the path to LS 4825 is stronger than for the paths to the disk O stars in the Carina OB1 association. The ISM at ~ 1 kpc below the galactic center has more gas containing C IV and N V than the gas probed into the Carina OB1 association at z ~ 0 kpc.

Table 10 also lists log [N(C IV)$_{total}$/N(Si IV)$_{toal}$] for LS 4825 and the other lines of sight with the largest observed high ion column densities. Since all the lines of sight have complex C IV and Si IV profiles with variations in log [N(C IV)/N(Si IV)] from component to component the derived ratio for the total column density only gives a broad impression of the relative behavior of C IV and Si IV. The value of log [N(C IV)$_{total}$/N(Si IV)$_{toal}$] for LS 4825 of 0.60±0.13 lies close to the values for all the objects listed in Table 9 except for the lines of sight through the Carina OB1 association that have an excess of Si IV leading to log [N(C IV)$_{total}$/N(Si IV)$_{toal}$] < 0.26. The Si IV excess in Carina is explained by the local production in the nebula of Si IV by the ionizing UV radiation from the association O stars (Lehner et al. 2011).

**Table 10. Total High Ion Column Densities and the Absorption Velocity Range for LS 4825
and Other Galactic and Extragalactic Lines of Sight**

| object | l (°) | b (°) | d or z (kpc) | log N(N V) | log N(C IV)_total | log N(Si IV)_total | v range (C IV) (km s⁻¹) | Log N(C IV)/N(Si IV) | Ref. |
|---|---|---|---|---|---|---|---|---|---|
| LS 4825 | 1.67 | -6.63 | 22±5 | 14.35±0.08 | 15.26±0.08 | 14.66±0.10 | -223 to 107 | 0.60±0.13 | 1 |
| HD 167402 | 2.26 | -6.39 | 7.0±1.7 | 13.77±0.01 | 14.20±0.01 | 13.65±0.01 | -20 to 100 | 0.55±0.01 | 1 |
| PDS 456 | 10.39 | 11.16 | 0.184 | 14.09±0.05 | 14.79±0.05 | 14.14±0.05 | -235 to 130 | 0.65±0.07 | 2 |
| HD 156359 | 328.68 | -14.52 | 11.5 | 14.09±0.15 | 14.77±0.05 | 14.28±0.15 | -140 to 40[a] | 0.51±0.16 | 3 |
| HD 163522 | 349.57 | -9.09 | 9.4 | 14.00±0.09 | 14.55±0.02 | 14.00±0.04 | -80 to 40[a] | 0.55±0.05 | 3 |
| HD 177989 | 17.81 | -11.88 | 4.9 | 13.49±0.03 | 14.43±0.01 | 13.82±0.01 | -60 to 70[a] | 0.61±0.02 | 4 |
| 3C 273 | 289.95 | 64.36 | 0.1731 | 13.94±0.02 | 14.51±0.02 | 13.85±0.02 | -95 to 95[a] | 0.66±0.03 | 4 |
| Mrk 509 | 35.97 | -29.86 | 0.0350 | 13.81±0.04 | 14.51±0.03 | 13.89±0.02 | -80 to 110 | 0.62±0.04 | 4 |
| NGC 5548 | 31.96 | 70.50 | 0.0163 | 13.63±0.03 | 14.46±0.03 | 13.81±0.02 | -100 to 20 | 0.65±0.04 | 4 |
| CPD 59 2603 | 287.59 | -0.69 | 3.5 | 13.93±0.03 | 14.51±0.01 | >14.25 | -159 to 20 | < 0.26 | 5 |
| HD 93205 | 287.57 | -0.71 | 3.3 | 13.20±0.07 | 14.42±0.01 | >14.19 | -107 to 87 | < 0.23 | 5 |
| HDE 303305 | 287.59 | -0.61 | 3.8 | 13.75±0.04 | >14.63 | >14.31 | -95 to 54 | … | 5 |
| HD 93222 | 287.74 | -1.02 | 3.6 | 12.90±0.15 | 14.41±0.01 | >14.24 | -111 to -14 | <0.17 | 5 |
| HD 93206 | 287.67 | -0.94 | 2.6 | … | 14.56±0.01 | >14.56 | -96 to 14 | <0.00 | 5 |

[a]Component velocities were not listed for these observations. We instead list the approximate full observed velocity range where the absorption depth of C IV exceeds ~20% of the continuum.
References: (1) From Table 5 in this paper. (2) The galactic center AGN from Fox et al. (2015). (3). Three distant galactic center stars studied b
Sembach & Savage (1992) with the largest high ion column densities. (4) Three extragalactic lines of sight in Wakker et al. (2012) with largest
high ion column densities. (5) Five galactic lines of sight listed from Lehner et al. (2011) with the largest high ion column density probing disk
gas in the Carina OB1 association. The Si IV absorption in the association is strongly saturated in all cases. The Si IV is most likely produced b
ionizing radiation from the association O stars.

The ratio of high ion column densities provides insights into the ionization mechanisms operating to produce the high ions. Wakker et al. (2012) provide a comprehensive overview of the many different ionization mechanisms in their Section 5. Figure 10 displays an important diagnostic plot from their paper with the addition of the result for the LS 4825 absorber integrated over the velocity ranges from –250 to 150 km s⁻¹ listed in Table 5. The value includes the effects of line saturation. The figure shows log [N(Si IV)/N(C IV)] vs log [N(C IV)/N(N V)] for many ionization mechanism models along with observational results for the Milky Way halo (Wakker et al. 2012) and the Milky Way disk (Lehner et al. 2011).

The ionization processes denoted in Figure 10 include: Collisional ionization equilibrium (CIE), shock heating, conductive interfaces, thick disk supernova heating, static cooling with non-equilibrium ionization, a non-equilibrium ionization cooling flow, and turbulent mixing layers (TMLs). These models in some cases include the effects of ionizing radiation produced by the modeled structures but do not include the effects of ionizing radiation from the extragalactic background or from the nuclear region of the galaxy. The full details of all the models are discussed in Wakker et al. (2012) and the model sources are listed in the caption to Figure 10.

The LS 4825 observation displayed in Figure 10 with the black oval falls in the region of the plot occupied by the Milky Way halo plasma observations of Wakker et al. (2012) toward AGNs shown as the black circles and black limit triangles. The ionization models that come closest to the LS 4825 and the Milky Way halo gas observations are the non-equilibrium cooling flow model of Benjamin and Shapiro (shown as the thick blue line and described in the appendix of Wakker et al. 2012) and the non-equilibrium turbulent mixing layer model of Kwak & Shelton (2010) shown as the blue contours.

The radiative cooling flow model of Benjamin and Shapiro follows the thermal and ionization state of the gas in a one-dimensional planar steady-state flow including the effects of the radiative transfer of the ionizing EUV photons produced by the cooling gas. This helps to increase the amount of C IV and particularly Si IV produced in the cooling gas. The TML concept of the mixing of cool, warm and hot gas through turbulence was first introduced by Begelman & Fabian (1990) and subsequently applied to the ionization of the transition temperature (T $\sim 10^5$ to $10^6$ K) ions by Slavin et al. (1993), Kwak & Shelton (2010) and Kwak et al. (2015). Note that the suggested TML process is related to that required to explain the self-irradiated cooling condensations expected to occur in the hot cooling flows discussed by Voit & Donahue (1990) in order to explain the optical line emission from the centers of clusters of galaxies.

The solid dark blue contours labeled NEI TML represent the distribution of ion ratios in the Kwak & Shelton (2010) TML models obtained by integrating through an interface at 1 Myr intervals for ages between 20 and 80 Myr. The Si IV results for the NEI TML models were provided by Shelton (private Communication) to Wakker et al. (2012) and subsequently published in Kwak et al. (2015).

Some combination of non-equilibrium radiative cooling and turbulent mixing can explain the origins of the high ion ionization absorption observed 1 kpc below the galactic plane. An outflow from the galactic center driven by hot gas with entrained cool/warm gas is exactly the kind of situation where one would expect turbulent mixing of cool, warm and hot gas to produce the intermediate temperature highly ionized species we have observed.

## 6.2. *The Low Ionization Absorption*

The low ionization components listed in Table 11 at -290 to -230, -211, -191, -152, -114 to -96, -78, -62 to -45, and 91 to 95 km s$^{-1}$ are all situated beyond d $\sim$ 7.0 kpc and most are probably associated with gas below the galactic center at z $\sim$ -0.7 to -1.5 kpc. Detailed comments about these velocity components are provided in the footnotes to Table 11 and in the following sections.

H I 21 cm emission is not observed to be associated with the components at -290 to -230, -210, -191, and -155 km s$^{-1}$ with 3$\sigma$ detection limits of log N(H I) < 17.5 in each component for a lines with FWHM = 32.5 km s$^{-1}$ or b = 19.5 km s$^{-1}$. H I emission is observed at -105, -89, -70, 90 and 98 km s$^{-1}$ and may exist at -46 km s$^{-1}$ (see Table 6). However, the possible emission near -46 km s$^{-1}$ is confused with much stronger low velocity emission from near-by disk gas and the limiting column density limit is very uncertain. The H I emission at -89 and 98 km s$^{-1}$ is not clearly associated with absorption components.

The H I component at -104.9 km s$^{-1}$ with log N(H I) = 18.28±0.02 is probably associated with the complex absorption system extending from -96 to -114 km s$^{-1}$ containing many species discussed in section 6.2.2..

The H I emission at -70.3 km s$^{-1}$ with log N(H I) = 19.43±0.01 is probably associated with the absorption system near -78 km s$^{-1}$ discussed in section 6.2.3.

The H I emission at 89.6 km s$^{-1}$ with log N(H I) = 18.75±0.02 is probably associated with the double component system near 91 and 93 km s$^{-1}$ detected in O I, C II*, Fe II, Si II, Mg I, Mg II and Ca II discussed in section 6.2.4.

We warn the reader that there are many uncertainties associated with relating the 21 cm emission to the UV absorption. The biggest problem is probably introduced by the effects of radio antenna beam smearing produced by the 9.2′ angular resolution of the GBT compared to the extremely small angular size of LS 4825. Unless the ISM structure has a relatively large physical size, the problem can become very significant at the ~8 kpc GC distances we are studying where 9.2′ corresponds to 21.4 pc.

**Table 11. Galactic Center Low Ion Absorption[a]**

| Ion | $v_{LSR}$ | b (km s$^{-1}$) | logN(X) | log(N(X)/N(S II)) | [X/S II][b] | Notes |
|---|---|---|---|---|---|---|
| C II | -290±1 | 5.7±0.9 | 12.85±0.06 | | | 1 |
| Mg II | -280±3 | 16.7±4.0 | 11.85±0.08 | | | 1 |
| C II | -272±1 | 8.5±1.6 | 12.83±0.07 | | | 1 |
| C II | -245±1 | 5.6±1.1 | 12.77±0.06 | | | |
| Mg II | -239±2 | 8.7±2.9 | 11.72±0.11 | | | 1 |
| C II | -230±1 | 4.6±0.9 | 13.00±0.10 | | | 1 |
| H I | -280 | 20 | < 17.5 | | | 1 |
| | | | | | | |
| Al II | -212±2 | 12.9±2.1 | 12.12±0.06 | | | 2 |
| C II | -211±1 | 9.7±0.7 | 14.18±0.02 | | | 2 |
| Mg II | -211±1 | 12.8±0.5 | 12.82±0.01 | | | 2 |
| Fe II | -208±1 | 17.4±0.8 | 12.97±0.02 | | | 2 |
| Si II | -207±1 | 17.4±0.9 | 13.59±0.02 | | | 2 |
| O I | -205±2 | 18.6±2.5 | 13.86±0.05 | | | |
| H I | -210 | 20 | < 17.5 | | | |
| | | | | | | |
| C II | -190±1 | 8.4±1.0 | 13.64±0.05 | | | 3 |
| H I | -191 | 20 | < 17.5 | | | 3 |
| | | | | | | |
| N I | -164±1 | 4.7±0.8 | 13.70±0.10 | | | |
| O I | -155±1 | 14.2±05 | 14.86±0.03 | 0.62±0.06 | -0.95±0.06 | 4 |
| Mg I | -152±1 | 9.4±2.2 | 11.52±0.08 | | | 4 |
| Mg II | -152±1 | 11.1±0.3 | 13.63±0.03 | | | 4 |
| Si  II | -152±1 | 17.0±0.6 | 14.24±0.01 | | | |
| N I | -151±2 | 11.1±2.5 | 13.69±0.11 | | | |
| S II | -151±1 | 18.5±1.3 | 14.42±0.04 | | | 4 |
| Fe II | -149±1 | 12.6±0.2 | 13.69±0.01 | -0.73±0.04 | -1.11±0.04 | 4 |
| Al II | -149±1 | 12.0±0.8 | 12.82±0.03 | -1.60±0.05 | -0.93±0.05 | 4 |
| H I | -155 | 20 | <17.5 | | | 4 |
| | | | | | | |
| C I | -114±1 | 6.9±1.0 | 13.10±0.06 | | | 5 |
| C I* | -114f | 7.0±0.6 | 13.42±0.03 | | | 5 |
| C I** | -114f | 7.8±0.8 | 13.26±0.04 | | | 5 |
| Na I | -112 | 6.0 | 12.18 | | | 5,6 |
| Ca II | -113 | 7.0 | 12.31 | | | 5,6 |
| Al II | -107±1 | 17.5±1.6 | 13.27±0.03 | | | 5 |
| Fe II | -106±1 | 14.2±0.4 | 14.15±0.01 | -0.61±0.09 | -0.99±0.09 | 5 |
| Mg I | -105±1 | 14.76±1.2 | 12.22±0.03 | | | 5 |
| S II | -104±1 | 15.0±1.2 | 14.76±0.09 | | | 5 |
| H I | -104.9±0.6 | 13.6±0.8 | 18.28±0.02 | | | 5,7 |
| C I | -101±1 | 6.9±0.5 | 13.49±0.03 | | | 5 |
| C I* | -101f | 6.4±1.2 | 13.12±0.06 | | | 5 |
| C  I** | -101f | 6.7±4.8 | 12.40±0.27 | | | 5 |
| C II | -99±1 | 27.2±2.3 | 17.59±0.49 | | | 5 |
| Si II | -98±2 | 15.5±1.9 | 14.99±0.10 | | | 5 |
| Na I | -98 | 5.0 | 12.12 | | | 5,6 |
| Ca II | -96 | 5.0 | 11.88 | | | 5,6 |
| H I | -88.6±0.2 | 7.9±0.4 | 18.38±0.03 | | | 5 |
| | | | | | | |
| Fe II | -79±1 | 10.6±0.7 | 13.90±0.03 | -0.67±0.05 | -1.05±0.05 | 8 |
| S II | -78±1 | 6.9±0.6 | 14.57±0.05 | | | 8 |
| Mg I | -77±1 | 6.8±0.9 | 12.06±0.04 | | | 8 |
| Al II | -77±1 | 6.1±1.2 | 12.89±0.13 | | | 8 |
| C I | -78±1 | 6.9±0.2 | 13.79±0.01 | | | 8 |
| C I* | -78f | 7.0±0.6 | 13.41±0.02 | | | 8 |
| C I** | -78f | 5.5±0.7 | 12.98±0.04 | | | 8 |
| Na I | -77.0 | 6.0 | 12.40 | | | 8 |
| Ca II | -77.0 | 4.5 | 12.29 | | | 8 |
| H I | 70.3±0.2 | 29.1±0.2 | 19.43±0.01 | | | 8 |
| | | | | | | |
| Mg I | -62±2 | 4.6±2.4 | 11.46±0.22 | | | 9 |
| S II | -52±2 | 40.5±10.2 | 15.09±0.10 | | | 9 |
| Fe II | -45±1 | 21.5±1.4 | 14.01±0.02 | | | 9 |
| Ca II | -46 | 25 | 12.22 | | | 9,6 |
| | | | | | | |
| H I | 89.6±1.1 | 29.9±1.6 | 18.75±0.02 | | | 10 |
| O I | 90.9±0.6 | 6.3±1.0 | 14.04±014 | | | 10 |
| C II* | 91.2±0.6 | 3.9±1.0 | 13.12±0.15 | | | 10 |
| Fe II | 91.8±0.3 | 6.0±0.4 | 13.30±0.03 | | | 10 |
| Si II | 92.4±0.3 | 3.6±0.7 | 13.46±0.09 | | | 10 |
| Fe II | 92.7±0.6 | 18.0±1.3 | 13.15±0.04 | | | 10 |
| C II* | 93.0±1.8 | 11.7±3.2 | 13.10±0.15 | | | 10 |
| O I | 93.3±0.6 | 11.9±1.0 | 14.14±0.11 | | | 10 |
| Mg I | 93.6±0.3 | 6.8±1.1 | 11.51±0.05 | | | 10 |
| Mg II | 93.6±0.1 | 13.4±0.3 | 13.37±0.01 | | | 10 |
| Si II | 94.5±0.3 | 15.8±0.5 | 13.68±0.03 | | | 10 |
| Ca II | 93 | 5.0 | 11.60 | | | 10,6 |
| H I | 98.3±0.6 | 4.0±1.0 | 17.46±0.13 | | | 10 |

[a] Components with v between 0 and 40 km s$^{-1}$ mostly sampling foreground gas with d < 7.0 kpc are not included. The properties of the foreground gas are discussed in the appendix.

[b] [Fe II/S II] = log(N(Fe II)/N(S II)) − log(Fe/S)$_\odot$ with log(Fe/S)$_\odot$ = 0.38 dex.

[O I/S II] = log(N(O I)/N(S II)) − log(O/S)$_\odot$ with log(O/S)$_\odot$ = 1.57 dex.

[Al II/S II] = log(N(Al II)/N(S II)) − log(Al/S)$_\odot$ with log(Al/S)$_\odot$ = -0.67 dex.

Solar elemental abundances in this paper are from Asplung et al. (2009). On a logarithmic system where A(H) = 12, we adopt the following values: A(C) = 8.43, A(N) = 7.83, A(O) = 8.69, A(Mg) = 7.60, A(Al) = 6.45, A(Si) = 7.51, A(S) = 7.12 and A(Fe) = 7.50.

Notes:

(1) The weak HVCs between -290 and -230 km s$^{-1}$: These weak HVCs detected in C II and Mg II trace the low column density most highly negative velocity gas in the GC outflow. The three C II absorbers are all narrow with b ~ 5 km s$^{-1}$ while the two Mg II absorbers have b ~ 9 and 17 km s$^{-1}$. The stronger C II component at -230 km s$^{-1}$ implies log N(H I+H II) = 17.61 if C/H in the absorber is solar. No H I 21 cm emission is detected at these velocities with log N(H I) < 17.5.

(2) The HVC component near -210 km s$^{-1}$: The component is detected in C II, Mg II, O I, Fe II, Si II, N I and Al II. It is not seen in H I 21 cm emission with log N(H I) < 17.5. The amount of neutral H I should be small with log N(O I) = 13.40±0.15 implying log N(H I) = 16.71 for solar O/H. The amount of warm ionized gas in the absorber is larger with C II implying log N(H +H II) = 17.75. The larger b values for C II, Mg II, Fe II and Si II compared to O I imply the absorber has a denser core and more ionized outer boundary. The absorber has complex ionization conditions.

(3) The HVC component near -191 km s$^{-1}$: This HVC is only detected in C II. The C II column density implies log N(H I+H II) = 17.17 for solar C/H. No H I emission is detected at this velocity with log N(H I) < 17.5.

(4) The HVC component near -155 km s$^{-1}$: [O I/S II] = -0.97, which implies that most of the S II exists in ionized gas if [O/S] is solar. No H I 21 cm emission is observed at this velocity. log N(O I) = 14.86±0.04 and log N(H I) < 17.5) implies log(N(O I)/N(H I))> -2.64 and [O/H] > 0.67. The observations are consistent with a supersolar oxygen abundance and ~1 dex ionization corrections for S II and Fe II. The value of log (Fe II)/N(S II) implies a 0.95 dex depletion of Fe relative to S assuming these two elements approximately trace the same gas phases. Higher angular resolution H I observations are required to reliably probe the oxygen abundance.

(5) The absorber with HVC components from -114 to -96 km s$^{-1}$ is discussed in detail in section 6.2.2.

(6) The Na I and Ca II observations are from Ryans et al. (1993).

(7) The listed H I column density assumes the wing of the H I emission from -105 to -120 km s$^{-1}$ is due to an H I component centered near -105 km s$^{-1}$ and the emission is not from the wing of the H I component at -70 km s$^{-1}$.

(8) The absorber near -78 km s$^{-1}$ is discussed in detail in section 6.2.3.

(9) The components near -50 km s$^{-1}$ are not aligned in velocity and have different line widths making it difficult to determine useful information about the abundances or physical conditions in the absorbers. A distinct H I 21 cm emission component is not observed near -50 km s$^{-1}$ and a limit will be large given the presence of strong emission at by the lower velocity gas.

(10) The absorber near 92 km s$^{-1}$ is discussed in detail in section 6.2.4.

### 6.2.1. *The Composition of the Low Ionization Absorption Components*

The observations of log N(Fe II) and log N(S II) are particularly valuable because both species are reliably measured in many of the absorption components. Using the notation [Fe II/S II] = log(N(Fe II)/N(S II)) − log(Fe/S)$_\odot$ with log(Fe/S)$_\odot$ = 0.38 dex from Asplund (2009) we give the values of [Fe II/S II] in column 6 of Table 11. With first and second ionization potentials of 7.90 and 16.19 eV for iron and 10.36 and 23.34 for sulfur, Fe II and S II should be the dominant ionization state of Fe and S in H I absorbers where log N(H I) > 19.5 and [Fe II/S II] should be approximately equal to [Fe/S]. However, the observed absorption of Fe II and S II in the neutral gas can be contaminated by the existence of Fe II and S II in the warm ionized medium (WIM) along the line of sight if it absorbs at the same velocity. Sembach et al. (2000) have studied the impact of the existence of the WIM on elemental abundance studies and have developed a model for correcting for the contamination assuming the WIM is ionized by the radiation from hot stars. A detailed application of the Sembach et al. (2000) correction requires good observations of adjacent ion states such as Fe II /Fe III or S II / S III. Those measurements are not available for our line of sight. We could use measures of Si II / Si III but in all cases the Si III lines are highly saturated along the line of sight to LS 4825. We therefore need to treat inferences from the metal line observations with caution. We will see below that some of the detected absorbers have very serious WIM contamination problems.

[Fe II/S II] should be approximately equal to [Fe/S] in those absorbers where the WIM contamination is minor. When the WIM contamination is more substantial the measured values of [Fe II/S II] will be a column density weighted average of the ratio of Fe II/S II in the cold and warm neutral medium and in the WIM.

In diffuse ISM cloud absorbers, [Fe/H] is found to range from -2.4 to -1.0 dex with the smaller abundances found in denser disk gas and the larger abundances found in halo clouds (Savage & Sembach 1996). The deficiency of Fe from the gas phase (depletion) is attributed to the presence of Fe in interstellar dust. Since S mostly exists in the gas phase, [Fe/S] is a measure of the depletion of Fe in the different absorbing components.

The S II and Fe II absorption in the spectrum of LS 4825 allows the determination of [Fe II /S II] = -1.05±0.05, -0.99±0.09, and -1.11±0.04 in the absorption components near -78, - 106 and -150 km s$^{-1}$, respectively. These components are probably tracing outflowing neutral and ionized gas from the galactic center. The ~ -1 dex deficiency of Fe II with respect to S II in these components is probably due to survival of dust grain cores containing Fe in the outflowing gas.

### 6.2.2. *The Composition and Properties of the Components from -114 to -96 km s⁻¹*

The HVC absorption from -114 to -96 km s⁻¹ is complex.  In the metal lines of Fe II, S II, and Mg I, single absorption components with v ∼ -105 km s⁻¹ and b ∼ 15 km s⁻¹ adequately describe the observed line profiles.  The absorption by C II, Si II and Al II is stronger and more uncertain because of line saturation and overlap with adjacent absorption components.   As shown above the values of log N(Fe II) and log N(S II) imply that Fe is depleted by -0.99±0.09 dex and dust containing Fe likely exists in the outflowing gas.

If the S II absorption with log N(S II) = 14.76±0.09 mostly traces a region where the hydrogen is neutral, we would expect log N(H I) = 19.64±0.09  for solar sulfur abundances.  However, the 21 cm H I observations imply log N(H I) = 18.28±0.03 and b(H I) = 13.6±0.8 km s⁻¹  for a component at v = -105 km s⁻¹ if the wing of H I emission from -105 to -120 km s⁻¹ is due to a component centered near  -105 km s⁻¹.  If this H I column density is correct, it implies an extremely large sulfur abundance of [S/H] = 1.38. The absence of significantly stronger H I emission could be explained if the absorber has very small angular scale structure that has been diluted by H I radio antenna beam smearing and/or is very significantly affected by the presence of S II in ionized rather than neutral gas.

The weaker metal lines of C I, C I*, C I**, Na I and Ca II reveals the absorption is more complex with components near -114 and -101 km s⁻¹.  This double component structure is most evident in the C I absorption.  The fit to the weaker C I* and C I** absorption was performed by fixing their velocities to be the same as observed for the stronger C I absorption component as noted by the 'f' in the velocity columns of Table 11.

The observations of N(C I), N(C I*) and N(C I**)  at -114 km s⁻¹ listed in Table allow measures  of f1 = N(C I*)/N(C I$_{total}$) = 0.460± 0.073  and f2 = N(C I**)/N(C I$_{total}$) = 0.319±0.035 in the absorbing cloud.  Following Jenkins & Tripp (2011) we can use these values to estimate the pressure in the cloud.  Figure 11 displays the observed value of f1 and f2 as the brown cloud on the C I excitation figure adapted from Jenkins & Tripp (2011).  The cloud was constructed by taking random combinations of log N for C I, C I* and C I** that conformed to Gaussian distributions of the probabilities of log N whose widths were defined by the logarithmic column density uncertainties listed in Table 11.

The curves in Figure 11 display the expected level populations for three different temperatures for gas with different values of log (p/k) indicated with the dots with adjacent dots representing changes of 0.1 dex.  The large filled circles on the curves indicate integer values of log (p/k) ranging from 4 to 6.  Section 4.3 of Jenkins & Tripp (2011) discusses the assumed conditions in the gas that yields the three plotted curves.  The values of f1 and f2 for the component at -114 km s⁻¹ imply log p/k ∼ 5 dex.  This is an extremely large pressure for a cloud situated at ∼ - 1 kpc away from the galactic plane.   The possible origin of that high pressure is discussed in Section 8.

The values of N(C I), N(C I*) and N(C I**) for the absorber at -101 km s⁻¹  listed in Table 11 imply f1 = 0.43±0.07   and f2 ∼ 0.081 (+0.07, -0.04).  Note that N(C I**) is only a 1σ measurement.  Nevertheless, the observations displayed in Figure 11 with the blue cloud crudely constrain the pressure in the absorber at -101 km s⁻¹ to be log p/k ∼ 3.8 dex for T of the order of 80 K or less.   This value is similar to that found for many clouds in the diffuse ISM of the galactic disk where Jenkins & Tripp (2011) report for a sample of 89 sight lines a mean of log (p/k) = 3.6± 0.18 (rms) dex.

### 6.2.3.   The Composition and Properties of the -78 km s⁻¹ Component

We have the most complete information about abundances and physical conditions in the low ionization absorber near -78 km s⁻¹ (see Table 11). With log N(S II) = 14.57±0.05,  log N(Fe II) = 13.90±0.03, log N(Al II) = 12.89±0.13,  and log N(H I) = 19.43±0.01±0.15  from 21 cm emission at -70 km s⁻¹ we obtain   [S II/H I] = +0.02±0.05±0.15 ,  [Fe II/H I] = -1.05±0.04±0.15, and [Al II/H I)] = -0.90±0.13±0.15,  where the second errors allow approximately for radio antenna beam smearing uncertainty as estimated by Wakker et al. (2001).    Assuming no ionization corrections for S II, Fe II and Al II when compared to H I, these values imply [S/H] = 0.02±0.16, [Fe/H] = -1.05±0.15 and [Al/H] = -0.90±0.20.    The ionization assumption should be valid given the relatively large H I column density.  The abundances imply a solar abundance of S and ~ 1 dex depletions of Fe and Al.   Dust containing Fe and Al has evidently survived in the processes creating this absorber.

The observed H I emission is much broader than the S II absorption (b = 29.1 vs 6.9 km s⁻¹).  The b value difference implies T ~ 4.2x10⁴ K if the S II and H I exist in the same gas and the difference in line width is due to thermal line broadening.  H I this hot is difficult to understand.  It is possible H I antenna beam smearing is causing the observed H I profile to have a different shape than the H I profile that actually applies to the absorption line observations.  Also, the component structure of the H I emission may be more complex than our simple assumption of two Gaussian components.  Even if H I beam smearing problems exist, we can still conclude that Fe and Al are depleted from the gas phase by ~ 1 dex provided Fe II, Al II and S II are the dominant ionization states in the absorber.  That assumption seems reasonable given log N(H I) = 19.43±0.01±0.15.

The observations of N(C I), N(C I*) and N(C I**)  at -78 km s⁻¹ allow measures  of f1 = N(C I*)/N(C I_total) = 0.265±0.014 and f2 = N(C I**)/N(C I_total) = 0.099±0.0097  in the absorbing cloud.  The values of f1 and f2 are plotted in Figure 11 as the yellow data cloud.  The inferred value of log (p/k) ~ 4.1 dex is 0.5 dex larger than that found for many clouds in the diffuse ISM of the galactic disk where Jenkins & Tripp (2011) report for a sample of 89 sight lines a mean value of log (p/k) = 3.6± 0.18 (rms) dex.

### 6.2.4.   The Composition and Properties of the 93 km s⁻¹ Components

There are overlapping narrow and broad absorption components in the system containing O I, Fe II, Si II, C II* and other ions near 93 km s⁻¹.  There is a small (~2 km s⁻¹) velocity offset between the two sets of components with one system having an average velocity of 91.6 km s⁻¹ and the second 93.4 km s⁻¹.  Although the STIS resolution is 6.6 km s⁻¹ the 2 km s⁻¹ offset is clearly revealed through the profile fitting process for all of the detected species.

If we combine the column densities for O I, Fe II, and Si II in the two sets of components we find:  log N(O I)_total = 14.39±0.18,  log N(Fe II)_total = 13.53±0.05,  and log N(Si II)_total = 13.88±0.10, and log N(Mg II) = 13.37±0.01.  The total observed amount of O I implies log N(H I) = 17.17 if the absorber contains a solar abundance of O I.  However, we observe a much larger value of log N(H I) = 18.75±0.02.    The column densities imply [O I/H I] = -1.06±0.18, [Fe II/H I] = -0.72±0.05, [Si II/H I] = -0.38±0.10, and [Mg II/H I] = -0.98±0.02.  With log N(H I) = 18.75, the photoionization modeling of Bordoloi et al. (2016)  implies the ionization correction for O I should be small so [O I/H I] should be equal to  [O/H].  This is a consequence of the large charge-exchange cross section between O and H (Stancil et al. (1999).   However, for such a small value of log N(H I) the ionization corrections for Si II, Fe II, and Mg II could  be as large as 0.5 to 1.0 dex because substantial amounts of these ions may exist in the ionized medium associated with the absorber.  The observations imply [O/H] = -1.06±0.18 and that Si II, Fe II and Mg II require large and uncertain ionization corrections due to the likely existence of these species in ionized gas not traced by the H I emission.

The H I line width of 29.9±1.6 km s$^{-1}$ is much larger than for O I estimated for the two components (~12 km s$^{-1}$).  If H I and O I occur in the same gas and the difference in line width is from thermal broadening, the implied temperature is ~4x10$^4$ K which is much hotter than expected for gas in photoionization equilibrium.  This is similar to the result found for the -78 km s$^{-1}$ absorber and could be the result of beam smearing of an absorber with a complex spatial structure or because the simple assumed component structure fitted to the H I emission is incorrect.  In the case of O I and Mg II, the large line strengths imply the derived column densities may be significantly affected by line saturation if there are subcomponents with small Doppler line widths.  Because of these problems we believe the derived abundances referenced to hydrogen for this absorber are very uncertain.

C II* is detected in both components of the 93 km s$^{-1}$ absorber with log N(C II*)$_{total}$ = 13.44±0.21.  C II*  to C II  157.7μ emission is a major source of cooling of the diffuse ISM (Lehner et al 2004).  From measures of N(C II*) and N(H I) it is possible to estimate the C II cooling rate per neutral H  atom.  The result is $l_C$ = 2.89x10$^{-20}$ N(C II*)/N(H I)  ergs s$^{-1}$ (H atom)$^{-1}$ (see Lehner et al. 2004) for a gas temperature between 10$^3$ and  10$^4$ K.  For the 92 km s$^{-1}$ absorber we obtain a C II cooling rate of 1.4x10$^{-25}$ erg s$^{-1}$ (H atom)$^{-1}$.  This cooling rate is ~7 times larger than found for low velocity clouds in the Galactic disk with H I column densities larger than 10$^{20}$ cm$^{-2}$ where Lehner et al. (2004) obtain a mean cooling rate of 2x10$^{-26}$ erg s$^{-1}$ (H atom)$^{-1}$.  Most of the IVCs studied by Lehner et al. (2004) have cooling rates about 2x smaller than for the low velocity clouds.  The very high C II cooling rate for the 92 km s$^{-1}$ absorber is probably associated with its location in the extreme environment of the GC.  However, the low but uncertain oxygen abundance in this absorber suggests that it does not arise in the GC.  High angular resolution and higher S/N 21 cm H I observations of the H I column density are required to resolve the contradiction.

## 7.  THE KINEMATICAL SIGNATURE OF AN OUTFLOW IN MULTIPLE GAS PHASES

Toward the AGN PDS 456 with l, b = 10.4° and +11.2°, Fox et al. (2015) observed absorption components at -235, -5, +130 and +250 km s$^{-1}$.  The high velocity absorption was interpreted as being the result of the detection of the front and backside of an expanding biconical outflow emanating from the GC.

The line of sight to LS 4825 with l, b = 1.67°, -6.63° passes closer to the GC than for PDS 456 (see Figure 1).  The complex and multicomponent absorption toward LS 4825 is very likely sampling outflow from processes operating in the GC region.  The components we identify in the spectrum of LS 4825 that occur beyond the foreground star HD 167402 at d = 7.0 kpc have LSR velocities spanning the range from -290 to +107 km s$^{-1}$.   It appears that for the absorbers from -40 to -290 km s$^{-1}$ we are sampling multiphase gas on the front side of the GC outflow.  The Si III, Si IV and C IV near 107 km s$^{-1}$ and the complex low ion component near 93 km s$^{-1}$ may be tracing outflowing gas on the far side of the outflow and powered by the blast wave responsible for the extremely high pressure in the component at -114 km s$^{-1}$  (see section 8).

Using $v_{GSR} = v_{LSR} + (254$ km s$^{-1}$) sin(l) cos(b) where 254 km s$^{-1}$ is the rotation velocity of the LSR at the Sun's distance from the GC (Reid et al. 2009), we find the very large outflow velocity for the direction to LS 4825 traced by C II of $v_{LSR} = -290$ km s$^{-1}$ corresponds to $v_{GSR} = -283$ km s$^{-1}$. The C II observations for LS 4825 are consistent with the outflow velocity versus galactic latitude trend established by the five lines of sight through the northern Fermi Bubble displayed in Figure 3 of Bordoloi et al. (2017).

Of the 5 sight lines through the northern Fermi Bubble only one exhibited intermediate or high velocity redshifted gas (Bordoloi et al. 2016). That was for PDS 456 with high velocity absorption components at $v_{LSR} = -235$, 130 and 235 km s$^{-1}$. For LS 4825 we are possibly detecting the front and back side of swept gas in the complex low ion absorption components from -114 to -100 km s$^{-1}$ and at 93 km s$^{-1}$. However, given the low galactic latitude of LS 4825 of -6.63° compared to PDS 456 of +11.2°, it is surprising we did not also record a component with $v_{LSR} > 250$ km s$^{-1}$ in the spectrum of LS 4825. If the biconical outflow model is correct for these high velocity components, and the outflow is symmetric, we would expect to see both high velocity blueshifted and redshifted absorbers for a line of sight with a latitude so close to the GC. The non-detection of a $v_{LSR} > 250$ km s$^{-1}$ absorption component in the LS 4825 spectrum might be hinting at a non-unity filling factor of entrained cooler material inside a hot outflow. This sightline-to- sightline variation in similar latitude ranges can be very useful to quantify the covering fraction of such entrained gas in an outflow. However, the study of the entrained gas filling factors and the far side of the outflow will need to wait for an analysis of the complete set of southern Fermi Bubble observations.

## 8. ORIGIN OF THE HIGH PRESSURE ABSORBER AT -114 km s$^{-1}$

We propose that the components with low ionization that are seen at intermediate and high negative velocities are clumps of gas that have been accelerated by the ram pressure produced by the surrounding, more rapid flow of a lower density medium that originates from the Galactic center. One of these components at -114 km s$^{-1}$ exhibits an extraordinarily large thermal pressure log $(p/k) = 5$ which is well above the characteristic pressures log $(p/k) = 3.6$ (±0.18 dex rms dispersion) found by Jenkins & Tripp (2011) for quiescent material in the nearby ISM. Under normal conditions, it would be surprising to find a pressure well above this amount at any large distance from the Galactic plane.

We follow an argument similar to one presented earlier by Jenkins, Jura, & Loewenstein (1983) for the C I excitation observed in the slightly different context of gas clumps immersed in a stellar wind. After a time scale that is longer than the sound crossing time within the cloud, a hydrostatic pressure gradient will be established, and the pressure should decrease exponentially with distance from the upwind side of the cloud that has a pressure $p_{max} = \rho_{ext} v^2$, where $\rho_{ext}$ is the density of the external medium that is moving at a velocity $v$ with respect to the clump. The acceleration $a$ (in units of cm s$^{-2}$) is given by the relation,

$$a = p_{max}/[N(H)\mu] = 6.4 \times 10^7 \text{ K}^{-1} \text{ cm } (p_{max}/k)/N(H)$$

where the mean atomic mass $\mu$ is based on having $n(\text{He})/n(\text{H}) = 0.1$.

The average pressure that we sense in the cloud is weighted by the local densities of C I, which scale in proportion to $p^2$, since they are governed by the product of the local gas densities and the ionization equilibria between neutral and ionized carbon atoms. Hence, this pressure outcome should equal $2p_{max}/3$

Our value for the measured pressure log(p/$k$) = 5.0 can be replicated with reasonable guesses for the flow parameters, for which we can adopt the values $\rho_{ext}$ = 2.3×10$^{-28}$ gm cm$^{-3}$ [which corresponds to n(H) = 10$^{-4}$ cm$^{-3}$] and v = 3×10$^8$ cm s$^{-1}$. As long as the external medium's temperature is of order 10$^7$ K, the flow past the cloud will be supersonic, which reduces the influence of Kelvin-Helmholtz instabilities in breaking up the cloud (Bruggen & Scannapieco 2016).

An evaluation of the equation above and our earlier derivation N(H) = 4.4×10$^{19}$ cm$^{-2}$ yields an outcome for the acceleration $a$ = 2.2×10$^{-7}$ cm s$^{-2}$. This value is larger than a representative Galactic gravitational acceleration 1×10$^{-7}$ cm s$^{-2}$ at the location of the cloud. If this acceleration were operating in a steady fashion over an interval of 2 Myr, the initially stationary clump would reach a velocity of up to about 160 km s$^{-1}$, which could agree with our observed radial velocity if one accounts for a reasonable cos θ projection factor for our sight line. Over this time interval, the clump would have traveled over a distance of only 170 pc, which means the original location must still have been of order 1 kpc below the plane, much like the considerably larger cold clouds observed by Ford et al. (2008).

One might question whether or not a small cloud could survive without being disrupted by instabilities arising from interactions with the surrounding flow (Cooper et al. 2009). While this may pose a challenge to our interpretation of cloud compression and acceleration, we are aware that there is empirical evidence from high velocity absorption features appearing in the spectra of stars within or behind the Vela supernova remnant that small clouds pressurized by a blast wave can survive over a time interval at least comparable to the age of the remnant of about 10$^4$ yr (Jenkins et al. 1981, Jenkins & Wallerstein 1995, Jenkins et al. 1998).

The blast wave interpretation for the high pressure -114 km s$^{-1}$ absorption components may also explain the entire collection of low ionization components from -114 to -98 km s$^{-1}$. Gas on the far side of the blast wave may be responsible for the collection of low ion absorption components near 93 km s$^{-1}$.

## 9. THE UV OBSERVATIONS AND THE ENERGETIC EVENTS IN THE GC

Our UV absorption line observations are revealing the consequences of the GC energetic events on the ISM along the line of sight to LS 4825. Important results establishing a possible connection with the UV absorption and those energetic events include the following:

1. The kinematically complex set of UV absorption line profiles spanning an extremely large velocity range from -290 to 107 km s$^{-1}$ implies a line of sight passing through a highly disturbed medium.

2. The high ionization plasma traced by N V, C IV and Si IV has a complex absorption pattern extending from -223 to 107 km s$^{-1}$. The total C IV column density is larger than observed along any other line of sight in the Galaxy by 0.44 dex including lines of sight through the Carina OB association at z = 0 kpc. The high ion observations are consistent with non-equilibrium cooling of the plasma with possible additional contributions from turbulent mixing.

3. The extraordinary high pressure (p/k ~ 10$^5$ cm$^{-3}$ k) estimated from the C I excitation in the component near -114 km s$^{-1}$ suggests the component is being accelerated and compressed by the ram pressure of the low density hot gas on the near side of the GC outflow. The far side of the outflow may be associated with the low ionization components near 93 km s$^{-1}$.

4. The S II and Fe II absorption in components near -78, -106 and -150 km s$^{-1}$ are probably tracing outflowing gas from the galactic center. The -1.11 to -0.99 dex deficiency of Fe with respect to S in these components is probably due to survival of dust grain cores containing Fe in the outflowing medium.   In the component near -78 km s$^{-1}$ where [S/H] = 0.02±0.16, the near solar sulfur abundance is consistent with a GC outflow origin.

The GC line of sight is complex. It is therefore difficult to assign a specific location for the different absorption components or to discriminate among the different models for the explosive processes occurring at or near the GC.   In a series of papers summarized by Sofue (2017a), it is proposed that there are two expanding structures in the inner galaxy.  The inner most structure is associated with the Fermi Bubble which is proposed to be produced by an AGN explosion that occurred a few My ago.  The outermost structure at about 3 kpc from the GC is associated with the H I hole in the inner galaxy extending from l = 340 to 20 observed by Lockman & McClure-Griffiths (2016).  At low latitude the expansion velocity of this structure is 53 km s$^{-1}$.   At higher latitudes the expansion velocity could be larger.  The swept up ISM associated with both structures should be evident in UV absorption spectra.   The lack of a detection of the both structures  in absorption toward HD 167402 places their distances at d > 7 kpc corresponding to GC distances of  < 1.4  kpc.  However, the large 1.7 kpc error on the distance to HD 167402 implies the larger structure could be at a GC distance of < 3.1 kpc which is consistent with the Sofue (2017a) estimate.

The origin of the outer structure is unclear because it requires a huge explosion with E ~ 3x10$^{58}$ ergs (Sanders & Prenderast 1974).  The Galaxy is a bared spiral  (Bland-Hawthorn  & Gerhard 2016). Therefore, oval flow in the bar's potential could be strongly influencing gas motions in the inner galaxy.  An additional complication includes the presence of the 1.5 kpc radius tilted H I disk structure in the inner galaxy observed by Burton & Liszt (1978) and discussed in section 5.   With such a large set of possible phenomena producing gas with intermediate and high velocity and high pressure we feel it is highly speculative to assign specific observed UV absorption components to specific structures.  However, it is clear that some combination of the mentioned structures must be playing a role to create the highly disturbed ISM observed in UV absorption toward LS 4825.

## 10. SUMMARY

Observations of the UV ISM absorption in the spectrum of LS 4825 at a distance of 21±5 kpc in the direction l = 1.67° and b = -6.63° sample the multiphase gas ~ 1 kpc below the galactic center.  Comparisons with the UV absorption toward a foreground star at 7.0±1.7 kpc allows a separation based on kinematics of the absorption in the foreground gas at d < 7.0 kpc versus the more distant gas.

1. The LS 4825 spectrum reveals extremely strong and complex low and high ionization absorption extending over a velocity range from -290 to 107 km s$^{-1}$.  The foreground star reveals that the absorption with d < 7.0±1.7 kpc is mostly confined to the velocities from -7 to +100 km s$^{-1}$.

2. The absorption components in the spectrum of LS 4825 with -290 < $v_{LSR}$ < -40 and $v_{LSR}$ > 100 km s$^{-1}$ are probably associated with the highly disturbed multiphase plasma tracing the outflow of gas from the galactic center.

3. The low ionization plasma is traced by lines of O I, C I, C I*, C I**, C II, C II*, N I, Si II, Fe II, Mg II, S II, Mn II, Zn II, Al II, and many ions from rarer elements in the strongest absorber near 5 km s$^{-1}$ that is mostly produced in close-by cold/warm neutral gas.

4. The S II and Fe II absorption allows the determination of [Fe II/S II] in absorption components near -78, - 106 and -150 km s$^{-1}$ that are probably tracing outflowing gas from the galactic center. The -1.11 to -0.99 dex deficiency of Fe with respect to S in these components is probably due to survival of dust grain cores containing Fe in the outflowing medium.

5. We find [S/H] = 0.02±0.16, [Fe/H] = -1.03±0.15 and [Al/H] = -0.90±0.20 in the strong component near -77 km s$^{-1}$ that has associated H I emission with log N(H I) = 19.43±0.15. The near solar sulfur abundance is consistent with the GC outflow origin.

6. We propose that the low ionization components at intermediate and high negative velocity are being accelerated and compressed by the ram pressure of the low density hot gas in the front side of the GC outflow. This idea is supported by the extraordinary high pressure (p/k ∼ 10$^5$ cm$^{-3}$ k) estimated from the C I excitation in the component near -114 km s$^{-1}$. The pressure in this absorber is ∼ 25 times larger than found for the quiescent ISM of the galactic disk.

7. The high ionization plasma traced by N V, C IV and Si IV has a complex absorption pattern extending from -223 to 107 km s$^{-1}$. The C IV column density is larger than observed along any other line of sight in the Galaxy by 0.44 dex. The high ion column density ratios involving N V, C IV and Si IV in the absorption are similar to those found for gas in the lower Galactic halo within several kpc of the Galactic disk. The observations are consistent with non-equilibrium cooling of the plasma with possible contributions from turbulent mixing.

8. The observations provide strong evidence for significant mechanical feedback from some combination of hot stars, supernovae and the black hole that are driving the outflowing multiphase plasma at ∼ 1 kpc below the GC.

We thank Professor Francis Keenan for obtaining the observations of LS 4825 through HST GO program 8096 "The ISM Near to and Beyond the Galactic Center" and for letting us perform the scientific analysis of the data. Given its direction, distance, and brightness, LS 4825 is a very special star that has allowed us to study phenomena occurring ∼1 kpc below the GC in the UV. We thank Bob Benjamin for comments regarding the 1.5 kpc radius tilted H I disk in the GC. We continue to be extremely impressed by the successful scientific operation of the HST 27 years after its launch into space. We thank the many people involved with designing and building STIS and in determining its performance characteristics. Some of the spectra were retrieved from the Multi-Mission Archive (MAST) at the Space Telescope Science Institute. The Green Bank Observatory is a facility of the National Science Foundation operated under a cooperative agreement by Associated Universities, Inc. TSK is supported by the European Research Council grant titled "Cosmology with the IGM" through grant GA-257670 and the NSF-AST118913.

*Facilities: HST (STIS), FUSE, GBT*

## APPENDIX

Table 12 provides the complete line list ordered by LSR velocity measured through the profile fit process to the STIS observations of LS 4825 and HD 167402. Figures 12 to 16 display the continuum normalized intermediate and low ionization absorption observations and profile fit results for both stars. Contaminating absorption lines are displayed with the light grey lines. The component velocities are marked by the vertical lines with the profiles displayed with the color lines. For species with large numbers of absorption lines such as C I, C I*, C I**, Si II, Fe II, and Ni II we do not show all the absorption fits even though all the lines were used in the simultaneous fit process. Note that for some species such as O I we only have access to a very strong line (O I $\lambda$1302) and a very weak line (O I $\lambda$1355). For such species in the case of LS 4825 we can only obtain good column densities for the principal absorber near 5 km s$^{-1}$ and the low column density high velocity absorbers near -205, -155 and 92 km s$^{-1}$. The O I absorption from -140 to 80 km s$^{-1}$ in the strong O I $\lambda$1302 line is completely saturated and is only recorded in the weak O I $\lambda$1355 line near 5 km s$^{-1}$. For C II only the single strong $\lambda$1334 line is detected and the absorption toward LS 4825 is totally saturated from -170 to 120 km s$^{-1}$. In this case, the only useful column density and component structure information is for the HVCs from -290 to -190 km s$^{-1}$. In the case of Fe II with very small, moderate, and large oscillator strength lines observed, the component structure and column densities in the LS 4825 spectrum can be well measured over the full velocity range of the absorption from -208 to 93 km s$^{-1}$. For the intermediate strength lines of S II toward LS 4825 we can obtain good column densities and velocities for the absorption components at -151, -104, -78, -52 km s$^{-1}$ and 40 km s$^{-1}$. However, the S II absorption is strongly saturated near 0 km s$^{-1}$ and S II lines are not strong enough to detect S II in the higher negative and positive velocity components revealed in the more abundant species. Because of these problems, great care must be exercised when trying to relate the absorption properties of one species to the next. In some cases, it is necessary to compare high quality absorption line information from a species like Fe II to moderately saturated absorption to species like Si II. When such comparisons are made, it is important to verify that the line width and velocity determined for the moderately saturated absorber is similar to that found for the unsaturated absorber. In Table 12 we have identified all absorbers with absorption parameters determined from saturated lines by including the derived column densities and their errors in parenthesis. For strongly saturated absorption the errors in v, b and log N can be very large and uncertain because the true component structure of the absorption will generally be more complex than the component structure assumed in the fit process.

The body of this paper discusses the absorption in the spectrum of LS 4825 at d = 21±5 kpc that likely occurs beyond the foreground star HD 167402 at d = 7.0±1.7 kpc. In this appendix we briefly comment on the foreground absorption. The low ionization absorption near 5 and 38 km s$^{-1}$ observed toward LS 4825 is similar in strength and velocity structure to that seen toward HD 167402. Column densities toward both stars in the components near 5 km s$^{-1}$ are listed in Table 13 for H I derived from the damped Ly $\alpha$ line and for metal lines based on unsaturated absorption. The metal line column densities listed in Table 13 therefore should be reliable. There is slightly more (0.05 dex) H I toward LS 4825 than toward HD 167402 in the low velocity gas. This result is also reflected in the low ion absorbers of the dominant ion states in neutral gas where the logarithmic column density difference toward LS 4825 and HD 167402 is given in column 8 of Table 13. For Mg II, Fe II, Mn II, Ni II Cu II and Ge II the difference is ranges from -0.21 to 0.21 dex. The exception is O I where the column density is 0.41 dex larger toward LS 4825. The similarity in all but one of the low ion column densities for each line of sight implies that most of the low ion absorption near 5 km s$^{-1}$ toward each star occurs at d < 7.0±1.7 kpc.

In Table 14 we list the logarithmic elemental abundances found in the 5 km s$^{-1}$ absorbers toward LS 4825 and HD 167402 using for the reference to H values of log N(H I) from this paper along with preliminary values of log N(H$_2$) as discussed in the note to Table 14. Oxygen has [O/H] = -0.05±0.06 dex toward LS 4825 and [O/H] = -0.45±0.07 dex toward HD 167402. The normally highly depleted elements Fe, Ni and Mn have abundances of [X/H] ~ -1.3 to -1.6 dex.  Elements that usually have moderate depletions including Zn, Cu, Ge, and Ga have abundances ranging from -0.62 dex (Zn) to -1.02 dex (Ga).  These values are representative of those normally found for sight lines passing through diffuse clouds in the low galactic halo (Savage & Sembach 1996) with the exception of O toward HD 167402 which  is depleted by 0.45  dex.

**Table 12.  Profile Fit Line List Ordered by LSR Velocity for LS 4825 and HD 167402**

| star | ion | $v_{LSR}$ (km s$^{-1}$) | $\sigma$ | b (km s$^{-1}$) | $\sigma$ | log N | $\sigma$ | note |
|------|-----|------|------|------|------|-------|------|------|
| ls4825 | Si III | -283.0 | 2.9 | 4.0 | 2.0 | 12.57 | 0.29 | 1 |
| ls4825 | C II | -290.0 | 0.6 | 5.7 | 0.9 | 12.85 | 0.06 | |
| ls4825 | Mg II | -280.5 | 2.7 | 16.7 | 4.0 | 11.85 | 0.08 | |
| ls4825 | Si IV | -273.6 | 7.2 | 28.4 | 10.9 | 12.39 | 0.14 | |
| ls4825 | C II | -271.8 | 0.9 | 8.5 | 1.6 | 12.83 | 0.07 | |
| ls4825 | C II | -244.6 | 0.6 | 5.6 | 1.1 | 12.77 | 0.06 | |
| ls4825 | Mg II | -238.6 | 2.1 | 8.7 | 2.9 | 11.72 | 0.11 | |
| ls4825 | Si IV | -235.3 | 1.5 | 8.3 | 2.7 | 12.34 | 0.13 | |
| ls4825 | C II | -229.6 | 0.6 | 4.6 | 0.9 | 13.00 | 0.10 | |
| ls4825 | C IV | -226.9 | 1.8 | 15.9 | 3.3 | 13.41 | 0.15 | |
| ls4825 | Si III | -216.0 | 3.0 | 15.0 | 6.0 | 13.82 | 0.79 | 1 |
| ls4825 | Si IV | -212.6 | 0.6 | 9.2 | 1.3 | 12.86 | 0.04 | |
| ls4825 | Al II | -211.7 | 1.5 | 12.9 | 2.1 | 12.12 | 0.06 | |
| ls4825 | C II | -211.4 | 0.3 | 9.7 | 0.7 | 14.18 | 0.02 | 1 |
| ls4825 | Mg II | -210.5 | 0.3 | 12.8 | 0.5 | 12.82 | 0.01 | |
| ls4825 | Fe II | -208.4 | 0.6 | 17.4 | 0.8 | 12.97 | 0.02 | |
| ls4825 | Si II | -207.2 | 0.6 | 17.4 | 0.9 | 13.59 | 0.02 | |
| ls4825 | O I | -205.4 | 1.5 | 18.6 | 2.5 | 13.86 | 0.05 | |
| ls4825 | C II | -190.5 | 0.6 | 8.4 | 1.0 | 13.64 | 0.05 | |
| ls4825 | Si IV | -183.3 | 1.8 | 12.2 | 2.7 | 12.95 | 0.08 | |
| ls4825 | C IV | -179.7 | 3.9 | 45.0 | 4.9 | 14.23 | 0.05 | |
| ls4825 | N I | -164.2 | 0.6 | 4.7 | 0.8 | 13.70 | 0.10 | |
| ls4825 | Si IV | -160.9 | 1.2 | 9.4 | 1.9 | 12.94 | 0.09 | |
| ls4825 | O I | -155.2 | 0.3 | 14.2 | 0.5 | 14.86 | 0.03 | 1 |
| ls4825 | Mg II | -151.9 | f | 11.1 | 0.3 | 13.63 | 0.03 | 1, 3 |
| ls4825 | Mg I | -151.9 | 0.3 | 9.4 | 2.2 | 11.52 | 0.08 | |
| ls4825 | Si II | -151.6 | 0.3 | 17.0 | 0.6 | 14.24 | 0.01 | |
| ls4825 | N I | -151.3 | 2.4 | 11.1 | 2.5 | 13.69 | 0.11 | |
| ls4825 | S II | -151.3 | 0.9 | 18.5 | 1.3 | 14.42 | 0.04 | |
| ls4825 | Al II | -148.9 | 0.6 | 12.0 | 0.8 | 12.82 | 0.03 | 1 |
| ls4825 | Fe II | -148.9 | 0.3 | 12.6 | 0.2 | 13.69 | 0.01 | |
| ls4825 | C I** | -114.2 | f | 7.8 | 0.8 | 13.26 | 0.04 | 3 |
| ls4825 | C I | -114.2 | 0.3 | 6.9 | 1.0 | 13.10 | 0.06 | |
| ls4825 | C I* | -114.2 | f | 7.0 | 0.6 | 13.42 | 0.03 | 3 |
| ls4825 | Al II | -106.7 | 0.9 | 17.5 | 1.6 | 13.27 | 0.03 | 1 |
| ls4825 | Fe II | -106.4 | 0.3 | 14.2 | 0.4 | 14.15 | 0.01 | |
| ls4825 | Mg I | -105.0 | 0.9 | 14.8 | 1.2 | 12.22 | 0.03 | |
| ls4825 | S II | -104.4 | 0.6 | 15.0 | 1.2 | 14.76 | 0.09 | |
| ls4825 | Si IV | -102.9 | 0.6 | 29.4 | 1.2 | 14.22 | 0.02 | 2, 4 |
| ls4825 | C I* | -100.8 | f | 6.4 | 1.2 | 13.12 | 0.06 | 3 |
| ls4825 | C I** | -100.8 | f | 6.7 | 4.8 | 12.40 | 0.27 | 3 |
| ls4825 | C I | -100.8 | 0.3 | 6.9 | 0.5 | 13.49 | 0.03 | |
| ls4825 | C II | -99.3 | 1.2 | 27.2 | 2.3 | 17.59 | 0.49 | 1 |
| ls4825 | Si II | -98.4 | 2.1 | 15.5 | 1.9 | 14.99 | 0.10 | 1 |
| ls4825 | C II* | -94.2 | 0.6 | 22.1 | 1.6 | 14.91 | 0.07 | 1 |
| ls4825 | N V | -90.9 | 0.6 | 34.7 | 0.8 | 14.21 | 0.01 | |
| ls4825 | C IV | -89.7 | 1.2 | 37.9 | 1.9 | 14.91 | 0.03 | 2, 4 |
| ls4825 | Fe II | -78.9 | 0.3 | 10.6 | 0.7 | 13.90 | 0.03 | |
| ls4825 | S II | -78.3 | 0.3 | 6.9 | 0.6 | 14.57 | 0.05 | |

| ls4825 | C I | -77.7 | 0.3 | 6.9 | 0.2 | 13.79 | 0.01 | |
|--------|-----|-------|-----|-----|-----|-------|------|---|
| ls4825 | C I* | -77.7 | f | 7.0 | 0.6 | 13.41 | 0.02 | 3 |
| ls4825 | C I** | -77.7 | f | 5.5 | 0.7 | 12.98 | 0.04 | 3 |
| ls4825 | Mg I | -77.4 | 0.6 | 6.8 | 0.9 | 12.06 | 0.04 | |
| ls4825 | Al II | -77.1 | 0.6 | 6.1 | 1.2 | 12.89 | 0.13 | |
| ls4825 | O I | -67.0 | 0.6 | 24.6 | 1.4 | 16.80 | 0.22 | 1 |
| ls4825 | Mg I | -61.6 | 1.5 | 4.6 | 2.4 | 11.46 | 0.22 | |
| ls4825 | Al II | -52.0 | 1.5 | 13.6 | 2.3 | 12.94 | 0.08 | 1 |
| ls4825 | S II | -52.0 | 2.4 | 40.5 | 10.2 | 15.09 | 0.10 | |
| ls4825 | Si IV | -47.5 | 0.9 | 11.5 | 1.1 | 13.23 | 0.04 | |
| ls4825 | Mg II | -46.1 | f | 44.6 | 2.2 | 15.17 | 0.10 | 1, 3 |
| ls4825 | Mg I | -46.1 | 3.0 | 11.9 | 3.9 | 11.74 | 0.12 | |
| ls4825 | Si II | -45.8 | 16.5 | 37.2 | 37.1 | 14.70 | 0.43 | 1 |
| ls4825 | Fe II | -45.5 | 0.6 | 21.5 | 1.4 | 14.01 | 0.02 | |
| ls4825 | C II* | -41.6 | 1.2 | 14.2 | 2.2 | 13.91 | 0.06 | |
| ls4825 | Mg I | -17.0 | 1.5 | 3.6 | 2.0 | 11.66 | 0.26 | |
| ls4825 | Al II | -5.4 | 3.9 | 25.2 | 6.1 | 13.47 | 0.11 | 1 |
| ls4825 | S II | 0.0 | 0.9 | 15.0 | 0.9 | 15.79 | 0.04 | |
| ls4825 | Si II | 1.2 | 1.5 | 16.5 | 1.8 | 15.95 | 0.06 | |
| ls4825 | Mg I | 1.5 | 0.9 | 12.2 | 2.1 | 12.79 | 0.06 | |
| ls4825 | Zn II | 1.8 | 0.6 | 9.9 | 0.6 | 13.18 | 0.03 | |
| ls4825 | Mn II | 2.1 | 1.5 | 3.2 | 1.9 | 13.05 | 0.21 | |
| ls4825 | Mg II | 3.3 | 0.3 | 8.9 | 0.5 | 15.94 | 0.03 | |
| ls4825 | Cu II | 3.9 | 2.1 | 9.7 | 3.1 | 12.80 | 0.11 | |
| ls4825 | Ge II | 4.2 | 0.9 | 6.2 | 1.1 | 12.21 | 0.06 | |
| ls4825 | CO J0 | 4.5 | 0.9 | 5.7 | 1.3 | 12.98 | 0.07 | |
| ls4825 | O I | 4.5 | 0.3 | 5.3 | 0.2 | 17.88 | 0.04 | |
| ls4825 | Fe II | 4.8 | 0.3 | 14.5 | 0.3 | 15.32 | 0.01 | |
| ls4825 | H I | 5.0 | f | 25.0 | 4.8 | 21.20 | 0.04 | 5 |
| ls4825 | Cl I | 5.1 | 0.3 | 5.3 | 0.5 | 13.37 | 0.05 | |
| ls4825 | C I | 5.4 | 0.1 | 5.4 | 0.2 | 13.96 | 0.02 | |
| ls4825 | C I* | 5.4 | 0.3 | 5.9 | 0.3 | 13.53 | 0.02 | |
| ls4825 | Ni II | 5.7 | 0.3 | 16.1 | 0.6 | 14.15 | 0.01 | |
| ls4825 | Ni II | 6.3 | 0.6 | 3.6 | 0.8 | 13.17 | 0.10 | |
| ls4825 | P II | 9.6 | 1.2 | 33.6 | 1.7 | 14.09 | 0.02 | |
| ls4825 | N V | 12.9 | 2.1 | 41.7 | 3.2 | 13.67 | 0.03 | |
| ls4825 | C II* | 16.7 | 0.9 | 27.1 | 1.2 | 14.94 | 0.05 | 1 |
| ls4825 | Si IV | 21.5 | 0.6 | 29.8 | 0.8 | 13.94 | 0.01 | 2, 4 |
| ls4825 | C IV | 21.8 | 0.9 | 33.9 | 1.0 | 14.50 | 0.01 | 2, 4 |
| ls4825 | Mg II | 24.8 | 2.4 | 19.0 | 1.0 | 15.27 | 0.11 | 1 |
| ls4825 | Mg I | 29.6 | 1.8 | 18.9 | 1.3 | 12.73 | 0.05 | |
| ls4825 | S II | 29.9 | 1.8 | 17.5 | 1.1 | 15.42 | 0.05 | |
| ls4825 | Si II | 37.1 | 3.0 | 14.4 | 1.3 | 15.27 | 0.11 | |
| ls4825 | Fe II | 39.5 | 0.6 | 13.5 | 0.4 | 14.58 | 0.03 | |
| ls4825 | Ni II | 39.8 | 0.6 | 14.8 | 0.9 | 13.70 | 0.03 | |
| ls4825 | O I | 41.9 | 0.6 | 15.5 | 0.6 | 15.07 | 0.03 | 1 |
| ls4825 | C II | 47.2 | 1.2 | 30.9 | 2.5 | 16.70 | 0.33 | |
| ls4825 | O I | 90.9 | 0.6 | 6.3 | 1.0 | 14.04 | 0.14 | 2 |
| ls4825 | C II* | 91.2 | 0.6 | 3.9 | 1.0 | 13.12 | 0.15 | |
| ls4825 | Fe II | 91.8 | 0.3 | 6.0 | 0.4 | 13.30 | 0.03 | |
| ls4825 | Si II | 92.4 | 0.3 | 3.6 | 0.7 | 13.46 | 0.09 | |
| ls4825 | Fe II | 92.7 | 0.6 | 18.0 | 1.3 | 13.15 | 0.04 | |
| ls4825 | C II* | 93.0 | 1.8 | 11.7 | 3.2 | 13.10 | 0.15 | |
| ls4825 | O I | 93.3 | 0.6 | 11.9 | 1.0 | 14.14 | 0.11 | |
| ls4825 | Mg I | 93.6 | 0.3 | 6.8 | 1.1 | 11.51 | 0.05 | |
| ls4825 | Mg II | 93.6 | 0.1 | 13.4 | 0.3 | 13.37 | 0.01 | 1 |

| | | | | | | | |
|---|---|---|---|---|---|---|---|
| ls4825 | Si II | 94.5 | 0.3 | 15.8 | 0.5 | 13.68 | 0.03 | |
| ls4825 | Si IV | 104.4 | 0.9 | 13.8 | 1.3 | 12.78 | 0.03 | |
| ls4825 | Si III | 106.0 | 1.0 | 12.0 | 2.0 | 13.32 | 0.22 | 2 |
| ls4825 | C IV | 107.3 | 0.9 | 18.3 | 1.1 | 13.53 | 0.02 | |
| hd167402 | Si III | -6.9 | 7.5 | 22.2 | 5.6 | 13.23 | 0.20 | 1 |
| hd167402 | N I | -6.6 | 1.5 | 14.5 | 1.1 | 15.41 | 0.14 | 1 |
| hd167402 | C II* | 0.0 | f | 18.7 | 2.0 | 14.34 | 0.13 | 1,3 |
| hd167402 | C II | 0.0 | 3.9 | 12.0 | 1.0 | 17.70 | 0.15 | 1 |
| hd167402 | Si II | 0.9 | 0.6 | 13.5 | 0.2 | 16.07 | 0.03 | 1 |
| hd167402 | Si IV | 2.7 | 1.5 | 11.3 | 2.1 | 12.79 | 0.18 | |
| hd167402 | Mn II | 3.3 | 0.6 | 3.9 | 0.8 | 13.17 | 0.06 | |
| hd167402 | Mg II | 3.6 | 0.3 | 5.8 | 0.5 | 15.84 | 0.04 | |
| hd167402 | Ge II | 4.2 | 0.9 | 5.9 | 1.9 | 12.06 | 0.14 | |
| hd167402 | Cu II | 4.2 | 0.6 | 7.2 | 0.9 | 12.61 | 0.03 | |
| hd167402 | S II | 4.5 | 0.6 | 16.5 | 0.3 | 16.01 | 0.03 | 1 |
| hd167402 | Ni II | 4.8 | 0.6 | 11.7 | 1.1 | 13.94 | 0.12 | |
| hd167402 | Ga II | 4.8 | 0.9 | 2.8 | 1.2 | 11.25 | 0.10 | |
| hd167402 | H I | 5.0 | f | 20.9 | 6.0 | 21.15 | 0.03 | 5 |
| hd167402 | C I | 5.1 | 0.2 | 4.6 | 0.2 | 13.96 | 0.02 | |
| hd167402 | O I | 5.4 | 0.6 | 3.0 | 1.3 | 17.44 | 0.08 | |
| hd167402 | Fe II | 6.0 | 0.9 | 16.6 | 0.5 | 15.25 | 0.02 | |
| hd167402 | C I* | 6.0 | 0.2 | 3.6 | 0.2 | 13.73 | 0.01 | |
| hd167402 | CO J0 | 6.3 | 0.3 | 3.1 | 0.4 | 13.31 | 0.03 | |
| hd167402 | CO J1 | 6.3 | 0.6 | 3.3 | 0.7 | 13.20 | 0.04 | |
| hd167402 | C I** | 6.6 | 0.3 | 3.3 | 0.6 | 13.00 | 0.03 | |
| hd167402 | S I | 6.6 | 0.3 | 3.9 | 0.4 | 12.82 | 0.02 | |
| hd167402 | Cl I | 6.9 | 0.3 | 3.7 | 0.3 | 13.49 | 0.03 | |
| hd167402 | Ni II | 10.8 | 25.7 | 30.9 | 9.7 | 13.76 | 0.66 | |
| hd167402 | C I | 21.8 | 0.3 | 2.9 | 0.5 | 13.02 | 0.03 | |
| hd167402 | C II* | 24.2 | f | 21.5 | 0.8 | 15.00 | 0.07 | 1, 3 |
| hd167402 | C IV | 24.2 | 0.9 | 29.6 | 1.0 | 14.13 | 0.01 | |
| hd167402 | C II | 24.2 | 1.5 | 20.9 | 1.3 | 17.34 | 0.45 | 1 |
| hd167402 | Si IV | 29.9 | 1.5 | 22.5 | 2.3 | 13.49 | 0.04 | |
| hd167402 | N V | 31.4 | 1.5 | 42.1 | 2.3 | 13.73 | 0.02 | |
| hd167402 | S II | 37.7 | 0.9 | 15.2 | 0.6 | 15.34 | 0.04 | |
| hd167402 | Ni II | 38.0 | 4.8 | 17.9 | 5.6 | 13.77 | 0.49 | |
| hd167402 | N I | 39.2 | 1.2 | 10.3 | 1.4 | 16.47 | 0.64 | 1 |
| hd167402 | Fe II | 39.5 | 1.5 | 14.8 | 1.0 | 14.79 | 0.04 | |
| hd167402 | Si II | 46.9 | 0.6 | 11.9 | 0.4 | 14.92 | 0.05 | 1 |
| hd167402 | Si III | 48.1 | 4.5 | 22.7 | 10.7 | 14.30 | 0.83 | 1 |
| hd167402 | Si IV | 72.7 | 1.2 | 12.2 | 2.2 | 12.75 | 0.09 | |
| hd167402 | C IV | 78.6 | 1.8 | 21.0 | 2.1 | 13.40 | 0.05 | |
| hd167402 | N V | 96.6 | 2.7 | 16.3 | 3.8 | 12.88 | 0.10 | |
| hd167402 | Si IV | 99.0 | 4.2 | 14.8 | 4.6 | 12.39 | 0.14 | |
| hd167402 | Si III | 100.5 | 2.1 | 7.5 | 2.0 | 12.69 | 0.16 | |

Notes: (1) Strongly saturated absorption. The VPFIT column densities approximately correct for the saturation but if the assumed profile structure is not correct the errors can be very large.

(2) Saturated absorption. The listed VPFIT errors are uncertain. (3) Velocity fixed to agree with that for another absorber. The velocity error is listed as f. (4) The saturation corrections for the high ion Si IV and C IV absorption is discussed in section 4.1. (5) log N(H I) is derived from the fit to the damping wings of the strongly saturated Ly $\alpha$ absorption. The H I velocity is fixed to that for the weak O I and C I absorption.

**Table 13. Weak Line Column Densities for the Low Ions in the Low Velocity Absorption[a]**

| ion | LS 4825 v (km s⁻¹) | b (km s⁻¹) | log N(X) | HD 167402 v (km s⁻¹) | b (km s⁻¹) | log N(X) | ΔlogN[b] | Note |
|---|---|---|---|---|---|---|---|---|
| H I | = 5 | 25.0±4.8 | 21.20±0.04 | = 5 | 20.9±5.0 | 21.15±0.03 | 0.05 | |
| O I | 4.5±0.3 | 5.3±0.2 | 17.88±0.04 | 5.4±0.6 | 3.3±1.0 | 17.47±0.06 | 0.41 | 1 |
| C I | 5.4±0.1 | 5.4±0.2 | 13.96±0.02 | 5.1±0.1 | 4.6±0.2 | 13.96±0.02 | -0.01 | |
| C I* | 5.4±0.3 | 5.9±0.3 | 13.53±0.02 | 6.0±0.1 | 3.6±0.2 | 13.73±0.01 | -0.10 | |
| C I** | … | … | … | 6.6±0.3 | 3.3±0.6 | 13.00±0.03 | | 2 |
| Cl I | 5.1±0.3 | 5.3±0.5 | 13.37±0.03 | 6.9±0.3 | 3.7±0.3 | 13.49±0.03 | -0.12 | |
| S I | … | … | … | 6.6±0.3 | 3.9±0.4 | 12.82±0.02 | | 2 |
| Mg II | 3.3±0.3 | 8.9±0.5 | 15.94±0.03 | 3.6±0.3 | 5.8±0.47 | 15.84±0.04 | 0.10 | |
| Fe II | 4.8±0.3 | 14.5±0.3 | 15.32±0.01 | 6.0±0.9 | 16.6±0.5 | 15.25±0.02 | 0.07 | |
| Mn II | 2.1±1.5 | 3.2±1.9 | 13.05±0.21 | 3.3±0.6 | 3.9±0.8 | 13.17±0.06 | -0.12 | |
| Ni II | 5.7±0.3 | 16.1±0.6 | 14.15±0.01 | 4.8±0.6 | 11.7±0.6 | 13.94±0.06 | 0.21 | |
| Ni II | 6.3±0.6 | 3.6±0.8 | 13.17±0.10 | … | … | … | | |
| Zn II | 1.8±0.6 | 9.9±0.6 | 13.18±0.03 | | | | | |
| Cu II | 3.9±2.1 | 9.7±3.1 | 12.80±0.11 | 4.2±0.6 | 7.2±0.9 | 12.61±0.03 | 0.19 | |
| Ga II | … | … | … | 4.8±0.9 | 2.8±1.2 | 11.25±0.10 | | |
| Ge II | 4.2±0.9 | 6.2±1.1 | 12.21±0.06 | 4.2±0.6 | 5.9±1.9 | 12.06±0.14 | 0.15 | |
| CO J= 0 | 4.5±0.9 | 5.7±1.3 | 12.98±0.07 | 6.3±0.3 | 3.1±0.4 | 13.31±0.03 | -0.33 | |
| CO J =1 | … | … | … | 6.3±0.3 | 3.3±0.7 | 13.20±0.04 | | 2 |

[a]Measurements for saturated absorption lines are not included with the exception of the H I damped Ly α absorption.

[b]Δlog N = log N(X)$_{LS\ 4825}$ −log N(X)$_{HD167402}$ reveals the relatively small differences between the low ion absorption in the dominant absorber near 5 km s⁻¹ toward LS 4825 at d = 21 kpc and HD 167402 at d = 7.0 kpc. The exception is O I where log N(O I) is 0.41 dex larger toward LS 4825.

Notes: (1) The large difference for O I is possibly associated with oxygen depletion onto dust grain cores toward HD 167402.
(2) These species are not detected in the lower S/N STIS spectrum of LS 4825.

**Table 14. Elemental Abundances in the 5 km s⁻¹ AbsorbersToward LS 4825 and HD 167402[a]**

| Element | LS 4825 [X/H] | HD 167402 [X/H] |
|---|---|---|
| O | -0.05±0.06 | -0.45±0.07 |
| Mg | -0.90±0.05 | -0.99±0.05 |
| Fe | -1.43±0.04 | -1.48±0.04 |
| Mn | -1.62±0.21 | -1.49±0.07 |
| Ni | -1.31±0.15 | -1.51±0.07 |
| Zn | -0.62±0.05 | …….. |
| Cu | -0.63±0.12 | -0.81±0.04 |
| Ge | -0.68±0.07 | -0.82±0.07 |
| Ga | …….. | -1.02±0.10 |

[a] Abundances in the logarithmic system, [X/H] = log [N(X)/N(H)]- log [N(X)/N(H)]$_{\odot}$, are determined from log N(X II) assuming X II is the dominant ion in the neutral gas. For N(H) = N(H I) +2N(H₂) we use the values of log N(H I) listed in Table 13 along with preliminary values of log N(H₂) from an analysis of the FUSE spectra for LS 4825 and HD 167402. The full results of that analysis will be presented in a future paper. The H₂ absorption toward each star produces strongly damped profiles for the H₂ J = 0 and J = 1 lines with log N(H₂, J= 0) and log N(H₂, J = 1) having values of 19.73±0.02 and 19.60±0.02 toward LS 4825, and 19.93±0.02 and 19.83±0.02 toward HD 167402. With N(H₂) = N(H₂, J = 0) + N(H₂, J = 1) we obtain log N(H) = 21.24 and 21.23 toward LS 4825 and HD 167402, respectively. log [N(X)/N(H)]$_{\odot}$ is taken from Asplund et al. (2009). The normally depleted elements have abundances similar to those found in the gas of the lower galactic halo. The value of [O/H] = -0.45±0.07 toward HD 167402 is lower than expected. It possibly implies the depletion of oxygen onto dust grain mantles in the form of ice. If this is true the 3.1μm ice feature might be evident in absorption toward the star.

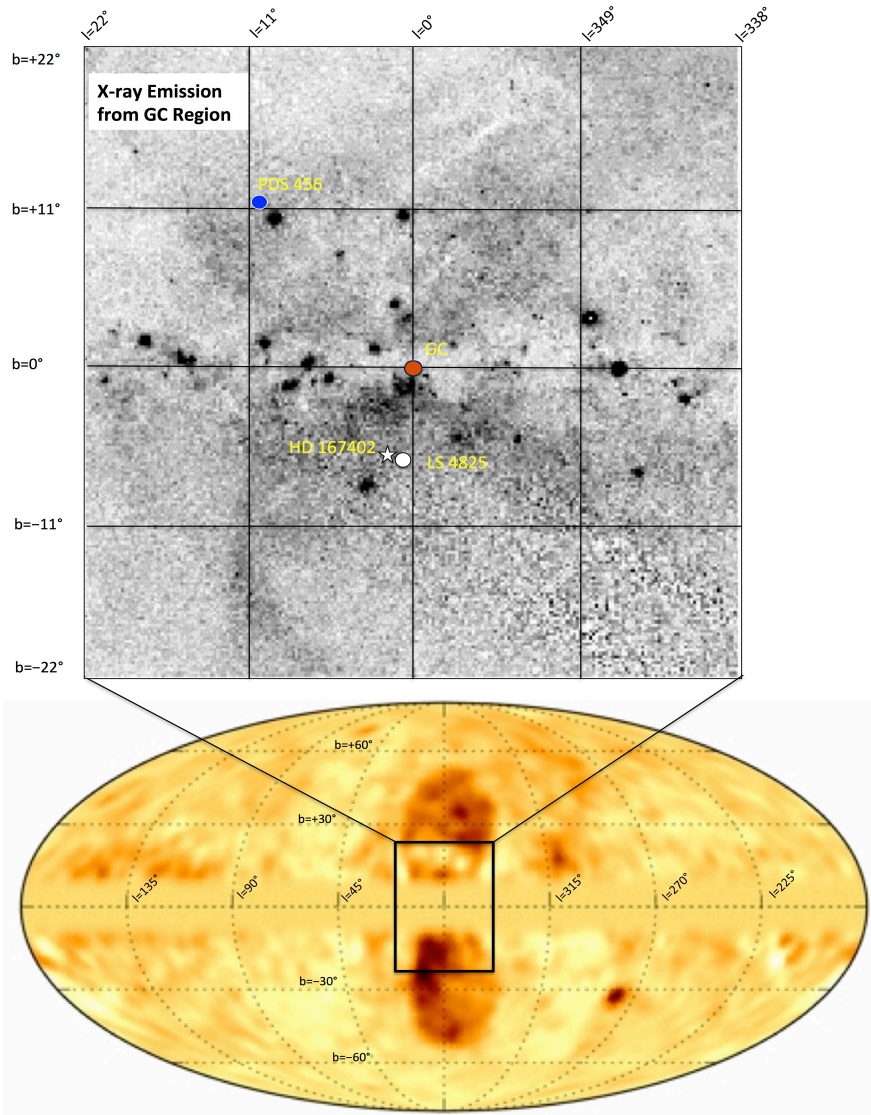

FIG.1. The upper image is the ROSAT 1.5 keV X-ray grey scale 44°x44° view of the GC originally obtained by Snowden et al. (1997) and reprocessed by Wang (2002) revealing the bi-conical emission from hot gas first discussed by Bland-Hawthorn & Cohen (2003). The red circle denotes l = 0.00° and b = 0.00°. The white circle denotes the direction to LS 4825 (l = 1.67° and b = -6.63°) at a distance of 21±5 kpc. The white star denotes the direction to the foreground star HD 167402 (l = 2.26° and b = -6.39°) at a distance of 7.0±1.7 kpc. The blue circle denotes the direction to the QSO PDS 456 (l = 10.39° and b = 11.16°). The lower image is the all sky Fermi image of the residual gamma-ray intensity in the 3-10 GeV range in Galactic coordinates adapted from Ackermann et al. (2014). The γ−ray emission for |b| < 10° is contaminated by bright Galactic disk emission. The bright emission has been replaced with the faint yellow band. The relationship between the γ−ray Fermi bubble and X-ray emission for |b| > 10° is clearly revealed through the overlap of the biconical structures.

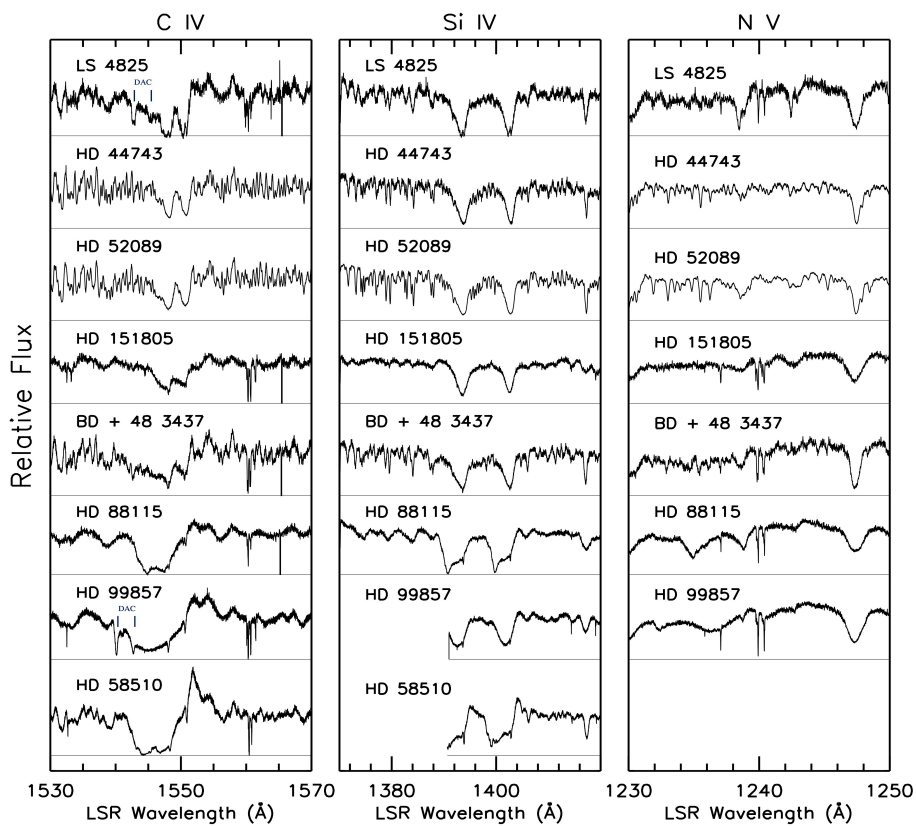

FIG. 2. UV spectra of LS 4825 and seven early-B type comparison stars in the vicinity of the high ionization resonance lines of C IV (left panel, 1530 to 1570 Å), Si IV (center panel, 1370 to 1420 Å) and N V (right panel, 1230 to 1250 Å). The spectrum of LS 4825 is at the top of each panel and the spectra of the early B comparison stars are ordered from top to bottom according to the increasing strength of the C IV stellar wind line. The stellar spectrum of LS 4825 most closely resembles that of the B1 Iab star BD +48 3437 (5[th] from the top). Note that the high ionization ISM lines for LS 4825 are extremely strong and of moderate strength in the spectrum of BD +48 3437. High velocity DACs are marked in the C IV spectra for LS 4825 and HD 99857.

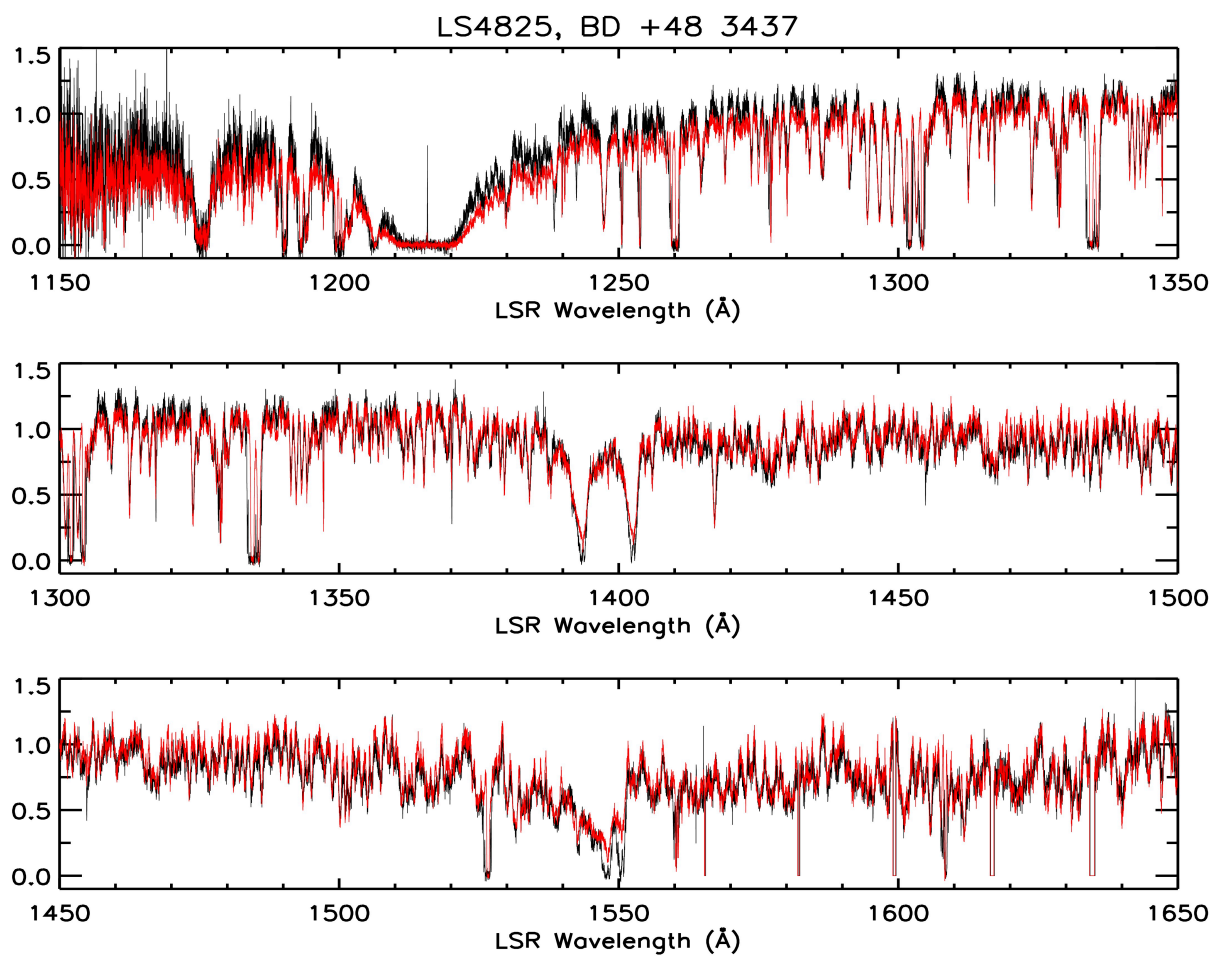

FIG. 3. STIS E140M spectra of LS 4825 (in black) and BD +48 3437 (in red) from 1150 to 1650 Å are overplotted to show their similarity. BD +48 3437 is classified as B1 Iab. The ISM lines in LS 4825 are much stronger than in BD +48 3437. The C IV λ1550 stellar wind line and the Si III λ1417 photospheric absorption line are sensitive to the luminosity of early B stars.

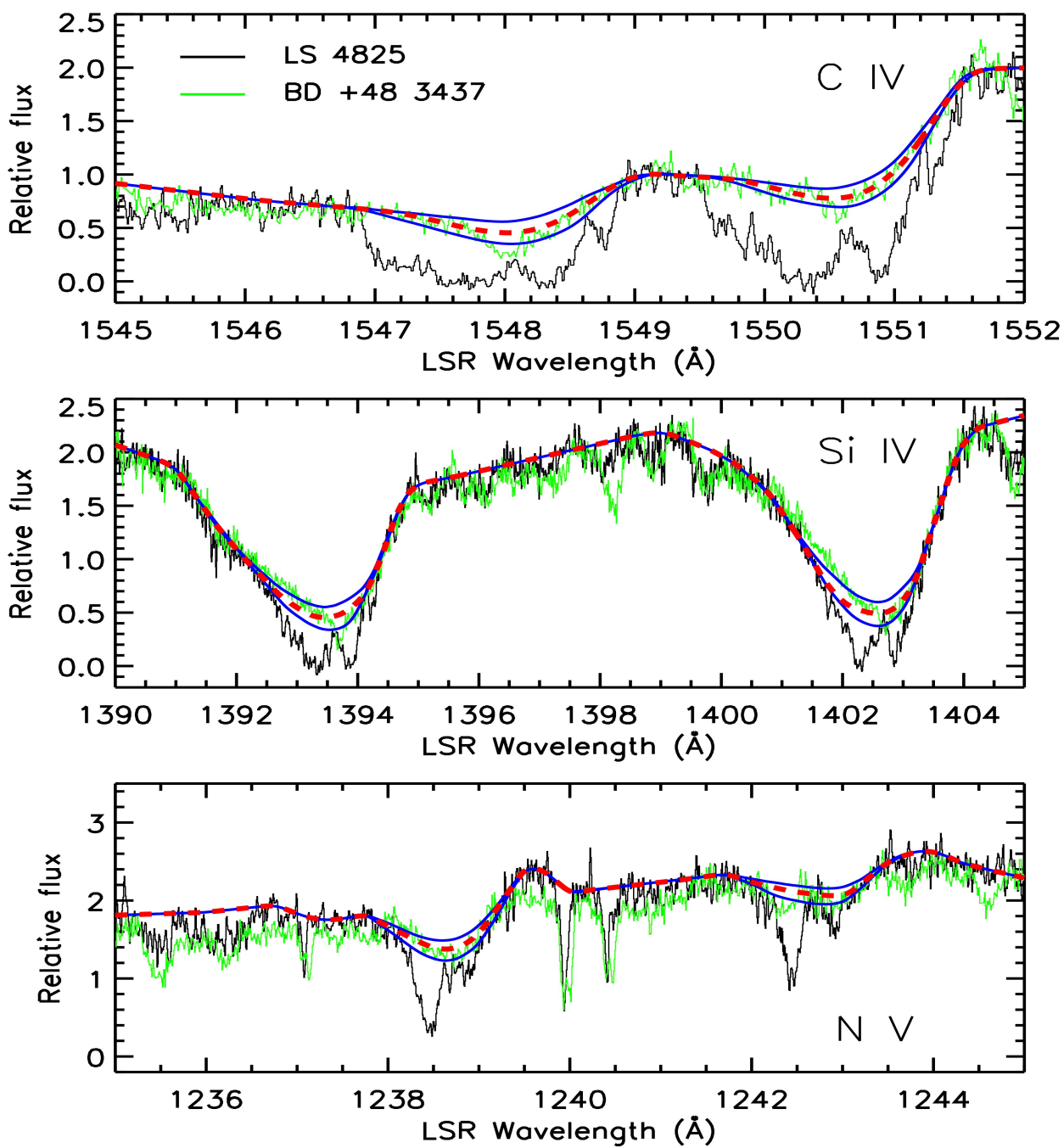

FIG. 4. The observed C IV, Si IV and N V spectra for LS 4825 and the well matched comparison star BD +48 3437. The red dashed lines show the continuum we have adopted when fitting the high ion absorption toward LS 4825. The upper and lower lines show alternate higher and lower continua that are used to estimate the continuum placement errors for the high ion absorption. The continuum for C IV is relatively well defined. The best continuum for Si IV is very similar to that for the comparison star except at negative velocities where it lies below that for the comparison star. This adjustment makes the negative velocity extent of the Si IV ISM absorption agree with that observed in the much stronger C IV absorption

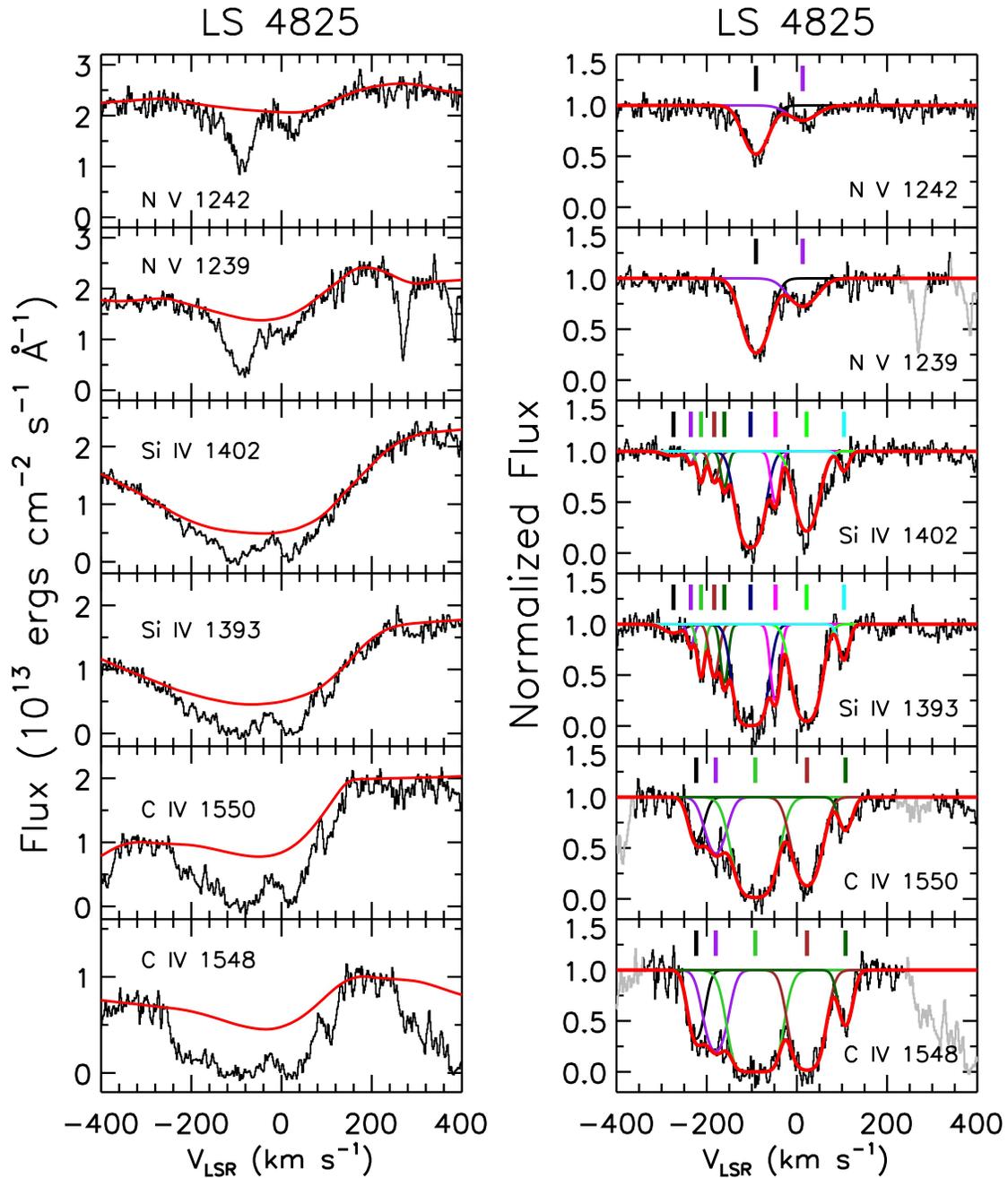

FIG. 5. The high ion absorption in the spectrum of LS 4825 at d = 21±5 kpc plotted against LSR velocity. The left panel shows the observed flux and the best assumed continuum for each high ion line. The right panel shows the continuum normalized intensity along with the profile fit results for each ion listed in Table 4.

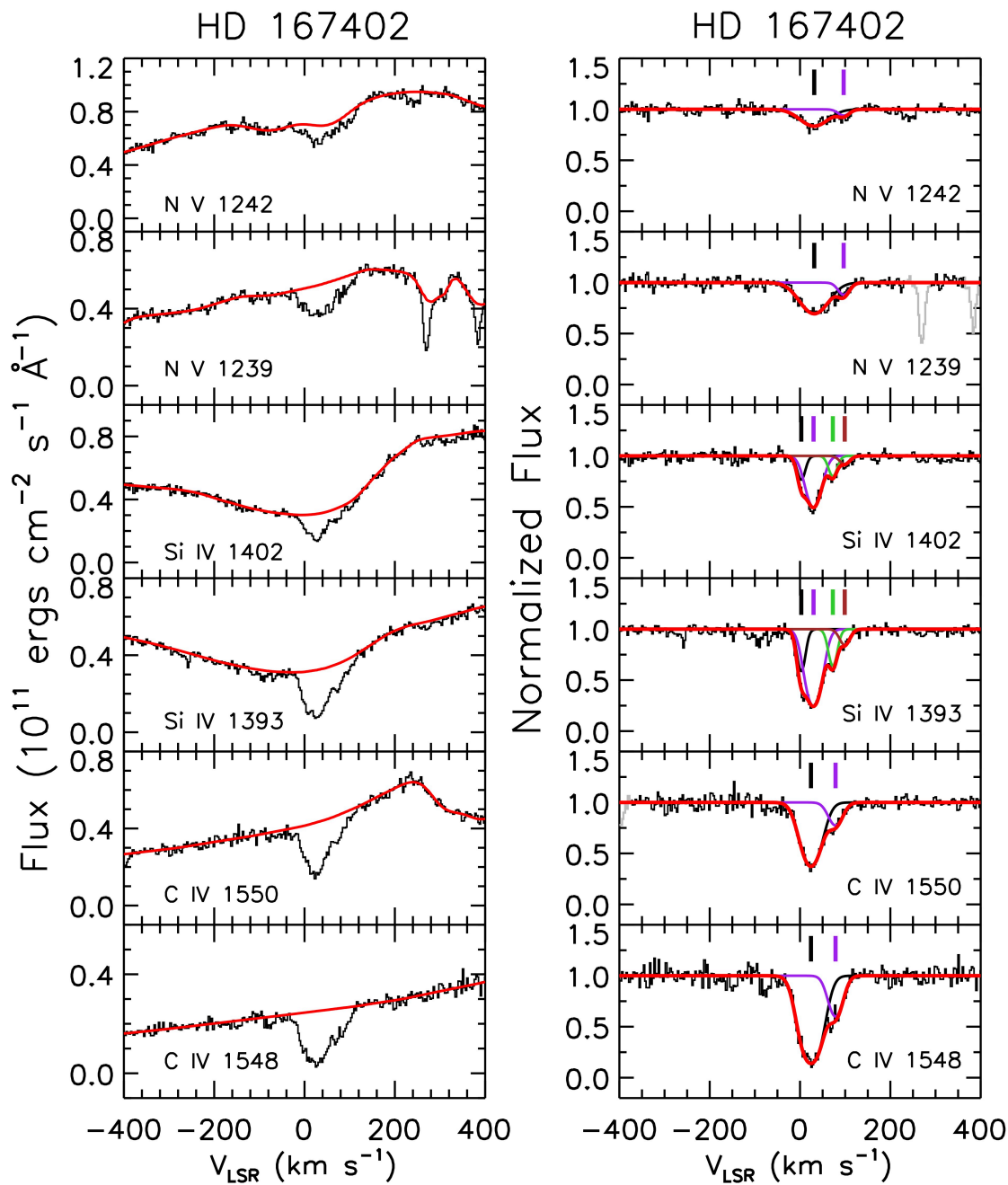

FIG. 6. The high ion absorption in the spectrum of the foreground star HD 167402 at d = 7.0±1.7 kpc plotted against LSR velocity. The left panel shows the observed flux and the best assumed continuum for each high ion line. The right panel shows the continuum normalized intensity along with the profile fit results for each ion listed in Table 5.

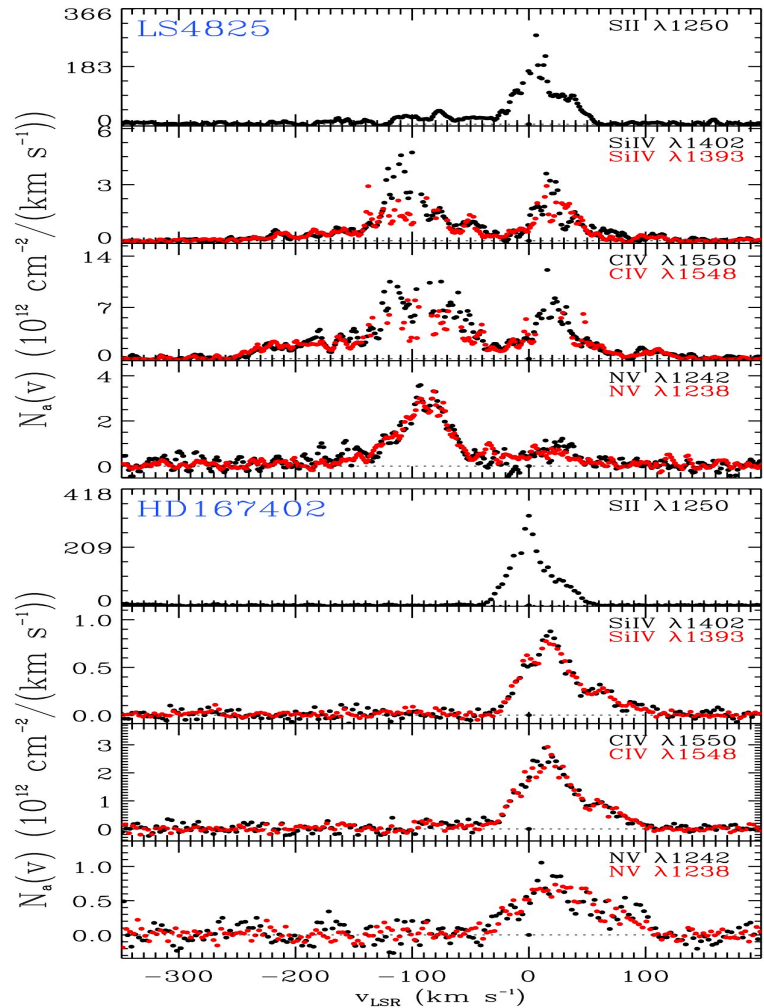

FIG.7. Apparent optical depth (AOD) profiles for the weak and strong lines of Si IV, C IV and N V for LS 4825(upper panel) and the foreground star HD 167402 (lower panel) plotted versus LSR velocity. AOD profiles are also shown for the low ionization S II λ1250 absorption line. Note the different Y axis scale for the two stars because the high ion absorption toward LS 4825 is much stronger than toward HD 167402. The weaker high ion lines are plotted as black dots while the stronger high ion lies are plotted as the red dots. A comparison of the AOD profiles for the weak and strong lines reveals the level of line saturation in the observations. For the foreground star HD 167402 the weak and strong lines yield very similar results implying no line saturation. For LS 4825 the AOD profiles of the weaker Si IV λ1402 and C IV λ1550 lines lie higher than for the stronger lines over the velocity ranges from 10 to 40 and from -130 to -60 km s$^{-1}$ implying saturation in those velocity ranges. No saturation is evident in the N V profiles. The much stronger absorption toward LS 4825 spanning a much wider velocity range is clearly evident when comparing the AOD profiles for LS 4825 at d = 21±5 kpc and the foreground star HD 167402 at d = 7.0±1.7 kpc.

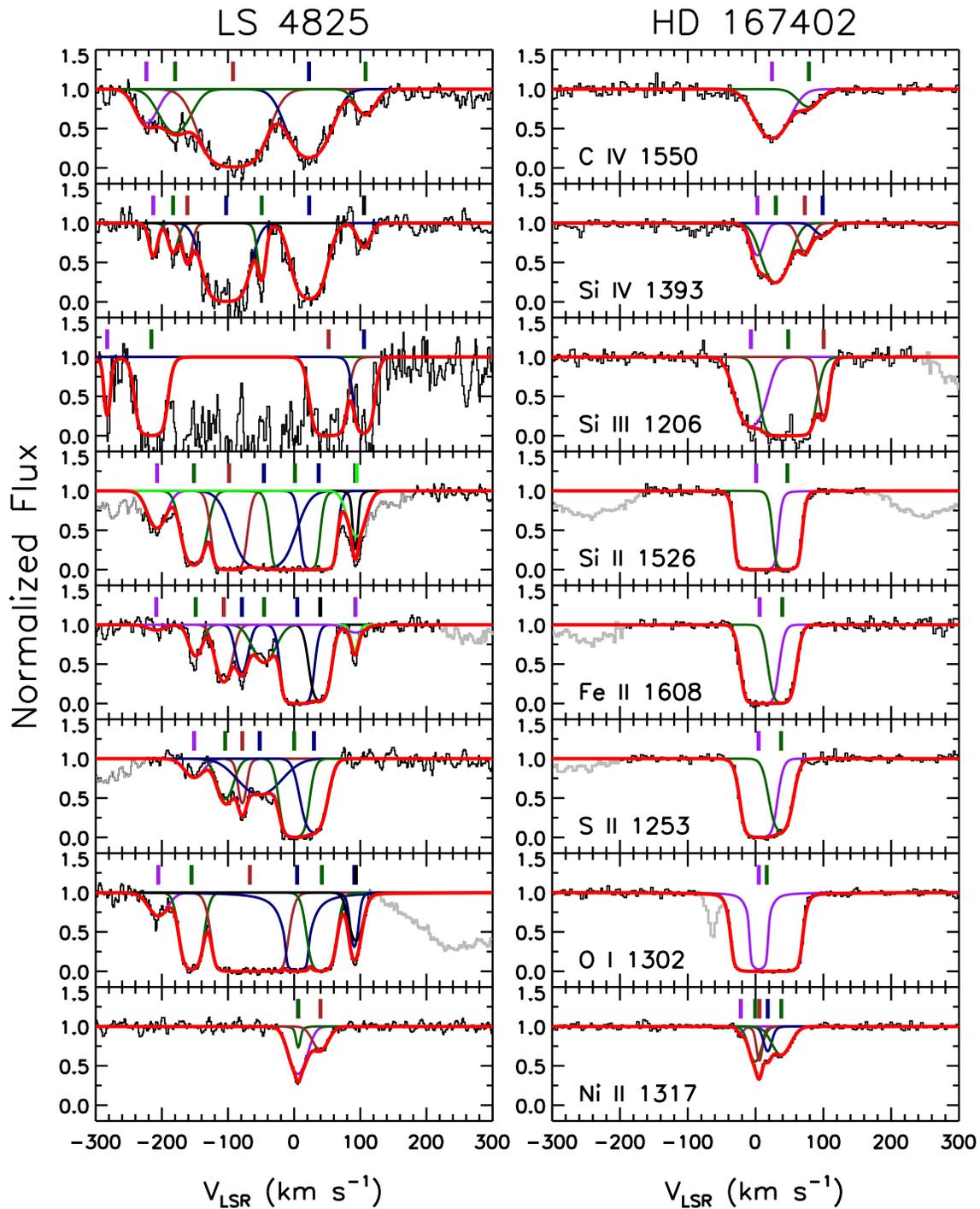

FIG. 8. Selected continuum normalized absorption lines plotted against LSR velocity in the spectrum of LS 4825 at d = 21±5 kpc and for HD 167402 at d = 7.0±1.7 kpc are shown in the left and right panels, respectively. Profile fit results as listed in the appendix are shown for each detected absorption component. A more extensive set of low ionization absorption line observations for both stars are displayed in the appendix.

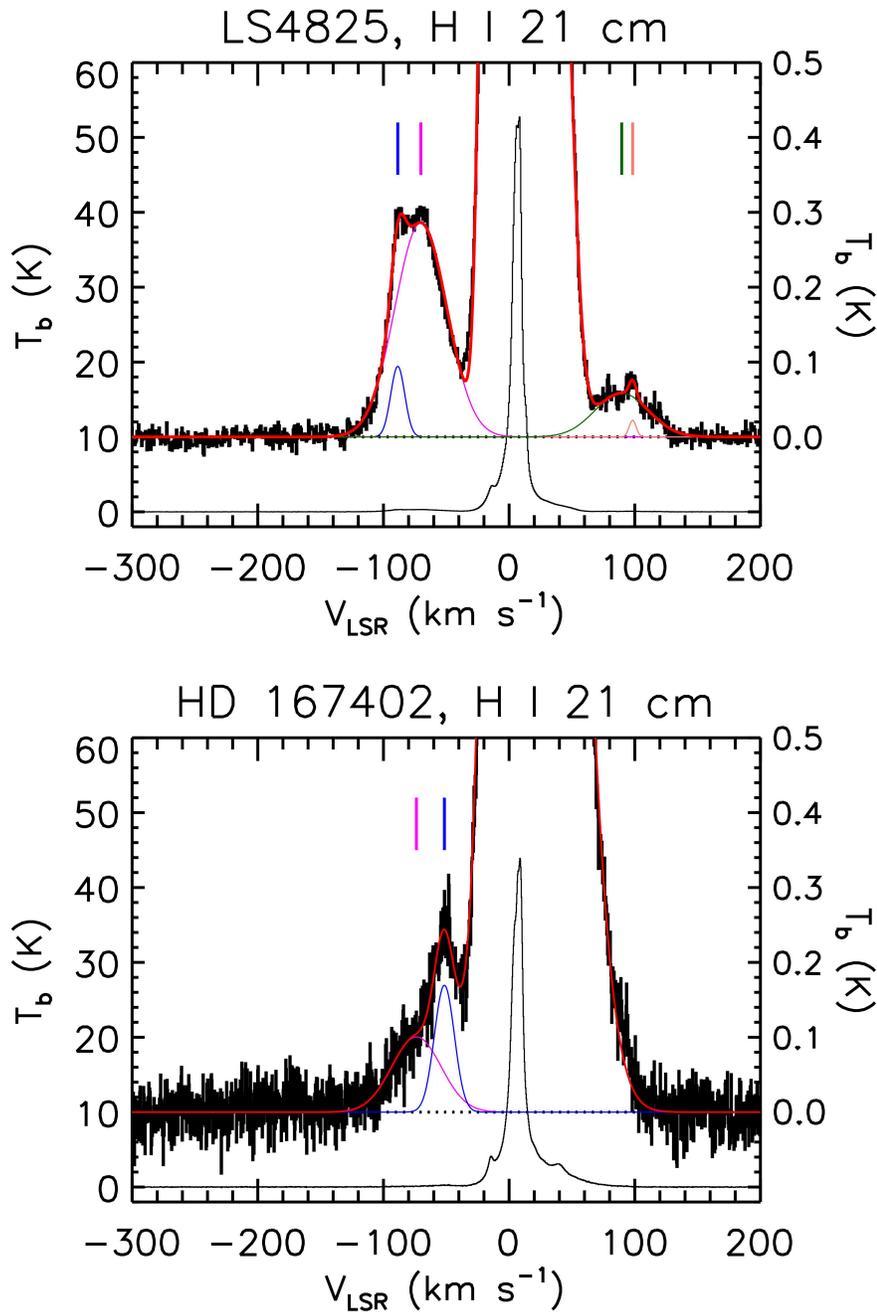

FIG. 9. H I 21 cm emission line profiles with an angular resolution of 9.2′ obtained by the Green Bank Telescope over the velocity range from -200 to 200 km s⁻¹ are shown for the direction to LS 4825 (upper panel) and HD 167402 (lower panel). The dotted lines show the individual profiles for Gaussian fits to the intermediate and high velocity components. The solid red line shows the fit to the total emission. A minimum number of components was fitted to the higher velocity emission in order to achieve $\chi_\nu^2 \sim 1$. The full spectra cover the velocity range from approximately -400 to +400 km s⁻¹. However, we only display the narrower velocity range where there is significant emission.

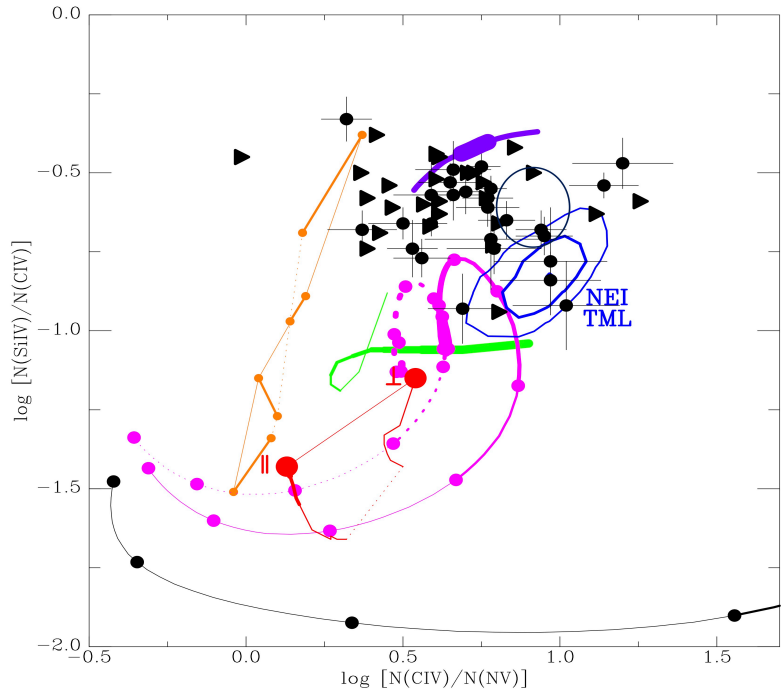

FIG.10. Ion ratio, ion ratio plot displaying log [N(Si IV)/N(C IV)] versus log [N(C IV)/N(N V)] for different theoretical models for the production of the high ions in the ISM and IGM. The open black oval just above the NEI-TML contours is the result for the total column densities along the line of sight to LS 4825. The size of the oval indicates the range of the ±1 sigma errors for the observation. The various models are displayed with the different colored curves as follows: CIE from Gnat & Sternberg (2007) for log T ranging from 5.4(on the left) to 4.4 (on the right) with the black line. The static non-equilibrium radiative cooling models from Gnat & Sternberg (2007) for log T = 5.4 to 4.4 for solar metallicity (magenta line) and for two times solar metallicity (dashed magenta line). These lines get thicker at lower temperatures. Shock models from Dopita & Sutherland (1996) for shock velocities of 200 and 300 km s$^{-1}$ (orange line). The conductive interface models of Borkowski et al. (1990) for a magnetic field perpendicular and parallel to the interface (red lines). The thick disk supernova model of Shelton (1998) is shown with the green line with the thickness proportional to the amount of time spent in each region of the plot. The cooling flow model of Benjamin & Shapiro described in the appendix of Wakker et al. (2012; thick blue line). The non-equilibrium turbulent mixing layer model of Kwak & Shelton (2010) and Kwak et al. (2015; solid blue contours labeled NEI TML). Black circles and black limit triangles are for thick disk absorption by the Milky Way halo gas observed by Wakker et al. (2012) toward AGNs.

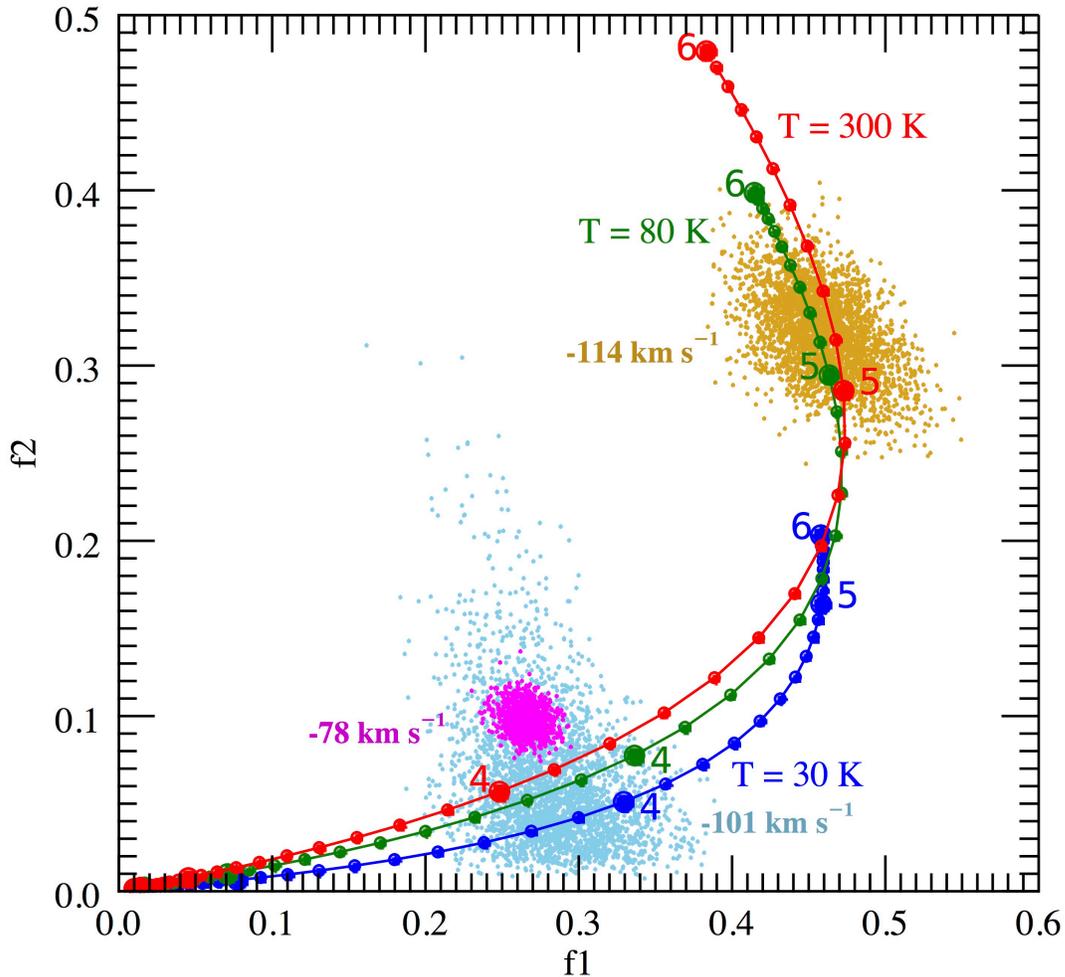

FIG.11. C I excitation diagnostic f1 vs f2 plot adapted from Jenkins & Tripp (2011) where f1 = N(C I*)/N(C I$_{total}$) and f2 = N(C I**)/N(C I$_{total}$). The three curves display the expected C I level populations for three different gas temperatures with different values of log (p/k) indicated on the curves with dots spaced by differences of 0.1 dex. The accompanying numbers on the curves give integer values of log (p/k) ranging from 4 to 6. The characteristic pressures in the ISM of the Galactic disk are found to be log (p/k) = 3.6±0.18(rms) dex. The purple and brown observation clouds imply the component at -78 km s$^{-1}$ has a disk cloud pressure log (p/k) ~ 4 while the component at -114 km s$^{-1}$ has an extremely high pressure of log (p/k) ~ 5. The component at -101 km s$^{-1}$ displayed with the blue observation cloud implies a pressure log(p/k) ~ 3.8 dex.

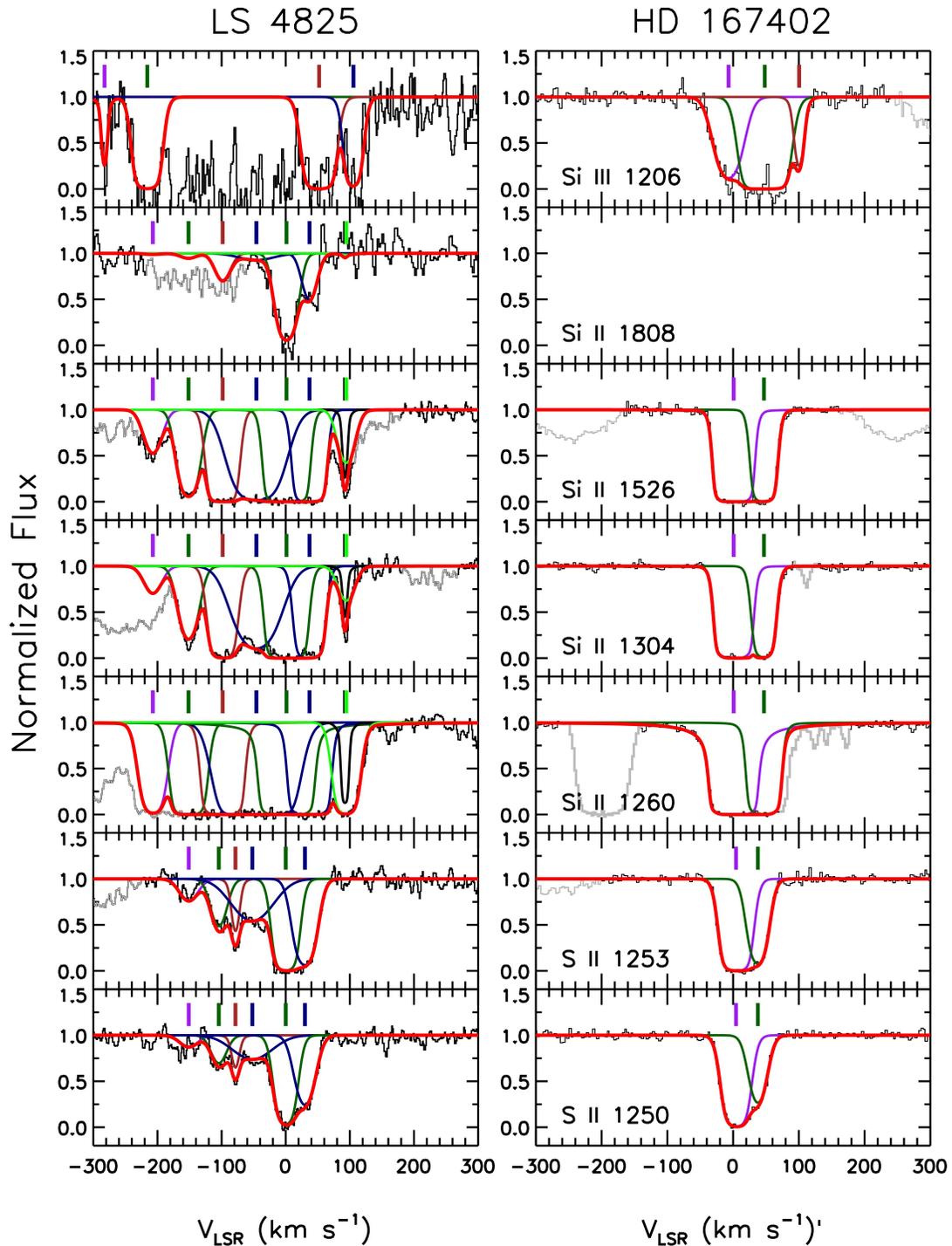

FIG. 12. Continuum normalized intermediate and low ionization absorption observed toward LS 4825 at d = 21±5 kpc (left panels) and the foreground star HD 167402 at d = 7.0±1.7 kpc are displayed with the black lines. Contaminating absorption lines are displayed with the light grey lines. Profile fit results listed in the appendix Table 12 are displayed with the color lines with the component velocities marked by the vertical lines. The lines displayed include Si III λ1206, Si II λ1808, Si II λ1526, Si II λ1304, Si II λ1260, S II λ1253 and S II λ1250.

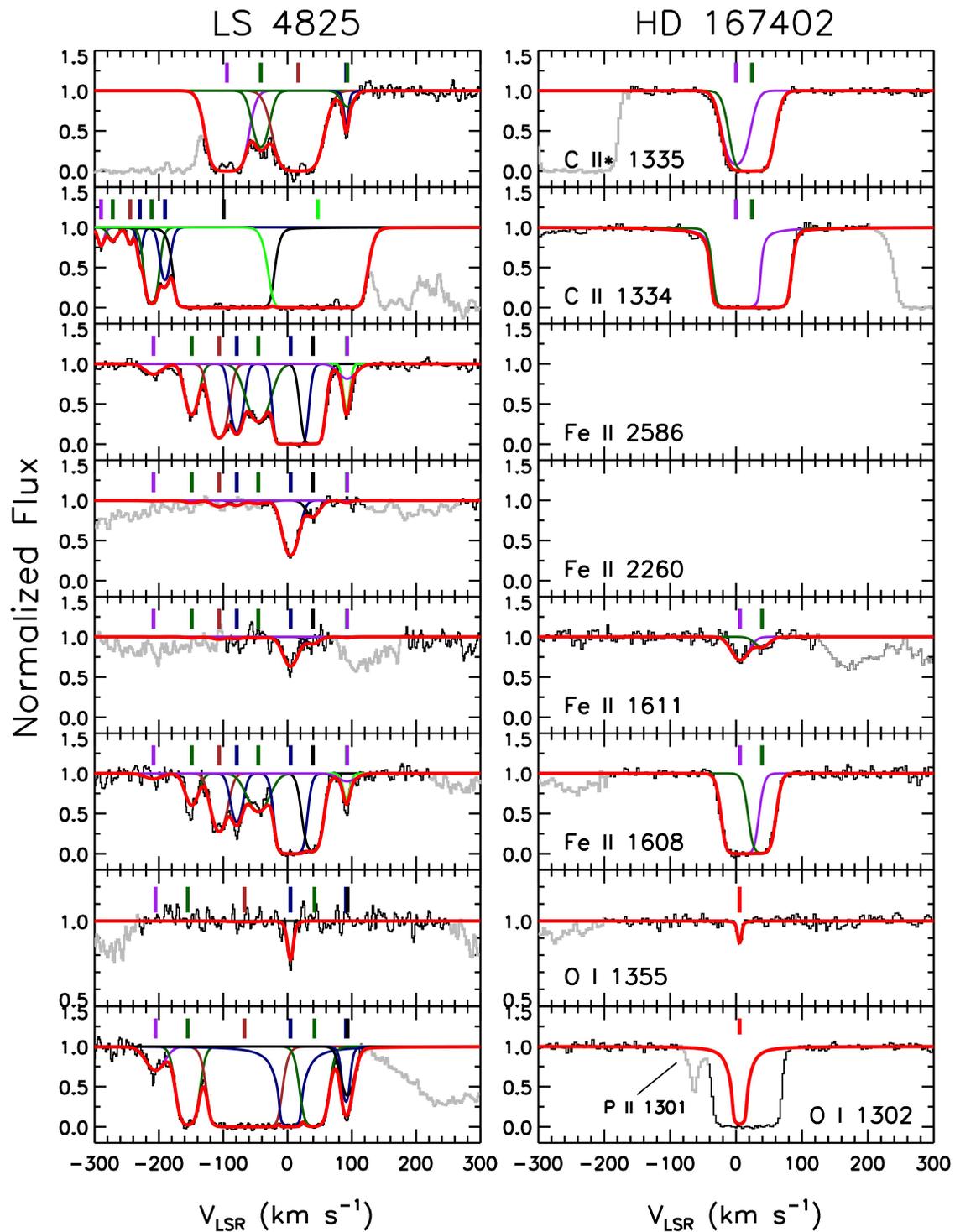

FIG. 13. Same as Figure 12 except the lines displayed are C II* λ1335, C II λ1334, Fe II λ2586, Fe II λ2260, Fe II λ1611, Fe II λ1608, O I λ1355, and O I λ1302.

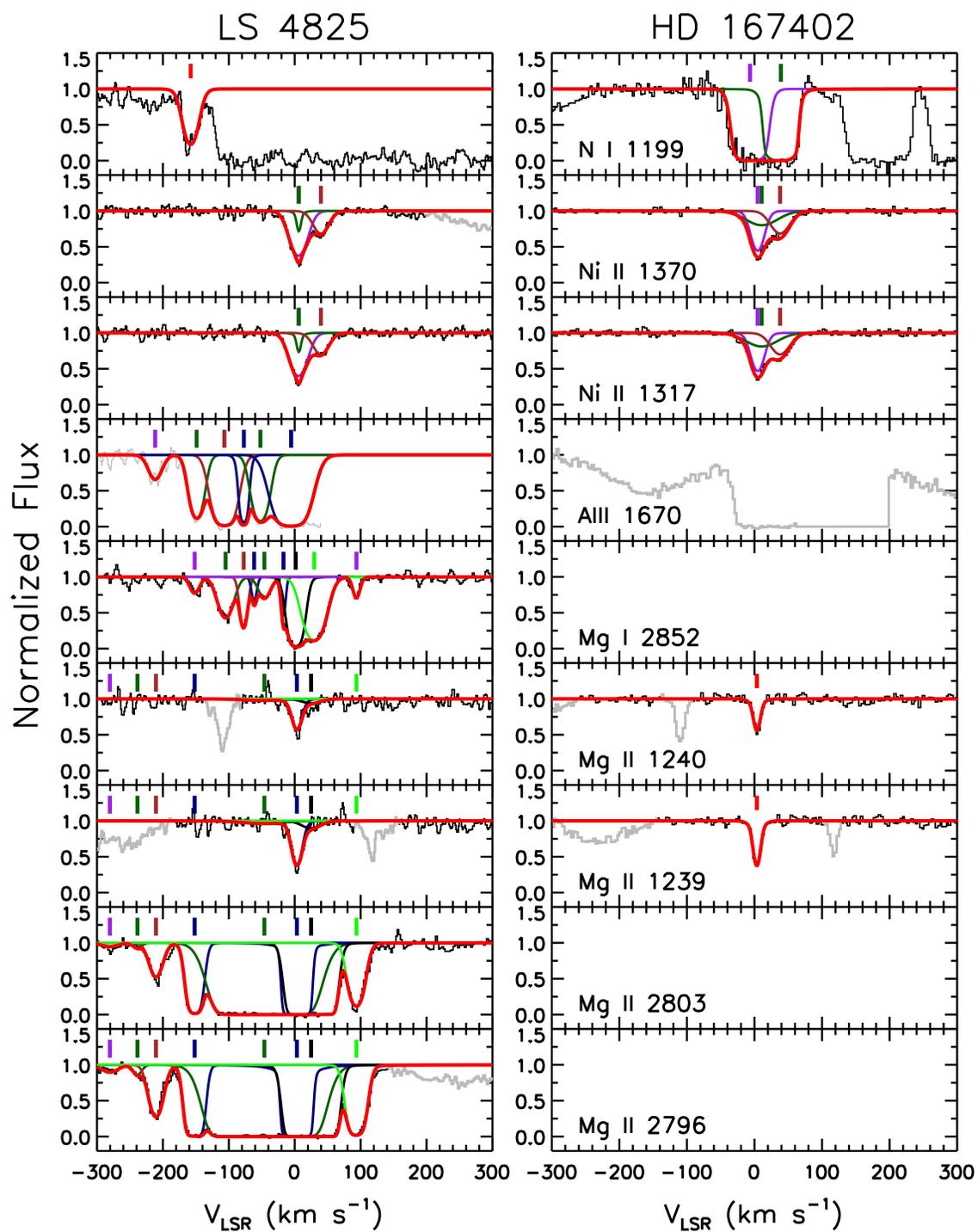

FIG. 14. Same as Figure 12 except the lines displayed are N I λ1199, Ni II λ1370, Ni II λ1317, Al II λ1670, Mg I λ2852, Mg II λ1240, Mg II λ1239, Mg II λ2803, and Mg II λ2796.

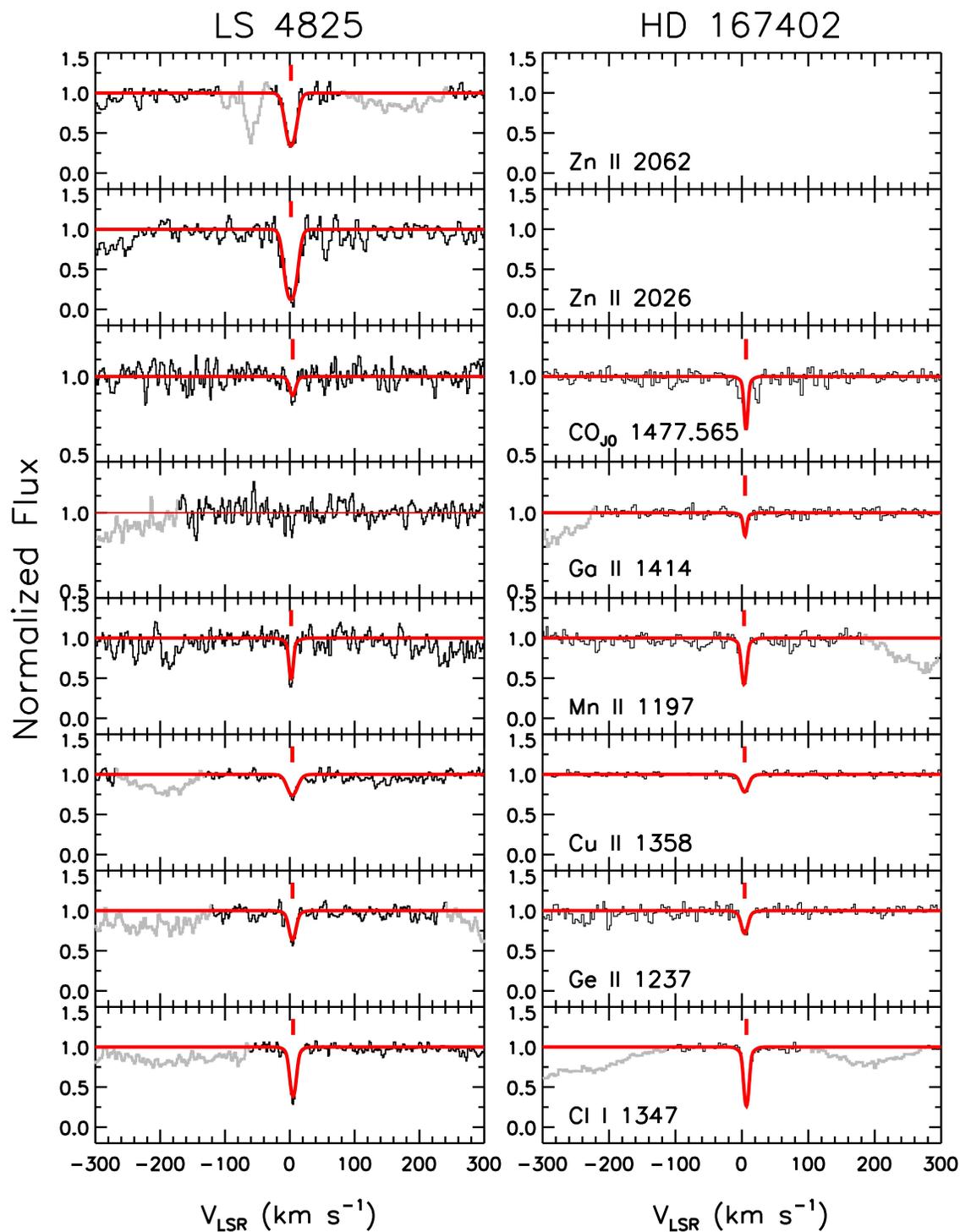

FIG. 15. Same as Figure 12 except the lines displayed are Zn II λ2062, Zn II λ2026, CO J = 0 λ1477.565, Ga II λ1414, Mn II λ1197, Cu II λ1358, Ge II λ1237, and Cl I λ1347.

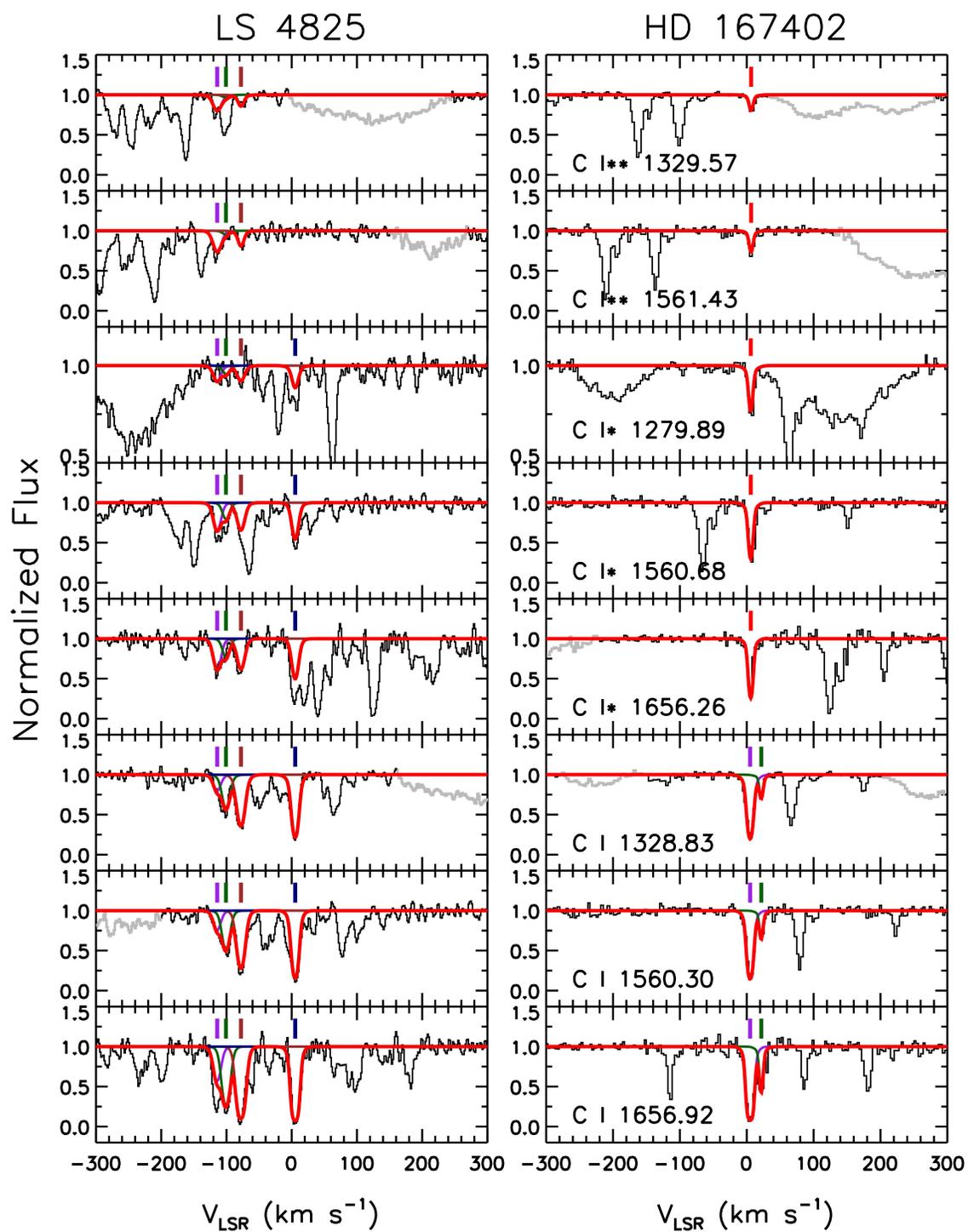

FIG. 16. Same as Figure 12 except the lines displayed are C I** λ1329.57, C I** λ1561.43, C I* λ1278,89, C I* λ1560.68, C I* λ1656.26, C I λ1328.83, C I λ1560.30, and C I λ1656.92.